\input harvmac.tex
\input epsf.tex
\input amssym
\input ulem.sty
\input graphicx.tex

\let\includefigures=\iftrue
\let\useblackboard=\iftrue
\newfam\black

\def\figin{\epsfcheck\figin}\def\figins{\epsfcheck\figins}
\def\epsfcheck{\ifx\epsfbox\UnDeFiNeD
\message{(NO epsf.tex, FIGURES WILL BE IGNORED)}
\gdef\figin##1{\vskip2in}\gdef\figins##1{\hskip.5in}
\else\message{(FIGURES WILL BE INCLUDED)}%
\gdef\figin##1{##1}\gdef\figins##1{##1}\fi}
\def\DefWarn#1{}
\def\figinsert{\goodbreak\midinsert}
\def\ifig#1#2#3{\DefWarn#1\xdef#1{fig.~\the\figno}
\writedef{#1\leftbracket fig.\noexpand~\the\figno} %
\figinsert\figin{\centerline{#3}}\medskip\centerline{\vbox{\baselineskip12pt
\advance\hsize by -1truein\noindent\footnotefont{\bf
Fig.~\the\figno:} #2}}
\bigskip\endinsert\global\advance\figno by1}


\includefigures
\message{If you do not have epsf.tex (to include figures),}
\message{change the option at the top of the tex file.}
\input epsf
\def\figin{\epsfcheck\figin}\def\figins{\epsfcheck\figins}
\def\epsfcheck{\ifx\epsfbox\UnDeFiNeD
\message{(NO epsf.tex, FIGURES WILL BE IGNORED)}
\gdef\figin##1{\vskip2in}\gdef\figins##1{\hskip.5in}
\else\message{(FIGURES WILL BE INCLUDED)}%
\gdef\figin##1{##1}\gdef\figins##1{##1}\fi}
\def\DefWarn#1{}
\def\figinsert{\goodbreak\midinsert}
\def\ifig#1#2#3{\DefWarn#1\xdef#1{fig.~\the\figno}
\writedef{#1\leftbracket fig.\noexpand~\the\figno}%
\figinsert\figin{\centerline{#3}}\medskip\centerline{\vbox{
\baselineskip12pt\advance\hsize by -1truein
\noindent\footnotefont{\bf Fig.~\the\figno:} #2}}
\endinsert\global\advance\figno by1}
\else
\def\ifig#1#2#3{\xdef#1{fig.~\the\figno}
\writedef{#1\leftbracket fig.\noexpand~\the\figno}%
\global\advance\figno by1} \fi

\def\figin{\epsfcheck\figin}\def\figins{\epsfcheck\figins}
\def\epsfcheck{\ifx\epsfbox\UnDeFiNeD
\message{(NO epsf.tex, FIGURES WILL BE IGNORED)}
\gdef\figin##1{\vskip2in}\gdef\figins##1{\hskip.5in}
\else\message{(FIGURES WILL BE INCLUDED)}%
\gdef\figin##1{##1}\gdef\figins##1{##1}\fi}
\def\DefWarn#1{}
\def\figinsert{\goodbreak\midinsert}
\def\ifig#1#2#3{\DefWarn#1\xdef#1{fig.~\the\figno}
\writedef{#1\leftbracket fig.\noexpand~\the\figno} %
\figinsert\figin{\centerline{#3}}\medskip\centerline{\vbox{\baselineskip12pt
\advance\hsize by -1truein\noindent\footnotefont{\bf
Fig.~\the\figno:} #2}}
\bigskip\endinsert\global\advance\figno by1}


\def\unlockat{\catcode`\@=11}
\def\lockat{\catcode`\@=12}

\unlockat

\def\newsec#1{\global\advance\secno by1\message{(\the\secno. #1)}
\global\subsecno=0\global\subsubsecno=0\eqnres@t\noindent
{\bf\the\secno. #1}
\writetoca{{\secsym} {#1}}\par\nobreak\medskip\nobreak}
\global\newcount\subsecno \global\subsecno=0
\def\subsec#1{\global\advance\subsecno
by1\message{(\secsym\the\subsecno. #1)}
\ifnum\lastpenalty>9000\else\bigbreak\fi\global\subsubsecno=0
\noindent{\it\secsym\the\subsecno. #1}
\writetoca{\string\quad {\secsym\the\subsecno.} {#1}}
\par\nobreak\medskip\nobreak}
\global\newcount\subsubsecno \global\subsubsecno=0
\def\subsubsec#1{\global\advance\subsubsecno by1
\message{(\secsym\the\subsecno.\the\subsubsecno. #1)}
\ifnum\lastpenalty>9000\else\bigbreak\fi
\noindent\quad{\secsym\the\subsecno.\the\subsubsecno.}{#1}
\writetoca{\string\qquad{\secsym\the\subsecno.\the\subsubsecno.}{#1}}
\par\nobreak\medskip\nobreak}

\def\subsubseclab#1{\DefWarn#1\xdef
#1{\noexpand\hyperref{}{subsubsection}%
{\secsym\the\subsecno.\the\subsubsecno}%
{\secsym\the\subsecno.\the\subsubsecno}}%
\writedef{#1\leftbracket#1}\wrlabeL{#1=#1}}
\lockat


\def\lzbarone{{\underrel{\approx}\over{\bar{z}\rightarrow 1}}}
\def\btoinf{{\underrel{\approx}\over{b\rightarrow \infty}}}
\def\lntoinf{{\underrel{\approx}\over{l,n\rightarrow \infty}}}

\def \pa {\partial}

\def\vev#1{\left\langle #1 \right\rangle}
\def\OO{{\cal OO}}
\def \tmu {\tilde{\mu}}
\def \lgb {\lambda_{\rm GB}}

\catcode`\@=11
\def\slash#1{\mathord{\mathpalette\c@ncel{#1}}}
\overfullrule=0pt

\def\BB{{\cal B}}

\def\GG{{\cal G}}

\def\II{{\cal I}}

\def\NN{{\cal N}}
\def\OO{{\cal O}}

\def\II{{\cal I}}

\def\underrel#1\over#2{\mathrel{\mathop{\kern\z@#1}\limits_{#2}}}

\catcode`\@=12


\def\vev#1{\left\langle #1 \right\rangle}

\def \sinh{{\rm sinh}}
\def \cosh{{\rm cosh}}


\def\zbar{{\bar z}}
\def\wbar{{\bar w}}

\def\p{{\partial}}

\def\tmu{{\tilde \mu}}

\def\ft{{\tilde f}}

\def\tmu{{\tilde{\mu}}}

\def\hbar{{\bar h}}

\def\DL{{\Delta_{L}}}
\def\DH{{\Delta_{H}}}
\def\DHL{{\Delta_{HL}}}
\def\OL{{\OO_{L}}}
\def\OH{{\OO_{H}}}


\lref\MaldacenaRE{
  J.~M.~Maldacena,
  ``The Large N limit of superconformal field theories and supergravity,''
Int.\ J.\ Theor.\ Phys.\  {\bf 38}, 1113 (1999), [Adv.\ Theor.\ Math.\ Phys.\  {\bf 2}, 231 (1998)].
[hep-th/9711200].
}
\lref\WittenQJ{
  E.~Witten,
  ``Anti-de Sitter space and holography,''
Adv.\ Theor.\ Math.\ Phys.\  {\bf 2}, 253 (1998).
[hep-th/9802150].
}
\lref\GubserBC{
  S.~S.~Gubser, I.~R.~Klebanov and A.~M.~Polyakov,
  ``Gauge theory correlators from noncritical string theory,''
Phys.\ Lett.\ B {\bf 428}, 105 (1998).
[hep-th/9802109].
}

\lref\WittenZW{
  E.~Witten,
  ``Anti-de Sitter space, thermal phase transition, and confinement in gauge theories,''
Adv.\ Theor.\ Math.\ Phys.\  {\bf 2}, 505 (1998).
[hep-th/9803131].
}

\lref\FerraraVZ{
  S.~Ferrara, A.~F.~Grillo, G.~Parisi and R.~Gatto,
  ``Covariant expansion of the conformal four-point function,''
Nucl.\ Phys.\ B {\bf 49}, 77 (1972), Erratum: [Nucl.\ Phys.\ B {\bf 53}, 643 (1973)]..
}

\lref\FerraraNF{
  S.~Ferrara, A.~F.~Grillo, R.~Gatto and G.~Parisi,
  ``Analyticity properties and asymptotic expansions of conformal covariant green's functions,''
Nuovo Cim.\ A {\bf 19}, 667 (1974)..
}

\lref\DobrevRU{
  V.~K.~Dobrev, V.~B.~Petkova, S.~G.~Petrova and I.~T.~Todorov,
  ``Dynamical Derivation of Vacuum Operator Product Expansion in Euclidean Conformal Quantum Field Theory,''
Phys.\ Rev.\ D {\bf 13}, 887 (1976)..
}

\lref\ArutyunovKU{
  G.~Arutyunov, S.~Frolov and A.~C.~Petkou,
  ``Operator product expansion of the lowest weight CPOs in $\NN=4$ SYM$_4$ at strong coupling,''
Nucl.\ Phys.\ B {\bf 586}, 547 (2000), Erratum: [Nucl.\ Phys.\ B {\bf 609}, 539 (2001)].
[hep-th/0005182].
}

\lref\DolanUT{
  F.~A.~Dolan and H.~Osborn,
  ``Conformal four point functions and the operator product expansion,''
Nucl.\ Phys.\ B {\bf 599}, 459 (2001).
[hep-th/0011040].
}

\lref\CostaDW{
  M.~S.~Costa, J.~Penedones, D.~Poland and S.~Rychkov,
  ``Spinning Conformal Blocks,''
JHEP {\bf 1111}, 154 (2011).
[arXiv:1109.6321 [hep-th]].
}

\lref\CostaMG{
  M.~S.~Costa, J.~Penedones, D.~Poland and S.~Rychkov,
  ``Spinning Conformal Correlators,''
JHEP {\bf 1111}, 071 (2011).
[arXiv:1107.3554 [hep-th]].
}


\lref\ZamolodchikovGT{
  A.~B.~Zamolodchikov,
  ``Irreversibility of the Flux of the Renormalization Group in a 2D Field Theory,''
JETP Lett.\  {\bf 43}, 730 (1986), [Pisma Zh.\ Eksp.\ Teor.\ Fiz.\  {\bf 43}, 565 (1986)]..
}

\lref\CardyCWA{
  J.~L.~Cardy,
  ``Is There a c Theorem in Four-Dimensions?,''
Phys.\ Lett.\ B {\bf 215}, 749 (1988)..
}

\lref\KomargodskiVJ{
  Z.~Komargodski and A.~Schwimmer,
  ``On Renormalization Group Flows in Four Dimensions,''
JHEP {\bf 1112}, 099 (2011).
[arXiv:1107.3987 [hep-th]].
}



\lref\HofmanAR{
  D.~M.~Hofman and J.~Maldacena,
  ``Conformal collider physics: Energy and charge correlations,''
JHEP {\bf 0805}, 012 (2008).
[arXiv:0803.1467 [hep-th]].
}

\lref\HofmanAWC{
  D.~M.~Hofman, D.~Li, D.~Meltzer, D.~Poland and F.~Rejon-Barrera,
  ``A Proof of the Conformal Collider Bounds,''
JHEP {\bf 1606}, 111 (2016).
[arXiv:1603.03771 [hep-th]].
}

\lref\HartmanLFA{
  T.~Hartman, S.~Jain and S.~Kundu,
  ``Causality Constraints in Conformal Field Theory,''
JHEP {\bf 1605}, 099 (2016).
[arXiv:1509.00014 [hep-th]].
}

\lref\LiITL{
  D.~Li, D.~Meltzer and D.~Poland,
  ``Conformal Collider Physics from the Lightcone Bootstrap,''
JHEP {\bf 1602}, 143 (2016).
[arXiv:1511.08025 [hep-th]].
}

\lref\HartmanDXC{
  T.~Hartman, S.~Jain and S.~Kundu,
  ``A New Spin on Causality Constraints,''
JHEP {\bf 1610}, 141 (2016).
[arXiv:1601.07904 [hep-th]].
}

\lref\FaulknerMZT{
  T.~Faulkner, R.~G.~Leigh, O.~Parrikar and H.~Wang,
  ``Modular Hamiltonians for Deformed Half-Spaces and the Averaged Null Energy Condition,''
JHEP {\bf 1609}, 038 (2016).
[arXiv:1605.08072 [hep-th]].
}

\lref\KomargodskiGCI{
  Z.~Komargodski, M.~Kulaxizi, A.~Parnachev and A.~Zhiboedov,
  ``Conformal Field Theories and Deep Inelastic Scattering,''
Phys.\ Rev.\ D {\bf 95}, no. 6, 065011 (2017).
[arXiv:1601.05453 [hep-th]].
}

\lref\KologluMFZ{
  M.~Kologlu, P.~Kravchuk, D.~Simmons-Duffin and A.~Zhiboedov,
  ``The light-ray OPE and conformal colliders,''
[arXiv:1905.01311 [hep-th]].
}


\lref\FitzpatrickEFK{
  A.~L.~Fitzpatrick, K.~W.~Huang and D.~Li,
  ``Probing universalities in d > 2 CFTs: from black holes to shockwaves,''
JHEP {\bf 1911}, 139 (2019).
[arXiv:1907.10810 [hep-th]].
}

\lref\CaronHuotVEP{
  S.~Caron-Huot,
  ``Analyticity in Spin in Conformal Theories,''
JHEP {\bf 1709}, 078 (2017).
[arXiv:1703.00278 [hep-th]].
}

\lref\SimmonsDuffinNUB{
  D.~Simmons-Duffin, D.~Stanford and E.~Witten,
  ``A spacetime derivation of the Lorentzian OPE inversion formula,''
JHEP {\bf 1807}, 085 (2018).
[arXiv:1711.03816 [hep-th]].
}

\lref\LiZBA{
  Y.~Z.~Li,
  ``Heavy-light Bootstrap from Lorentzian Inversion Formula,''
[arXiv:1910.06357 [hep-th]].
}


\lref\RattazziPE{
  R.~Rattazzi, V.~S.~Rychkov, E.~Tonni and A.~Vichi,
  ``Bounding scalar operator dimensions in 4D CFT,''
JHEP {\bf 0812}, 031 (2008).
[arXiv:0807.0004 [hep-th]].
}
\lref\KulaxiziTKD{
  M.~Kulaxizi, G.~S.~Ng and A.~Parnachev,
  ``Subleading Eikonal, AdS/CFT and Double Stress Tensors,''
JHEP {\bf 1910}, 107 (2019).
[arXiv:1907.00867 [hep-th]].
}
\lref\FitzpatrickZQZ{
  A.~L.~Fitzpatrick and K.~W.~Huang,
  ``Universal Lowest-Twist in CFTs from Holography,''
[arXiv:1903.05306 [hep-th]].
}
\lref\KulaxiziDXO{
  M.~Kulaxizi, G.~S.~Ng and A.~Parnachev,
  ``Black Holes, Heavy States, Phase Shift and Anomalous Dimensions,''
SciPost Phys.\  {\bf 6}, no. 6, 065 (2019).
[arXiv:1812.03120 [hep-th]].
}
\lref\KarlssonQFI{
  R.~Karlsson, M.~Kulaxizi, A.~Parnachev and P.~Tadi\'c,
  ``Black Holes and Conformal Regge Bootstrap,''
JHEP {\bf 1910}, 046 (2019).
[arXiv:1904.00060 [hep-th]].
}
\lref\BuchelSK{
  A.~Buchel, J.~Escobedo, R.~C.~Myers, M.~F.~Paulos, A.~Sinha and M.~Smolkin,
  ``Holographic GB gravity in arbitrary dimensions,''
  JHEP {\bf 1003}, 111 (2010).
  [arXiv:0911.4257 [hep-th]].
}
\lref\KarlssonDBD{
  R.~Karlsson, M.~Kulaxizi, A.~Parnachev and P.~Tadi\' c,
  ``Leading Multi-Stress Tensors and Conformal Bootstrap,''
JHEP {\bf 2001}, 076 (2020).
[arXiv:1909.05775 [hep-th]].
}
\lref\FitzpatrickYX{
  A.~L.~Fitzpatrick, J.~Kaplan, D.~Poland and D.~Simmons-Duffin,
  ``The Analytic Bootstrap and AdS Superhorizon Locality,''
JHEP {\bf 1312}, 004 (2013).
[arXiv:1212.3616 [hep-th]].
}
\lref\KomargodskiEK{
  Z.~Komargodski and A.~Zhiboedov,
  ``Convexity and Liberation at Large Spin,''
JHEP {\bf 1311}, 140 (2013).
[arXiv:1212.4103 [hep-th]].
}
\lref\FitzpatrickDM{
  A.~L.~Fitzpatrick and J.~Kaplan,
  ``Unitarity and the Holographic S-Matrix,''
JHEP {\bf 1210}, 032 (2012).
[arXiv:1112.4845 [hep-th]].
}

\lref\HuotLIF{
  S.~Caron-Huot,
  ``Analyticity in Spin in Conformal Theories,''
[arXiv:1703.00278 [hep-th]].
}
\lref\WittenLIF{
  D.~Simmons-Duffin, D.~Stanford, E.~Witten,
  ``A spacetime derivation of the Lorentzian OPE inversion formula,''
[arXiv:1711.03816 [hep-th]].
}
\lref\ElShowkHT{
  S.~El-Showk, M.~F.~Paulos, D.~Poland, S.~Rychkov, D.~Simmons-Duffin and A.~Vichi,
  ``Solving the 3D Ising Model with the Conformal Bootstrap,''
Phys.\ Rev.\ D {\bf 86}, 025022 (2012).
[arXiv:1203.6064 [hep-th]].
}

\lref\MaldacenaRE{
  J.~M.~Maldacena,
  ``The Large N limit of superconformal field theories and supergravity,''
Int.\ J.\ Theor.\ Phys.\  {\bf 38}, 1113 (1999), [Adv.\ Theor.\ Math.\ Phys.\  {\bf 2}, 231 (1998)].
[hep-th/9711200].
}
\lref\WittenQJ{
  E.~Witten,
  ``Anti-de Sitter space and holography,''
Adv.\ Theor.\ Math.\ Phys.\  {\bf 2}, 253 (1998).
[hep-th/9802150].
}
\lref\GubserBC{
  S.~S.~Gubser, I.~R.~Klebanov and A.~M.~Polyakov,
  ``Gauge theory correlators from noncritical string theory,''
Phys.\ Lett.\ B {\bf 428}, 105 (1998).
[hep-th/9802109].
}
\lref\HeemskerkPN{
  I.~Heemskerk, J.~Penedones, J.~Polchinski and J.~Sully,
  ``Holography from Conformal Field Theory,''
JHEP {\bf 0910}, 079 (2009).
[arXiv:0907.0151 [hep-th]].
}
\lref\FerraraYT{
  S.~Ferrara, A.~F.~Grillo and R.~Gatto,
  ``Tensor representations of conformal algebra and conformally covariant operator product expansion,''
Annals Phys.\  {\bf 76}, 161 (1973)..
}
\lref\PolyakovGS{
  A.~M.~Polyakov,
  ``Nonhamiltonian approach to conformal quantum field theory,''
Zh.\ Eksp.\ Teor.\ Fiz.\  {\bf 66}, 23 (1974), [Sov.\ Phys.\ JETP {\bf 39}, 9 (1974)]..
}
\lref\RattazziPE{
  R.~Rattazzi, V.~S.~Rychkov, E.~Tonni and A.~Vichi,
  ``Bounding scalar operator dimensions in 4D CFT,''
JHEP {\bf 0812}, 031 (2008).
[arXiv:0807.0004 [hep-th]].
}
\lref\ElShowkHT{
  S.~El-Showk, M.~F.~Paulos, D.~Poland, S.~Rychkov, D.~Simmons-Duffin and A.~Vichi,
  ``Solving the 3D Ising Model with the Conformal Bootstrap,''
Phys.\ Rev.\ D {\bf 86}, 025022 (2012).
[arXiv:1203.6064 [hep-th]].
}
\lref\ElShowkHT{
  S.~El-Showk, M.~F.~Paulos, D.~Poland, S.~Rychkov, D.~Simmons-Duffin and A.~Vichi,
  ``Solving the 3D Ising Model with the Conformal Bootstrap,''
Phys.\ Rev.\ D {\bf 86}, 025022 (2012).
[arXiv:1203.6064 [hep-th]].
}
\lref\PolyakovS{
   A.~M.~Polyakov,
  ``Conformal  symmetry  of critical fluctuation,''
JETP\ Lett. {\bf 12}, 381-383 (1970).
}
\lref\KarlssonTXU{
  R.~Karlsson,
  ``Multi-stress tensors and next-to-leading singularities in the Regge limit,''
[arXiv:1912.01577 [hep-th]].
}


\lref\CornalbaXK{
  L.~Cornalba, M.~S.~Costa, J.~Penedones and R.~Schiappa,
  ``Eikonal Approximation in AdS/CFT: From Shock Waves to Four-Point Functions,''
JHEP {\bf 0708}, 019 (2007).
[hep-th/0611122].
}

\lref\CornalbaXM{
  L.~Cornalba, M.~S.~Costa, J.~Penedones and R.~Schiappa,
  ``Eikonal Approximation in AdS/CFT: Conformal Partial Waves and Finite N Four-Point Functions,''
Nucl.\ Phys.\ B {\bf 767}, 327 (2007).
[hep-th/0611123].
}

\lref\CornalbaZB{
  L.~Cornalba, M.~S.~Costa and J.~Penedones,
  ``Eikonal approximation in AdS/CFT: Resumming the gravitational loop expansion,''
JHEP {\bf 0709}, 037 (2007).
[arXiv:0707.0120 [hep-th]].
}

\lref\CornalbaFS{
  L.~Cornalba,
  ``Eikonal methods in AdS/CFT: Regge theory and multi-reggeon exchange,''
[arXiv:0710.5480 [hep-th]].
}

\lref\CostaCB{
  M.~S.~Costa, V.~Goncalves and J.~Penedones,
  ``Conformal Regge theory,''
JHEP {\bf 1212}, 091 (2012).
[arXiv:1209.4355 [hep-th]].
}

\lref\KulaxiziIXA{
  M.~Kulaxizi, A.~Parnachev and A.~Zhiboedov,
  ``Bulk Phase Shift, CFT Regge Limit and Einstein Gravity,''
JHEP {\bf 1806}, 121 (2018).
[arXiv:1705.02934 [hep-th]].
}

\lref\CostaTWZ{
  M.~S.~Costa, T.~Hansen and J.~Penedones,
  ``Bounds for OPE coefficients on the Regge trajectory,''
JHEP {\bf 1710}, 197 (2017).
[arXiv:1707.07689 [hep-th]].
}



\lref\BoulwareWK{
  D.~G.~Boulware and S.~Deser,
  ``String Generated Gravity Models,''
Phys.\ Rev.\ Lett.\  {\bf 55}, 2656 (1985).
}

\lref\CaiDZ{
  R.~G.~Cai,
  ``Gauss-Bonnet black holes in AdS spaces,''
Phys.\ Rev.\ D {\bf 65}, 084014 (2002).
[hep-th/0109133].
}


\lref\KologluBCO{
  M.~Kologlu, P.~Kravchuk, D.~Simmons-Duffin and A.~Zhiboedov,
  ``Shocks, Superconvergence, and a Stringy Equivalence Principle,''
[arXiv:1904.05905 [hep-th]].
}
\lref\KologluMFZ{
  M.~Kologlu, P.~Kravchuk, D.~Simmons-Duffin and A.~Zhiboedov,
  ``The light-ray OPE and conformal colliders,''
[arXiv:1905.01311 [hep-th]].
}
\lref\KravchukHTV{
  P.~Kravchuk and D.~Simmons-Duffin,
  ``Light-ray operators in conformal field theory,''
JHEP {\bf 1811}, 102 (2018).
[arXiv:1805.00098 [hep-th]].
}
\lref\DolanHV{
  F.~A.~Dolan and H.~Osborn,
  ``Conformal partial waves and the operator product expansion,''
Nucl.\ Phys.\ B {\bf 678}, 491 (2004).
[hep-th/0309180].
}
\lref\MeltzerPYL{
  D.~Meltzer,
  ``AdS/CFT Unitarity at Higher Loops: High-Energy String Scattering,''
[arXiv:1912.05580 [hep-th]].
}


\lref\CarmiCUB{
  D.~Carmi and S.~Caron-Huot,
  ``A Conformal Dispersion Relation: Correlations from Absorption,''
[arXiv:1910.12123 [hep-th]].
}

\lref\BissiKKX{
  A.~Bissi, P.~Dey and T.~Hansen,
  ``Dispersion Relation for CFT Four-Point Functions,''
[arXiv:1910.04661 [hep-th]].
}

\lref\LiTPF{
  Y.~Z.~Li, Z.~F.~Mai and H.~Lü,
  ``Holographic OPE Coefficients from AdS Black Holes with Matters,''
JHEP {\bf 1909}, 001 (2019).
[arXiv:1905.09302 [hep-th]].
}

\lref\HuangYCS{
  K.~W.~Huang,
  ``A Lightcone Commutator and Stress-Tensor Exchange in d=4 CFTs,''
[arXiv:2002.00110 [hep-th]].
}

\lref\FitzpatrickFOA{
  A.~L.~Fitzpatrick, J.~Kaplan, M.~T.~Walters and J.~Wang,
  ``Hawking from Catalan,''
JHEP {\bf 1605}, 069 (2016).
[arXiv:1510.00014 [hep-th]].
}

\lref\BianchiDES{
  M.~Bianchi, A.~Grillo and F.~Morales,
  ``Chaos at the rim of black hole and fuzzball shadows,''
[arXiv:2002.05574 [hep-th]].
}

\lref\BernGJJ{
  Z.~Bern, H.~Ita, J.~Parra-Martinez and M.~S.~Ruf,
  ``Universality in the classical limit of massless gravitational scattering,''
[arXiv:2002.02459 [hep-th]].
}

\lref\BrowerEA{
  R.~C.~Brower, J.~Polchinski, M.~J.~Strassler and C.~I.~Tan,
  ``The Pomeron and gauge/string duality,''
JHEP {\bf 0712}, 005 (2007).
[hep-th/0603115].
}

\lref\FitzpatrickZHA{
  A.~L.~Fitzpatrick, J.~Kaplan and M.~T.~Walters,
  ``Virasoro Conformal Blocks and Thermality from Classical Background Fields,''
JHEP {\bf 1511}, 200 (2015).
[arXiv:1501.05315 [hep-th]].
}

\lref\FitzpatrickVUA{
  A.~L.~Fitzpatrick, J.~Kaplan and M.~T.~Walters,
  ``Universality of Long-Distance AdS Physics from the CFT Bootstrap,''
JHEP {\bf 1408}, 145 (2014).
[arXiv:1403.6829 [hep-th]].
}

\lref\FitzpatrickHFC{
  A.~L.~Fitzpatrick, J.~Kaplan, M.~T.~Walters and J.~Wang,
  ``Hawking from Catalan,''
JHEP {\bf 1605}, 069 (2016).
[arXiv:1510.00014 [hep-th]].
}

\lref\HijanoRLA{
  E.~Hijano, P.~Kraus and R.~Snively,
  ``Worldline approach to semi-classical conformal blocks,''
JHEP {\bf 1507}, 131 (2015).
[arXiv:1501.02260 [hep-th]].
}

\lref\HijanoQJA{
  E.~Hijano, P.~Kraus, E.~Perlmutter and R.~Snively,
  ``Semiclassical Virasoro blocks from AdS$_{3}$ gravity,''
JHEP {\bf 1512}, 077 (2015).
[arXiv:1508.04987 [hep-th]].
}

\lref\CollierEXN{
  S.~Collier, Y.~Gobeil, H.~Maxfield and E.~Perlmutter,
  ``Quantum Regge Trajectories and the Virasoro Analytic Bootstrap,''
JHEP {\bf 1905}, 212 (2019).
[arXiv:1811.05710 [hep-th]].
}

\lref\BeskenJYW{
  M.~Beşken, S.~Datta and P.~Kraus,
  ``Semi-classical Virasoro blocks: proof of exponentiation,''
JHEP {\bf 2001}, 109 (2020).
[arXiv:1910.04169 [hep-th]].
}

\lref\LiTPF{
  Y.~Z.~Li, Z.~F.~Mai and H.~Lü,
  ``Holographic OPE Coefficients from AdS Black Holes with Matters,''
JHEP {\bf 1909}, 001 (2019).
[arXiv:1905.09302 [hep-th]].
}
\lref\AldayQRF{
  L.~F.~Alday and E.~Perlmutter,
  ``Growing Extra Dimensions in AdS/CFT,''
JHEP {\bf 1908}, 084 (2019).
[arXiv:1906.01477 [hep-th]].
}


\Title{
\vbox{\baselineskip8pt
}}
{\vbox{
\centerline{Stress Tensor Sector of Conformal Correlators }
}}

\vskip.1in
 \centerline{
Robin Karlsson, Manuela Kulaxizi, Andrei Parnachev and Petar Tadi\' c \footnote{}{karlsson, manuela, parnachev, tadicp $@$ maths.tcd.ie  } } \vskip.1in
\centerline{\it 
School of Mathematics, Trinity College Dublin, Dublin 2, Ireland}

\vskip.7in \centerline{\bf Abstract}{
\vskip.2in 
\noindent An important part of a CFT four-point function, the stress tensor sector,  comprises the exchanges of the stress tensor and its composites.
The OPE coefficients of these multi-stress tensor operators and consequently, the
complete stress tensor sector of four-point functions in CFTs with a large central charge, can be determined by computing a heavy-heavy-light-light correlator.
We show  how one can make substantial  progress in this direction
by  bootstrapping a certain ansatz for the stress tensor sector of the correlator, iteratively computing the OPE coefficients 
of multi-stress tensor operators with increasing twist.
Some parameters are not fixed by the bootstrap -- they correspond  
to the OPE coefficients of multi-stress tensors with spin zero and two.
We further show that  in holographic CFTs  one can use the phase shift computed in the dual gravitational theory to reduce 
the set of undetermined parameters to the OPE coefficients of multi-stress tensors with spin zero.
Finally, we verify some of these results using the Lorentzian OPE inversion formula and comment on its regime of applicability.
}

\Date{February 2020}

\listtoc\writetoc
\vskip 1.57in \noindent

\eject

\newsec{Introduction and Summary}

\subsec{Introduction}

Conformal field theories (CFTs) are the harmonic oscillators of our times; besides being significantly more amenable to analytic study compared to generic quantum field theories, they also provide a non-perturbative definition of gravity in negatively curved spacetimes via the AdS/CFT correspondence \refs{\MaldacenaRE\WittenQJ-\GubserBC}. 
Their robust structure bears many important consequences which have come to light in recent years due to the development
 of conformal
 bootstrap techniques following \refs{\FerraraYT\PolyakovGS\RattazziPE-\ElShowkHT}. This is especially pronounced in spacetime dimension $d>2$ which this article is focused upon.  

Conformal symmetry imposes highly non-trivial constraints on the theory. Two- and three-point correlation functions are fixed up to a handful of position-independent parameters \PolyakovS. Four- and higher-point functions \refs{\FerraraVZ\FerraraNF-\DobrevRU} are determined as long as the CFT spectrum of local operators and the respective OPE coefficients are known (for recent techniques see the original works of \refs{\ArutyunovKU-\DolanUT} and the modern approach developed in \refs{\CostaDW-\CostaMG}). 

While computing four-point correlation functions is possible in principle, the amount of necessary data makes it difficult in practice. Consistency principles, such as crossing symmetry and unitarity, come to rescue. In fact, the idea of the conformal bootstrap programme is to use these consistency conditions to place restrictions on the CFT data (spectrum of operators and OPE coefficients) and, if possible, solve the theory completely. 

One way to make use of crossing symmetry is to consider kinematic regimes which enhance the contribution of a limited number of operators in a given channel, and are typically reproduced by an infinite number of operators in another channel. A standard example is the lightcone limit where the initially spacelike separation between two operators is allowed to become null. Focusing on the lightcone limit of a four-point correlation function allows one to deduce the existence of double-twist operators at large spin in any CFT in dimensions $d>2$ \refs{\FitzpatrickYX-\KomargodskiEK}. 


A natural assumption when considering an arbitrary CFT is the existence of a stress tensor. The two-point function of the stress-tensor depends on a single parameter, the central charge $C_T$, which serves as a rough measure of the number of degrees of freedom in the theory. In this paper, we will consider local CFTs with a large number of degrees of freedom, a.k.a. large central charge $C_{T}\gg 1$.

Specifically, our goal herein is to  study the contribution of the stress-tensor sector in scalar CFT correlation functions, $\vev{\OO_1\OO_1\OO_2\OO_2}$. What we mean here by the ``stress-tensor sector'' is the set of  operators composed out of stress-tensors and derivatives\foot{The identity operator is considered as the first trivial entry of the stress-tensor sector.}, schematically denoted by $:T_{\mu_1\nu_1}\cdots T_{\mu_{p-1}\nu_{p-1}}\p^{2n} \p_{\lambda_1}\cdots\p_{\lambda_{q}}T_{\mu_{p}\nu_p}:$. 
Such operators are present in large $C_T$ CFTs, but their contribution to a correlation function is of particular interest in CFTs with holographic duals since it is related to the contribution of multiple gravitons in the corresponding Witten diagrams.

We consider the four-point function $\langle\OO_{H}\OO_{L}\OO_{L}\OO_{H}\rangle$ of two pairwise identical scalar operators labeled as ``light, L'', and ``heavy, H'', depending on whether their conformal dimension scales with the number of degrees of freedom, $\DH\propto\OO(C_T)$,  or not, $\DL \propto \OO(1)$.  The reason this correlator is well-suited to the exploration of the stress-tensor sector is the presence of an additional parameter, $\mu$, proportional to the ratio of the conformal dimension of the heavy operators with the central charge, $\mu \propto \DH/C_{T}$.  This parameter naturally counts the number of stress-tensors in a composite multi-stress tensor operator. To distinguish the contribution of such operators from the full HHLL correlator in what follows we will denote it as $\GG(z,\zbar)$, {\it i.e.},
\eqn\sts{\GG(z,\zbar)=\langle\OO_{H}(\infty)\OO_{L}(1)\OO_{L}(z,\zbar)\OO_{H}(0)\rangle\Big|_{{\rm multi-stress \ tensors}}.
}  
%
Note that from $\GG(z,\zbar)$ in  \sts\ one can read off the OPE coefficients of multi-stress tensor operators to leading order in $1/C_T$ but
exact in $\Delta_L$.

The HHLL correlator is interesting in its own right. 
In the limit of a large number of degrees of freedom, it is related to the thermal two-point function $\vev{\OO_L\OO_L}_T$ - as long as the average energy of the canonical ensemble is roughly equal to the conformal dimension of the heavy operator. 
When the CFT is additionally characterised by an infinite gap, $\Delta_{gap}\to\infty$, in the spectrum of primary single-trace (non-composite) operators with spin greater than two, the situation is even more interesting. In this case, the theory has an equivalent description in terms of a classical, local gravitational theory in AdS \HeemskerkPN. Such a CFT is called holographic as a minimally defined realisation of the holographic paradigm. 
When a holographic CFT is considered at finite temperature, the appropriate gravitational description is that of an asymptotically AdS black hole  \WittenZW . In this case, the HHLL correlator, in a certain kinematical regime, 
 is expected to describe the scattering of a light particle by the black hole in the dual gravitational theory
\KulaxiziDXO.

To study the stress tensor sector of the HHLL correlator we will employ crossing symmetry and the conformal bootstrap. Specifically, we consider the lightcone limit where the separation between the two $\OO_L$ operators is close to being null. In this limit, the dominant contribution in the direct channel (T-channel, where the pairwise identical operators approach each other) is coming from multi-stress tensor operators with low twist (where the twist $\tau$ is the difference between the conformal dimension $\Delta$ and the spin $s$ of a given operator, $\tau=\Delta-s$). 
In the cross-channel (S-channel), an infinite number of double-twist operators of the schematic form $:\OO_{H}\pa_{\mu_{1}}\ldots\pa_{\mu_{l}}\pa^{2n}\OO_{L}:$ with $l\gg 1$ should be considered. 

In \FitzpatrickZQZ, it was argued through a holographic calculation that the OPE coefficients of minimal-twist multi-stress tensors are ``universal'' in the sense that they are completely fixed in terms of just two CFT parameters: $\Delta_L$ and $1\over C_T$
(see also \LiTPF). 
 In \KulaxiziTKD , a formula for the OPE coefficients of the minimal twist double-stress tensors was written. In \KarlssonDBD , it was shown how one can, at least in principle, evaluate the contribution of the stress tensor sector to all orders in $\mu$ in arbitrary even number of spacetime dimensions $d$ in the lightcone limit. The strategy there was based on proposing an ansatz for $\GG$ with a few undetermined parameters and then fixing these parameters by means of the lightcone bootstrap. In the process, one can  extract the OPE coefficients of all multi-stress tensors with minimal twist. 
 A different approach based on the Lorentzian inversion formula \refs{\CaronHuotVEP-\SimmonsDuffinNUB} for extracting the minimal-twist double- and triple-stress tensor OPE coefficients was used in \LiZBA\foot{One should exercise caution when using the Lorentzian inversion formula in the context of the HHLL correlator as the Regge behaviour of the correlator has not been rigorously established.} and also appears to confirm the universality of the minimal-twist stress tensor sector.

In this paper, we investigate the stress tensor sector further by considering contributions from multi-stress tensors with non-minimal twist. Our goal is to determine the structure of the correlator to subleading orders in the lightcone limit and extract the relevant OPE coefficients. Once more, we motivate an ansatz similar to the one successfully describing the leading lightcone behavior of $\GG(z, \bar z)$  and show that most of the parameters in the ansatz can be fixed using lightcone bootstrap. A few parameters are, however, left undetermined and
might depend on the details of the theory. They correspond to the OPE coefficients of multi-stress tensors with spin $s=0,2$. 
Our approach can be employed to study the stress-tensor sector to arbitrary orders in $\mu$ and $(1-\zbar)$. 
In this paper, we completed this program for the $\OO(\mu^2)$ subleading, subsubleading and subsubsubleading terms as well as the $\OO(\mu^3)$ subleading and subsubleading terms.

We also investigate a complementary approach to computing the OPE data of the stress tensor sector using the Lorentzian inversion formula. As noted earlier, the validity of the Lorentzian inversion formula for the HHLL correlator has not been rigorously established. It is however natural to expect that it is applicable in the large-$C_T$ and small-$\mu$ expansion, as long as a Regge bound is observed. Here we assume that the Regge behavior of the correlator is given by $\sigma^{-k}$ at $\OO(\mu^k)$ in the large-$C_T$ limit, which is consistent with the behaviour of the scattering phase shift from a black hole (or a massive star) computed classically in AdS. We then find that whenever the Lorentzian inversion formula is applicable, {\it i.e.}, for operators of spin $s>k+1$ at $\OO(\mu^k)$, OPE data extracted with both methods are in perfect agreement. However, already at order $\OO(\mu^{3})$, our ansatz combined with the crossing symmetry or Lorentzian inversion formula is more powerful than the Lorentzian inversion formula alone. For instance, while the former procedure allows us to determine the OPE coefficient of a triple-stress tensor with spin $s=4$ and twist $\tau=8$, this is not possible using solely the Lorentzian inversion formula.

Finally, we explore the possibility of obtaining the unknown OPE data from the gravitational description of the CFT. 
We use the phase shift calculation in the dual gravitational theory. The scattering phase shift -- acquired by a highly energetic particle travelling 
in the background of the AdS black hole -- was first computed in the Regge limit in Einstein gravity in \KulaxiziDXO. To explicitly see how the presence of higher derivative gravitational terms affects the OPE data, we work in 
Einstein-Hilbert + Gauss-Bonnet gravity with small Gauss-Bonnet coupling $\lgb$. To combine the gravitational results with those of the CFT in the lightcone regime, we follow the approach first discussed in \KulaxiziTKD\ and further developed in \KarlssonDBD, which involves an analytic continuation of the lightcone results around $z=0$ and an expansion around $z=1$. Matching terms in the correlator obtained from the gravitational calculation to those obtained from the CFT enables us to 
completely fix the stress tensor sector of the HHLL correlator up to the OPE coefficients of  the spin-0 multi-stress tensors which are left undetermined. 
Non-universality is manifest by the presence of the Gauss-Bonnet coupling in the expressions for the OPE coefficients.

\subsec{Summary of results.}

In this paper, we show that the stress tensor sector of the HHLL correlator 
in $d=4$ can be written in terms of products of $f_{a}(z)$ functions defined as
\eqn\definicija{f_{a}(z)=(1-z)^{a}{}_{2}F_{1}(a,a,2a,1-z).
}

The stress tensor sector of the HHLL correlator can be expanded in powers of $\mu$ and then in powers of $(1-\zbar)$ as
\eqn\expn{\GG(z, \zbar)=\sum_{k=0}^{\infty}\mu^{k}\GG^{(k)}(z,\zbar)={{1}\over{((1-z)(1-\zbar))^{\DL}}}+\sum_{k=1}^{\infty}\sum_{m=0}^{\infty}\mu^{k}(1-\zbar)^{-\DL + k +m} \GG^{(k,m)}(z),
}
\noindent where we have explicitly separated the contribution of the identity operator.\foot{The contribution of the identity operator is denoted with $k=0$.} We explain how one can write $\GG^{(k,m)}(z)$ for arbitrary $k$ and $m$.

We write an ansatz for each $\GG^{(k,m)}(z)$ with a few unknown coefficients and fix all, but a handful of them, via lightcone bootstrap. 
The undetermined coefficients correspond to the OPE coefficients of spin-0 and spin-2 exchanged operators. 
We further show that  in holographic CFTs  one can use the phase shift computed in the dual gravitational theory to reduce 
the set of undetermined parameters to the OPE coefficients of multi-stress tensors with spin zero.

Operators of non-minimal twist give a subleading contribution in the lightcone limit, $1-\zbar\ll 1$, which can be expressed as a sum of products of the functions $f_a(z)$
(times an appropriate power of $(1-\bar z)$ ). This form is similar to the contribution of minimal-twist multi-stress tensor operators considered in \KarlssonDBD. While our method can be used to address the contribution of operators of arbitrary twist, here we focus on determining the specific contributions of operators with twist $\tau=6,8,10$, at $\OO(\mu^2)$ and $\tau=8,10$, at $\OO(\mu^{3})$.

At $\OO(\mu)$, the only operator that contributes to the stress tensor sector of the correlator is the stress tensor and its contribution is completely fixed by conformal symmetry. In $d=4$ its exact (to all orders in $\zbar$) contribution is given by
\eqn\ordermusumry{
    \GG^{(1)}(z,\zbar) = {{1}\over{[(1-z)(1-\zbar)]^{\DL-1}}}{\DL\over {120(\zbar -z)}}\Big(f_{3}(z)-f_{3}(\zbar)\Big).
    }

At $\OO(\mu^{2})$, the leading contribution in the lightcone limit, due to twist-four double-stress tensors, was evaluated in \KulaxiziTKD
\eqn\secondCorrsumry{\eqalign{
  \GG^{(2,0)}(z) = &{1\over (1-z)^{\DL}}\left({\DL\over 28800(\DL-2)}\right)\times\cr
      \Big[&(\DL-4)(\DL-3)f_3^2(z)+{15\over 7}(\DL-8)f_2(z)f_4(z)+{40\over 7}(\DL+1)f_1(z)f_5(z)\Big].
}}

 We show that the subleading contribution in the lightcone limit, due to twist-four and twist-six double-stress tensors, is given by
\eqn\subgsumry{\eqalign{
  \GG^{(2,1)}(z) = {1\over (1-z)^{\DL}}\Big[\left({{3-z}\over {2(1-z)}}\right)\left(a_{33}f_{3}(z)^{2}+a_{24}f_{2}(z)f_{4}(z)+a_{15}f_{1}(z)f_{5}(z)\right) \cr +\left(b_{14}f_{1}(z)f_{4}(z)+c_{16}f_{1}(z)f_{6}(z)+c_{25}f_{2}(z)f_{5}(z)+c_{34}f_{3}(z)f_{4}(z)\right)  \Big],
}}
\noindent with coefficients $a_{mn}$ and $c_{mn}$ given in (3.14). The coefficient $b_{14}$ is non-universal and generically depends on the details of the theory. It corresponds to the OPE coefficient of twist-six double-stress tensor with spin $s=2$ 
\eqn\aa{b_{14}=P^{(2)}_{8,2},
}  
obtained holographically in \FitzpatrickZQZ\ and here, via the gravitational phase-shift calculation in (5.48).

The subsubleading contribution in the lightcone limit, due to twist-four, six and eight double-stress tensor operators, is 
\eqn\gsubsub{\eqalign{& \GG^{(2,2)}(z) = {1\over{\left(1-z\right)^{\DL}}} \Bigg(\left({{z (2 z-7)+11}\over{6 (z-1)^2}}\right)(a_{33}f_{3}^{2}+a_{24}f_{2}f_{4}+a_{15}f_{1}f_{5}   )    \cr &+\left({{2-z}\over{1-z}}\right)(b_{14}f_{1}f_{4}+c_{16}f_{1}f_{6}+c_{25}f_{2}f_{6}+c_{34}f_{3}f_{4})+(d_{17}f_{1}f_{7}+d_{26}f_{2}f_{6}\cr &+d_{35}f_{3}f_{5}+d_{44}f_{4}^{2}+e_{15}f_{1}f_{5}+g_{13}f_{1}f_{3} )\Bigg),
}}
\noindent with coefficients $d_{mn}$ given in (3.19). By $f_{a}$ we mean $f_{a}(z)$ which we will use for brevity. The coefficients $g_{13}$ and $e_{15}$ are theory dependent and are related to the OPE coefficients of twist-eight double-stress tensors with spin $s=0,2$ by
\eqn\rellsumry{\eqalign{g_{13}&=P_{8,0}^{(2)},\cr
				   e_{15}&=P_{10,2}^{(2)}-{5\over 252}P_{8,0}^{(2)}.
}}
\noindent These coefficients were also obtained by a gravitational computation in \FitzpatrickZQZ . Here we have used the calculation of the phase shift in the dual gravitational theory to determine the OPE coefficient of the spin-2 operator, $P_{10,2}^{(2)}$, in (5.51).

The subsubsubleading contribution in the lightcone limit, due to double-stress tensors with twists $\tau=4,6,8,10$, is given by
\eqn\gsubsubsubsumry{\eqalign{\GG^{(2,3)}(z) &= {1\over{\left(1-z\right)^{\DL}}} \Bigg(\left(  {{z ((13-3 z) z-23)+25}\over{12 (1-z)^3}}\right)   (a_{33}f_{3}^{2}+a_{24}f_{2}f_{4}+a_{15}f_{1}f_{5}   )    \cr &+  \left({{1}\over{(1-z)^2}}+{{1}\over{1-z}}+{{9}\over{10}}\right)    (b_{14}f_{1}f_{4}+c_{16}f_{1}f_{6}+c_{25}f_{2}f_{5}+c_{34}f_{3}f_{4})\cr &+ \left(  {{1}\over{1-z}}+ {{3}\over{2}}  \right)     (d_{17}f_{1}f_{7}+d_{26}f_{2}f_{6}+d_{35}f_{3}f_{5}  +d_{44}f_{4}^{2}+e_{15}f_{1}f_{5} + g_{13}f_{1}f_{3} +)
\cr & +g_{13}f_{3}+ (h_{18}f_{1}f_{8}+h_{27}f_{2}f_{7}+h_{36}f_{3}f_{6}+h_{45}f_{4}f_{5}+j_{16}f_{1}f_{6}+i_{14}f_{1}f_{4})  \Bigg),
}}
\noindent with $h_{mn}$ given in (3.25). The non-universal coefficients here are $i_{14}$ and $j_{16}$ which are related to the OPE coefficients of twist-ten double-stress tensor operators with spin $s=0,2$
\eqn\relllsumry{\eqalign{i_{14}&=P_{10,0}^{(2)},\cr
					j_{16}&=P_{12,2}^{(2)}-{2\over 99}P_{10,0}^{(2)}.
}}
\noindent The OPE coefficient $P_{12,2}^{(2)}$ is determined in (5.52) using the phase shift calculation in the dual gravitational theory. Non-universality is manifest through dependence on the Gauss-Bonnet coupling. 

 Using the results above, we also extract the OPE coefficients $P_{\Delta ,s}^{(2)}$ of double-stress tensors of given twist. For $\tau=6$: 
\eqn\opecnewsumry{\eqalign{P_{10+2\ell, 4+2\ell}^{(2)}&={{ \sqrt{\pi} 2^{-4 \ell-17} \Gamma (2 n+7)}\over{ (\ell+4) (\ell+5) (\ell+6) (2 \ell+1) (2 \ell+3) (2 \ell+5) \Gamma \left(2 \ell+{{13}\over{2}}\right)}}\cr &\times {{\DL \over{(\Delta_{L}-3) (\Delta_{L}-2)}}}(a_{1,\ell} \Delta_L^3 + b_{1,\ell}\Delta_L^{2} + c_{1,\ell}\DL + d_{1,\ell}),
}}
\noindent where $a_{1,\ell}$, $b_{1,\ell}$, $c_{1,\ell}$, $d_{1,\ell}$ can be found in (3.17). For $\tau=8$:
\eqn\opeesumry{\eqalign{P_{12+2\ell, 4+2\ell}^{(2)}&={{\sqrt{\pi } \DL 2^{-4 \ell-19}\Gamma (2 \ell+7)}\over{3 (\DL -4) (\DL -3) (\DL -2) (\ell+4) (\ell+5) }}\cr &\times{{a_{2,\ell} \Delta_L^{4} + b_{2,\ell}\Delta_L^{3} +c_{2,\ell}\Delta_L^{2}+d_{2,\ell}\DL + e_{2,\ell}   }\over{(\ell+6) (\ell+7) (2 \ell+1) (2 \ell+3) (2 \ell+5) \Gamma \left(2 \ell+{{15}\over{2}}\right)}},
}}
\noindent with $a_{2,\ell}$, $b_{2,\ell}$, $c_{2,\ell}$, $d_{2,\ell}$ and $e_{2,\ell}$ given in (3.22). Similarly for $\tau=10$:
\eqn\opeeesumry{\eqalign{P_{14+2\ell ,4+2\ell}^{(2)}&={{\sqrt{\pi } 2^{-4 \ell -22}   \Gamma (2\ell +9)  }\over{5 (2 \ell +1) (2 \ell +3) (2 \ell +5) (2 \ell +7)\Gamma \left(2 \ell +{{17}\over{2}}\right)}} \cr
                              & \times {{\DL (\DL +1)(a_{3,\ell}\Delta_L^4 +b_{3,\ell} \Delta_L^{3} + c_{3,\ell} \Delta_L^{2} +d_{3,\ell} \DL  + e_{3,\ell}  )}\over{(\ell +5) (\ell +6) (\ell +7) (\ell +8)(\DL -5) (\DL -4) (\DL -3) (\DL -2)}},
}}
\noindent with $a_{3,\ell}$, $b_{3,\ell}$, $c_{3,\ell}$, $d_{3,\ell}$ and $e_{3,\ell}$ expressed in terms of $\Delta_L$ in (3.28). Note that in all of these formulas $\ell \geq 0$ and, therefore, the OPE coefficients of operators with spin $s=0,2$ are not included here. It appears that at $\OO(\mu^{2})$, the OPE coefficients of all operators with spin $s\geq 4$ are universal in the sense that they only depend on $\Delta_L$ and $C_T$. On the other hand, the OPE coefficients of double-stress tensors with $s=0,2$ are non-universal.

At $\OO(\mu^3)$, the leading contribution of twist-six triple-stress tensors in the lightcone limit, was computed in \KarlssonDBD
\eqn\gtsumry{\eqalign{
  \GG^{(3,0)}(z) =& {1\over{(1-z)^{\Delta_{L}}}} \Big(a_{117} f_{1}(z)^2 f_7(z) +a_{126}f_1(z)f_2(z)f_6(z)+a_{135}f_1(z)f_3(z)f_5(z) \cr 
  &   +a_{225}f_2(z)^2f_5(z)+a_{234}f_2(z)f_3(z)f_4(z)+a_{333}f_3(z)^3\Big),
}}
\noindent where the coefficients $a_{ijk}$ can be found in (4.2).

The subleading contribution to the correlator is due to twist-eight and twist-six triple-stress tensors
\eqn\gtsubsumry{\eqalign{
  \GG^{(3,1)}(z) =& {1\over{(1-z)^{\Delta_{L}}}} \Bigg( \left({{2-z}\over{1-z}}\right)  (a_{117} f_{1}^2 f_7 +a_{126}f_1 f_2 f_6  +a_{135}f_1 f_3 f_5  +a_{225}f_2 ^2 f_5\cr 
  &  +a_{234}f_2 f_3 f_4 +a_{333}f_3 ^3)+(b_{116} f_6 f_1^2+c_{118} f_8 f_1^2  +c_{145} f_4 f_5 f_1+c_{127} f_2 f_7 f_1\cr 
  & +c_{244} f_2 f_4^2+c_{334} f_3^2 f_4+c_{235} f_2 f_3 f_5+c_{226} f_2^2 f_6)     \Bigg),
}}
\noindent with $b_{ijk}$ and $c_{ijk}$ given in (B.1). Terms proportional to $a_{ijk}$ come from the subleading contribution due to the minimal-twist triple-stress tensors in \gtsumry. Note that all of these coefficients are non-universal, since they depend on $b_{14}$ from the $\OO(\mu^{2})$ result. Accordingly, no OPE coefficients of non-minimal-twist triple-stress tensors are universal.

A similar story holds for the subsubleading contribution to the correlator at $\OO(\mu^{3})$. This is due to multi-stress tensors with twist six, eight and ten and takes the following form 
\eqn\gtsubsubsmry{\eqalign{
  &\GG^{(3,2)}(z) = {{1}\over{(1-z)^{\Delta_{L}}}} \Bigg( \left({{144 z^2-448 z+464}\over{160 (z-1)^2}}\right)  (a_{117} f_{1}^2 f_7 +a_{126}f_1 f_2 f_6 +a_{135}f_1 f_3 f_5  \cr 
  & +a_{225}f_2 ^2 f_5  +a_{234}f_2 f_3 f_4 +a_{333}f_3 ^3)+\left({{1}\over{1-z}}+{{3}\over{2}}\right)(b_{116} f_6 f_1^2+c_{118} f_8 f_1^2  +c_{145} f_4 f_5 f_1\cr 
  &+c_{127} f_2 f_7 f_1+c_{244} f_2 f_4^2+c_{334} f_3^2 f_4+c_{235} f_2 f_3 f_5+c_{226} f_2^2 f_6)   +(d_{117}f_{1}^{2}f_{7}+e_{115}f_{1}^{2}f_{5}\cr
  &+g_{119}f_{1}^{2}f_{9}+g_{128}f_{1}f_{2}f_{8}+g_{155}f_{1}f_{5}^{2}+g_{227}f_{2}^{2}f_{7}+g_{236}f_{2}f_{3}f_{6}+g_{245}f_{2}f_{4}f_{5}+g_{335}f_{3}^{2}f_{5}\cr
  &+g_{344}f_{3}f_{4}^{2})  \Bigg),\cr 
}}
\noindent with $d_{117}$ and $g_{ijk}$ in (C.1)-(C.3) and $e_{115}$ in (5.56).

We further explain how one can write an ansatz for the correlator at arbitrary order in $\mu$ and the lightcone expansion.  All unknown coefficients in the ansatz, except those that correspond to OPE coefficients of spin-0 and spin-2 operators, can be fixed by means of the lightcone bootstrap. 
We further show that  in holographic CFTs  one can use the phase shift computed in the dual gravitational theory to reduce 
the set of undetermined parameters to the OPE coefficients of multi-stress tensors with spin zero.
Our results for these OPE coefficients precisely match those in \FitzpatrickZQZ\ whenever available in the latter.

The OPE coefficients of multi-stress tensors can also be calculated using the Lorentzian inversion formula as in \LiZBA. In order to determine for which operators the formula can be applied, one should consider the behavior of the correlation function in the Regge limit.  The Regge behavior of the correlator at $\OO(\mu^{k})$ is $1/\sigma^{k}$, implying that the Lorentzian inversion formula can be used to extract the OPE coefficients of the operators with spin $s>k+1$. Accordingly, already at $\OO(\mu^{3})$, fixing the relevant OPE coefficients by combining an ansatz with the lightcone bootstrap allows one to determine more OPE data compared to those obtained with the sole use of the Lorentzian inversion formula. We explicitly check that it is not possible to extract the OPE coefficient of a triple-stress tensor with spin $s=4$ and twist $\tau=8$ using the Lorentzian inversion formula. Note, however, that this coefficient is completely determined in this article (where an ansatz is additionally employed).

\subsec{Outline}

This paper is organized as follows. In Section 2, we set up the notation and review the S- and T-channel expansions of the HHLL correlator. In Section 3, we analyze the stress tensor sector of the correlator at $\OO(\mu^2)$, where we compute the subleading, subsubleading and subsubsubleading contributions in the lightcone expansion. We also compute the OPE coefficients of double-stress tensors with twist $\tau=6,8,10$ and spin $s>2$. In Section 4, we analyze the stress tensor sector of the correlator at $\OO(\mu^3)$, where we explicitly calculate the subleading and subsubleading contributions in the lightcone expansion. In Section 5, we investigate the Gauss-Bonnet dual gravitational theory and give additional evidence for the universality of the OPE coefficients of minimal-twist multi-stress tensors using the phase shift calculation. Furthermore, we calculate the OPE coefficients of double- and triple-stress tensors with spin $s=2$ (up to undetermined spin zero data). In Section 6, we show how one can use the Lorentzian inversion formula in order to extract the OPE coefficients of double-stress tensors with twist $\tau=4,6$. We discuss our results in Section 7. Appendix A contains certain relations that products of $f_a$ functions satisfy, while Appendices B and C contain explicit expressions for the coefficients which determine the correlator in subleading and subsubleading lightcone order at $\OO(\mu^3)$. Several OPE coefficients  of twist-eight triple-stress tensors are listed in Appendix D. In Appendix E we clarify the relationship between the scattering phase shift as defined in \KulaxiziDXO\ and the deflection angle and finally, in Appendix F we explicitly write some of the S-channel anomalous dimensions at $\OO(\mu^2)$ and we investigate their relation with the phase shift.

\newsec{Review of near lightcone heavy-heavy-light-light correlator}
In this Section, we review the procedure for extracting information about the stress tensor sector of a four-point correlation function between two pairwise identical scalars $\OH$, $\OL$, with scaling dimensions $\DH\propto \OO(C_T)$ and $\DL\propto \OO(1)$, respectively, via the lightcone bootstrap. 
We closely follow  Ref.  \KarlssonDBD. Using conformal transformations to fix the positions of three of the operators at $0,1,x_4\to\infty$, we define the stress tensor sector of the correlator by 
\eqn\defStressTensorSector{
	\GG(z,\zbar) = \lim_{x_4\to\infty} x_4^{2\DH}\langle \OH(x_4)\OL(1)\OL(z,\zbar)\OH(0)\rangle \Big|_{\rm multi-stress \ tensors}, 
}
where $(z,\zbar)$ are the invariant cross-ratios given by 
\eqn\crossRatios{\eqalign{
	z\zbar &= {x_{14}^2x_{23}^2\over x_{13}^2x_{24}^2},\cr
	(1-z)(1-\zbar) &= {x_{12}^2x_{34}^2\over x_{13}^2x_{24}^2}.
}}

\subsec{T-channel expansion}
The notion of the stress-tensor sector comes from expanding the correlator in the T-channel defined as $\OL(z,\zbar)\times \OL(1)\to \OO_{\tau,s}$:
\eqn\tChExp{
	\GG(z,\zbar) = {1\over [(1-z)(1-\zbar)]^\DL}\sum_{\OO_{\tau,s}} P^{(HH,LL)}_{\OO_{\tau,s}} g_{\tau,s}^{(0,0)}(1-z,1-\zbar),
}
where $s$ and $\tau=\Delta-s$ denote the spin and the twist of the exchanged primary operator $\OO_{\tau,s}$. $P^{(HH,LL)}_{\OO_{\tau,s}}$ denotes the product of OPE coefficients 
\eqn\oppe{P^{(HH,LL)}_{\OO_{\tau,s}}=\left(-{1\over 2}\right)^{s}\lambda_{\OO_{H}\OO_{H}\OO_{\tau,s}}\lambda_{\OO_{L}\OO_{L}\OO_{\tau,s}}
}
\noindent and $g_{\tau,s}^{(0,0)}(1-z,1-\zbar)$ the corresponding conformal block. 

Consider the T-channel expansion \tChExp\ in $d=4$. Conformal blocks in $d=4$ are given by \DolanHV
\eqn\blocksFourD{
	g^{(0,0)}_{\tau,s}(1-z,1-\zbar) = {(1-z)(1-\zbar)\over \zbar - z}\left(f_{\beta\over 2}(z)f_{\tau-2\over 2}(\zbar)-f_{\beta\over 2}(\zbar)f_{\tau-2\over 2}(z)\right),
}
with conformal spin, $\beta=\Delta+s$, and
\eqn\defFFunc{
	f_{a}(z) = (1-z)^a {}_2F_1(a, a, 2a, 1-z).
}
In the lightcone limit, defined by $\zbar\to 1$ and $z$ fixed, the leading contribution to the conformal blocks \blocksFourD\ comes from the first term in parenthesis in \blocksFourD 
\eqn\leadingBlock{
	g^{(0,0)}_{\tau,s}(1-z,1-\zbar)= (1-\zbar)^{\tau\over 2}\Big(f_{\beta\over 2}(z)+\OO((1-\zbar))\Big).
}
From \leadingBlock\ it is clear that the operators with the lowest twist in the T-channel dominate the correlator in the lightcone limit. In any unitary CFT in $d=4$ the operator with the lowest twist is the identity operator with twist $\tau=0$. 
Another operator with low twist present in any local CFT is the stress tensor operator with  $\tau=2$. In particular, the exchange of the stress tensor is completely fixed since the product of the relevant OPE coefficients is determined by Ward identities
\eqn\OPEcoeffStress{
	P^{(HH,LL)}_{T_{\mu\nu}} = \mu{\DL\over 120},
}
where 
\eqn\mudef{
	\mu = {160\over 3}{\DH\over C_T}.
}
\noindent The central charge $C_T$ is defined via the two-point function of the stress tensor
\eqn\defct{\langle T_{\mu\nu}(x)T_{\rho \sigma}(0) \rangle = {{C_{T}}\over{\Omega_{d-1}^{2} x^{2d}}}\II_{\mu\nu,\rho\sigma}(x),
}
\noindent where
\eqn\definicije{\eqalign{\II_{\mu\nu,\rho\sigma}(x)&={1\over 2}\left(I_{\mu\rho}(x)I_{\nu\sigma}(x)+I_{\mu\sigma}(x)I_{\nu\rho}(x)\right)-{1\over d}\eta_{\mu\nu}\eta_{\rho\sigma},\cr
I_{\mu\nu}&=\eta_{\mu\nu}-2{{x_{\mu}x_{\nu}}\over x^{2}}, \qquad \Omega_{d-1}={{2\pi^{d/2}}\over{\Gamma({d\over 2})}}.
}}

\noindent Note that the only single-trace primaries with twist equal to or lower than that of the stress tensor are scalars $\OO$ with dimension $1\leq \Delta_\OO \leq 2$, or conserved currents with twist $\tau=2$. In a theory without  supersymmetry there is no {\it a priori} reason for the contributions of these operators, even if they exist, to be enhanced by a factor of $\DH$, so generically we expect them to be subleading in $C_{T}\to \infty$ limit.\foot{Interestingly, in \AldayQRF\ it is conjectured that OPE coefficients $\lambda_{\phi \psi \psi}$ of operators $\phi$ with conformal dimension $\Delta_{\phi}\ll\Delta_{\rm gap}$ and $\psi$ with conformal dimension $\Delta_{\psi}$, such that $\Delta_{\phi}\ll\Delta_{\psi}\ll C_{T}^{\#>0}$, scale as $\lambda_{\phi \psi \psi}\propto{\Delta_{\psi}\over{\sqrt{C_T}}}$. Note however that here we are working in different regime, as $\Delta_{H}\propto \OO(C_T)$.}

The stress tensor sector of the correlator \defStressTensorSector\ admits a perturbative expansion in $\mu$ given by 
\eqn\pertG{
	\GG(z,\zbar) = \sum_{k=0}^\infty \mu^k\GG^{(k)}(z,\zbar),
}
where the cases $k=0$ and $k=1$ correspond to the exchange of the identity and the stress tensor, respectively. For higher $k$ we expect ``multi-stress tensors'' to contribute to $\GG(z,\zbar)$; 
the minimal-twist  multi-stress tensor primaries are of the schematic form 
\eqn\mintwist{
	[T^k]_{\tau_{k,{\rm min}},s} = :T_{\mu_1\nu_1}\ldots T_{\mu_{k-1}\nu_{k-1}}\pa_{\lambda_1}\ldots \pa_{\lambda_{2\ell}}T_{\mu_k\nu_k}:,
}
with  twist $\tau_{k,{\rm min}}$ and spin $s$ given by
\eqn\twistAndSpin{\eqalign{
	&\tau_{k,{\rm min}} = 2k,\cr
	&s = 2k+2\ell,
}}
with $\ell$ an integer. Since we are interested in the four-point function of pairwise identical scalar operators, only multi-stress tensor operators with even spin give a nonvanishing contribution. At $\OO(\mu^2)$, the contribution of these operators was explicitly calculated in \KulaxiziTKD . Following that, it was shown in \KarlssonDBD\ how one can write the contributions of these operators at arbitrary order in the $\mu$-expansion, in the lightcone limit $(1-\zbar)\ll 1$, using an appropriate ansatz and lightcone bootstrap. We briefly review this procedure here since the contribution from non-minimal-twist operators is obtained in a similar manner. 

At $\OO(\mu^k)$, there are infinitely many minimal-twist multi-stress tensors with twist $2k$ according to \twistAndSpin\ which are distinguished by their conformal spin 
$\beta=\Delta+s$ given by $\beta=6k+4\ell$ with $\ell=0,1,2,\ldots$. Inserting the leading behavior of the blocks \leadingBlock\ in \tChExp\ one finds 
\eqn\gkCont{
	\GG^{(k)}(z,\zbar)\lzbarone  {(1-\zbar)^{k}\over [(1-z)(1-\zbar)]^\DL}\sum_{\ell=0}P_{\Delta(\ell) ,s(\ell)}^{(k)} f_{{\beta (\ell)}\over{2}}(z),
}
\noindent with
\eqn\opeee{\mu^{k} P_{\Delta(\ell) ,s(\ell)}^{(k)}=P^{(HH,LL)}_{[T^k]_{\tau,s(\ell)}},
}

\noindent where $\Delta(\ell)={{\tau + \beta}\over{2}}$, $\tau=2k$, $s(\ell)=2k+2\ell$ and conformal spin $\beta = 6k+4\ell$. Here $\lzbarone$ means that only the leading contribution as $\zbar \to 1$ is kept. It was shown in \KarlssonDBD\ that the infinite sum in \gkCont\ takes a particular form 
\eqn\gkAnz{
	\GG^{(k)}(z,\zbar) \lzbarone {(1-\zbar)^{k}\over [(1-z)(1-\zbar)]^\DL}\sum_{\{i_p\}}a_{i_1 ... i_k} f_{i_1}(z) ...  f_{i_k}(z), \qquad \sum_{p=1}^k i_p = 3k,
}
\noindent with $i_{p}$ being integers and $a_{i_1\ldots i_k}$ are coefficients that can be determined via lightcone bootstrap. Furthermore, using an identity for the product of two $f_a$ functions (Eq.\ (A.1) in \KulaxiziTKD) one can  express the $\GG^{(k)}(z,\zbar)$ in the form of \gkCont\ to read off the OPE coefficients for the exchange of minimal-twist multi-stress tensors of arbitrary conformal spin. 

In this paper, we want to consider multi-stress tensors with non-minimal twist. These operators are obtained by contracting indices in \mintwist\ either between the derivatives or between the operators. At $\OO\left(\mu^k\right)$ there exist operators $[T^k]_{\tau_{k,m},s}$ with twist 
\eqn\generalMultiStress{
	\tau_{k,m} = \tau_{k,{\rm min}}+2 m,
}  
for any non-negative integer $m$. For $m\neq 0$, these  operators provide subleading contributions to the correlator in the lightcone limit. To consider these subleading contributions it is convenient to expand $\GG^{(k)}(z,\zbar)$ from \pertG\ as 
\eqn\gexp{\GG^{(k)}(z,\zbar)=\sum_{m=0}^{\infty}(1-\zbar)^{-\DL + k + m}\GG^{(k,m)}(z),
}
where $\GG^{(k,m)}(z)$ comes from operators of twists $\tau_{k,m}$ and less.

For illustration, let us consider the case $k=2$ with $m=1$. There exist two infinite families of operators with twist $\tau_{2,1}=6$ of the schematic form 
\eqn\exDouble{\eqalign{
	\OO_{6,2\ell_{1}+2} &\sim\, :T_{\mu\kappa}\pa_{\lambda_1}\ldots \pa_{\lambda_{2\ell_{1}}}{T^{\kappa}}_{\nu}:,\cr
	\OO'_{6,2\ell_{2}+4} &\sim\, :T_{\mu\nu}\pa_{\lambda_1}\ldots \pa_{\lambda_{2\ell_{2}}}\pa^2 T_{\rho\sigma}:.
}} 
These two families share the same twist and spin for $\ell_1 = \ell_2 +1$. Hence, they are indistinguishable for $\ell_1 \geq 1$ at order $1/C_{T}$ in the large $C_{T}$ expansion. A single operator stands out; it corresponds to $\ell_1 = 0$ and is of the schematic form $:T_{\mu \alpha}{T^{\alpha}}_{\nu}:$. Note that $:T_{\mu \alpha}{T^{\alpha}}_{\nu}:$ has minimal conformal spin $\beta=10$, among the ones in \exDouble , since $\beta_{\ell_1}=\beta_{\ell_2 +1}=10+4\ell_1$, for $\ell_1 \geq 1$. 

Let us now move on to the case $k=2$ and $m=2$. Here, there are three infinite families $\OO_{8,s}$, $\OO'_{8,s}$ and $\OO''_{8,s}$ with conformal spin $8+4\ell_{1}$, $12+4\ell_{2}$ and $16+4\ell_{3}$, respectively. Schematically, these families can be represented as
\eqn\confSpinCont{\eqalign{
\OO_{8,2\ell_1}  &\sim\, :T_{\alpha\beta}\pa_{\lambda_1}\ldots \pa_{\lambda_{2\ell_1}}{T^{\alpha\beta}}:,\cr
\OO'_{8,2\ell_2+2}  &\sim\, :T_{\mu\alpha}\pa_{\lambda_1}\ldots \pa_{\lambda_{2\ell_2}}\pa^2{T^{\alpha}}_{\nu}:,\cr
\OO''_{8,2\ell_3+4}  &\sim\, :T_{\mu\nu}\pa_{\lambda_1}\ldots \pa_{\lambda_{2\ell_3}}(\pa^2)^{2}{T_{\rho\sigma}}:.
}} 
Notice once more that the infinite families are indistinguishable for conformal spin $\beta \geq 16$. Here, operators with $\beta=8,12$ stand out. The operator with $\beta=8$ is of the schematic form $:T_{\alpha\beta}T^{\alpha\beta}:$. For $\beta = 12$, there are two indistinguishable operators of the schematic form $:T_{\mu \alpha}\pa^{2}{T^{\alpha}}_{\nu}:$ and $:T_{\alpha\beta}\pa_{\mu}\pa_{\nu}T^{\alpha\beta}:$.

The same holds for $m\geq 3$ (and $\tau\geq 10$) since there is no other independent way to contract stress tensor indices. The discussion above generalizes straightforwardly to $\OO(\mu^{k})$ with $k+1$ number of infinite families at high enough twist. 

\subsec{S-channel expansion}
The correlator \defStressTensorSector\ can also be expanded in the S-channel defined as $\OL(z,\zbar)\times\OH(0)\to \OO_{\tau',s'}$, 
\eqn\sChannel{
	\GG(z,\zbar) = (z\zbar)^{-{1\over 2}(\DH+\DL)}\sum_{\OO_{\tau',s'}}P^{(HL,HL)}_{\OO_{\tau',s'}}g_{\tau',s'}^{(\DHL, -\DHL)}(z,\zbar),
}
where $P^{(HL,HL)}_{\OO_{\tau',s'}}$ denotes the product of OPE coefficients in the S-channel, $\DHL=\DH-\DL$, and $g_{\tau',s'}^{(\DHL, -\DHL)}(z,\zbar)$ are the relevant conformal blocks. Operators contributing in the S-channel expansion are  ``heavy-light double-twist'' operators \refs{\KulaxiziDXO,\KarlssonQFI}\foot{In the lightcone limit of $\langle \OO_1\OO_2\OO_2\OO_1\rangle$, with $\OO_1,\OO_2$ both light, it was found in \refs{\KomargodskiEK,\FitzpatrickYX} that the there exists ``light-light double-twist'' operators $[\OO_1\OO_2]_{n,l}= :\OO_1(\pa^2)^n\pa_{\mu_1}\ldots\pa_{\mu_l}\OO_2:$ for $l\gg 1$. These are found by matching with the identity exchange in the S-channel. The same is true for for the heavy-heavy-light-light case.} of the schematic form $[\OH\OL]_{n,l} = :\OH(\pa^2)^n\pa_{\mu_1}\ldots\pa_{\mu_l}\OL:$, with conformal dimensions $\Delta=\DH+\DL+2n+l+\gamma$.  The conformal blocks for these heavy-light double-twist operators in $d=4$ are given by 
\eqn\SchBlocks{
	g^{(\DHL, -\DHL)}_{\DH+\DL+2n+\gamma,l}(z,\bar z) = {(z\zbar)^{{1\over 2}(\DH+\DL+2n+\gamma_{n,l})}\over \zbar-z}\left(\zbar^{l+1}-z^{l+1}\right) + \OO\left({1\over \DH}\right).
}
\noindent The anomalous dimensions and the product of OPE coefficients for heavy-light double-twist operators admit an expansion in powers of $\mu$:
\eqn\expansion{\eqalign{
	\gamma_{n,l} &= \sum_{k=1}^\infty \mu^k \gamma^{(k)}_{n,l},\cr 
	P^{(HL,HL)}_{n,l} &= P^{(HL,HL); {\rm MFT}}_{n,l}\sum_{k=0}^{\infty}\mu^k P^{(HL,HL); (k)}_{n,l},
}}
where $P^{(HL,HL); {\rm MFT}}_{n,l}$ are the Mean Field Theory coefficients \FitzpatrickDM, which can be found by matching with the exchange of the identity in the T-channel, and $P^{(HL,HL); (0)}_{n,l}=1$. Explicitly, in $d=4$ and for $\DH\gg 1$, 
\eqn\PMFT{
	P^{(HL,HL); {\rm MFT}}_{n,l} = {(\DL-1)_n (\DL)_{l+n}\over n!\,l!\,(l+2)_n}  +\OO\left({1\over \DH}\right),
}
where $(a)_n$ is the Pochhammer symbol defined by $(a)_n={\Gamma(a+n) \over \Gamma(n)}$.

We begin by briefly reviewing the calculation in the lightcone expansion, i.e.\ due to the multi-stress tensors in the T-channel. Inserting the blocks \SchBlocks\ in the S-channel expansion \sChannel\ one finds that 
\eqn\sChCont{
	\GG(z,\zbar) = \sum_{n=0}^\infty{(z\zbar)^n\over \zbar-z}\int_0^\infty dl P^{(HL,HL)}_{n,l}(z\zbar)^{{1\over 2}\gamma_{n,l}}(\zbar^{l+1}-z^{l+1}),
}
where the sum was approximated by an integral over $l$. 
Expanding the OPE data in \sChCont\ according to \expansion\ and noting that
\eqn\pfunc{
   (z\zbar)^{{1\over 2}\gamma_{n,l}}= \sum_{j=0}^\infty {1\over j!}\left({\gamma_{n,l}\log (z\zbar)\over 2}\right)^j,
}
\noindent it follows that terms proportional to $\log^i z$ at $\OO(\mu^k)$, with $i=2,3,\ldots k$, in \sChCont\ are determined by OPE data at $\OO(\mu^{k-1})$. These terms can therefore be matched with the T-channel in order to fix the coefficients in the ansatz.

In \KarlssonDBD, the leading contribution of the OPE data of heavy-light double-twist operators as $l\to\infty$, together with the leading contribution of the conformal blocks as $\zbar\to 1$, was used to determine the minimal-twist contributions in the stress tensor sector of the T-channel. This paper extends that analysis by considering subleading corrections in the lightcone expansion and therefore probing non-minimal-twist contributions in the T-channel. In particular, the S-channel OPE data have the following dependence on the spin $l$ as $l\to\infty$:
\eqn\SpinBehavior{\eqalign{
	\gamma^{(k)}_{n,l}  &= {1\over l^{k}}\sum_{p=0}^\infty {\gamma^{(k,p)}_{n}\over l^p},\cr
	P^{(HL,HL); (k)}_{n,l} &= {1\over l^{k}}\sum_{p=0}^\infty {P^{(HL,HL); (k,p)}_n\over l^p},
}}
which is necessary in order to reproduce the correct power of $(1-\bar z)$ as $\bar z \to 1$. 
This can be seen by substituting the expansion of \PMFT\ in the large-$l$ limit
\eqn\mfto{\eqalign{P^{(HL,LH);MFT}_{n,l}=l^{\Delta_{L}}\Bigg(&{{(\Delta_{L}-1)_{n}}\over{n! \Gamma(\Delta_{L})l}}+{{(2n(\Delta_{L}-2)+\Delta_{L}(\Delta_{L}-1))(\Delta_{L}-1)_{n}}\over{2 (n!) \Gamma(\Delta_{L})l^{2}}}\cr
& + \OO\left({1\over l^{3}}\right)\Bigg),
}}
and \SpinBehavior\ in \sChCont\ which result in integrals of the form
\eqn\SIntegral{
	\int_0^\infty dl \zbar^{l} l^{\DL-m-1} = {\Gamma(\DL-m)\over (-\log \zbar)^{\DL-m}},
}
\noindent where $m$ is a positive integer. Expanding \SIntegral\ for $\zbar\to 1$, the correct $\zbar$-behavior of the stress tensor sector in the T-channel is reproduced from the S-channel. 


\newsec{Double-stress tensors in four dimensions}

In this Section, we analyze the stress tensor sector of the HHLL correlator at $\OO(\mu^{2})$ in $d=4$. The operators that contribute at this order in the T-channel are the double-stress tensors. 
Here, we investigate the subleading contributions that are coming from families of operators with nonminimal twist, specifically, $\tau_{2,1}=6$, $\tau_{2,2}=8$ and $\tau_{2,3}=10$, according to \generalMultiStress .

The dominant contribution in the lightcone limit  at $\OO(\mu^{2})$ was calculated in \KulaxiziTKD . 
It comes from the operators with minimal twist $\tau_{2,{\rm min}}=4$ and they are of the schematic form $:T_{\mu \nu}\partial_{\alpha_{1}}\ldots\partial_{\alpha_{2\ell}}T_{\rho \sigma}:$. 
These operators have conformal dimension $\Delta = 8 + 2\ell$ and spin $s=4+2\ell$. 
The result is \KulaxiziTKD\
\eqn\secondCorrIntro{\eqalign{
  \GG^{(2,0)}(z) = &{1\over (1-z)^{\DL}}\left({\DL\over 28800(\DL-2)}\right)\times\cr
      \Big[&(\DL-4)(\DL-3)f_3^2(z)+{15\over 7}(\DL-8)f_2(z)f_4(z)+{40\over 7}(\DL+1)f_1(z)f_5(z)\Big],
}}
\noindent where $f_{a}(z)=(1-z)^{a}{}_{2}F_{1}(a,a,2a,1-z)$. 


\subsec{Twist-six double-stress tensors}
Twist-six double-stress tensors contribute at $\OO(\mu^{2})$ and at subleading order in the lightcone expansion $\sim (1-\zbar)^{-\DL +3}$ as $\zbar\to 1$. As shown in this section, this contribution again takes a particular form with a few undetermined coefficients which, except for a single one, can be fixed using lightcone bootstrap. The undetermined data is shown to correspond to a single OPE coefficient due to the exchange of the twist-six and spin-two double-stress tensor $:{T_\mu}^\rho T_{\rho\nu}:$. 

We will now motivative an ansatz for the subleading contribution to the stress tensor sector at $\OO(\mu^2)$.
Let us focus first on corrections due to the leading lightcone contribution of twist-four double-stress tensors. These corrections originate from subleading terms in the lightcone expansion of the conformal blocks in \leadingBlock. Note however that they are purely kinematical and do not contain any new data. Explicitly, the subleading corrections to the blocks of twist-four double-stress tensors are given by
\eqn\conblotwo{\eqalign{g_{4, s}^{(0,0)}(1-z,1-\zbar)\lzbarone &(1-\zbar)^{2}\left(1+(1-\zbar) \left({{3-z}\over{2(1-z)}}\right) + \OO\left((1-\zbar)^{2}\right)\right)f_{{{\beta}\over 2}}(z)\cr
-(1-\zbar)^{s+3}&\left(1+(1-\zbar)\left({{s+2 }\over{2}}+{{1}\over{1-z}}\right)+\OO(\left(1-\zbar\right)^{2})\right)f_{1}(z).
}} 
\noindent Since we are interested in the subleading contribution, i.e.\ terms that behave as $(1-\zbar)^{3}$ as $\zbar\to 1$ in \conblotwo, only the first line in \conblotwo\ needs to be considered. (Note that $s\geq 4$ for minimal-twist double-stress tensors.)

Next, consider the contribution of twist-six double-stress tensors. Recall that the form of the minimal-twist double-stress tensors' contribution to \secondCorrIntro\ can be motivated by decomposing products of the type $f_a(z)f_b(z)$ in terms of the lightcone conformal blocks. This decomposition is explicitly given by \KulaxiziTKD:
\eqn\iden{f_{a}(z)f_{b}(z)=\sum_{\ell =0}^{\infty}p(a,b,\ell)f_{a+b+2\ell}(z),}
\noindent where 
\eqn\deffp{\eqalign{&p(a,b,\ell)=\cr
&{{2^{-4 \ell} \Gamma \left(a+{{1}\over{2}}\right) \Gamma \left(b+{{1}\over{2}}\right) \Gamma \left(\ell +{{1}\over{2}}\right) \Gamma (a+\ell) \Gamma (b+\ell) \Gamma \left(a+b+\ell -{{1}\over{2}}\right) \Gamma (a+b+2 \ell)}\over{\sqrt{\pi } \Gamma (a) \Gamma (b) \Gamma (\ell +1) \Gamma \left(a+\ell +{{1}\over{2}}\right) \Gamma \left(b+\ell +{{1}\over{2}}\right) \Gamma (a+b+\ell) \Gamma \left(a+b+2 \ell-{{1}\over{2}}\right)}}.
}}

Using the leading behavior of the conformal blocks \conblotwo\ in the lightcone limit, it was found that $a+b+2\ell$ should be identified with ${\beta\over 2}={\Delta+s\over 2}$. In order to reproduce twist-six double-stress tensors of the form $:T_{\mu \nu}\partial^{2}\partial_{\alpha_{1}}\ldots\partial_{\alpha_{2\ell}}T_{\rho \sigma}:$ we should therefore consider products $f_af_b$ with $a+b=7$. Likewise, to take into account operators of the form $:T_{\mu \beta}\partial_{\alpha_{1}}\ldots\partial_{\alpha_{2\ell}}{T^{\beta}}_{\nu}:$ we include products $f_af_b$ with $a+b=5$. 

From the arguments above, we make the following ansatz for the subleading correction in the lightcone expansion due to double-stress tensors:
\eqn\subg{\eqalign{
  \GG^{(2,1)}(z) = {1\over (1-z)^{\DL}}\Big[\left({{3-z}\over {2(1-z)}}\right)\left(a_{33}f_{3}(z)^{2}+a_{24}f_{2}(z)f_{4}(z)+a_{15}f_{1}(z)f_{5}(z)\right) \cr +\left(b_{14}f_{1}(z)f_{4}(z)+b_{23}f_{2}(z)f_{3}(z)+c_{16}f_{1}(z)f_{6}(z)+c_{25}f_{2}(z)f_{5}(z)+c_{34}f_{3}(z)f_{4}(z)\right)  \Big],
}}
where $b_{ij},c_{ij}$ are coefficients that will be determined using lightcone bootstrap and encode the contribution from twist-six double-stress tensors. Once $b_{ij}$ and $c_{ij}$ are determined, one can use the decomposition in \iden\ to read off the OPE coefficients of twist-six double-stress tensors with any given spin. Moreover, $a_{ij}$ in \subg\ are coefficients that can be read off from the minimal-twist contribution in \secondCorrIntro\ and do therefore not contain any new information.

We proceed with the S-channel calculation to fix the unknown coefficients in \subg .
Let us first mention that the products of $f_{a}(z)$ functions in the second line of \subg\ are not linearly independent as one can see from (A.1), so we set $b_{23}=0$. 
Moreover, the coefficients $a_{ij}$ must be the same as in \secondCorrIntro. 
We will momentarily keep them undetermined to have an extra consistency check of our calculation. 

In the S-channel we have double-twist operators of the form $:\OH\pa^{2n}\pa^{l}\OL:$ with conformal dimension $\Delta=\Delta_{H}+\Delta_{L}+2n+l+\gamma_{n,l}$. The relevant anomalous dimensions $\gamma_{n,l}$ and OPE coefficients are given in \expansion\ and \SpinBehavior\ ($k=2$ in this case). In the lightcone limit, the dominant contribution comes from operators with large spin $l$, $l \gg n$. The mean field theory OPE coefficients are given by \mfto .
The conformal blocks of these operators in the limit $1-\zbar \ll z \ll 1$ are
\eqn\cbsc{g_{n,l}^{(\Delta_{HL},-\Delta_{HL})}(z,\zbar) \approx {(z\zbar)^{{\Delta_{H}+\Delta_{L}+\gamma(n,l)}\over 2}\over {\zbar - z}} z^{n}\zbar^{l+n+1}. 
}

We first need to fix the OPE data at $\OO(\mu)$. Coefficients $\gamma^{(1,p)}_{n}$ and $P^{(1,p)}_{n}$ can be determined for every $p$ and $n$ by matching the S-channel correlator with the correlator in the T-channel at $\OO(\mu)$. This is just the stress tensor block times its OPE coefficient and it is known for arbitrary $z$ and $\zbar$. As we saw earlier
\eqn\tchanmu{(\zbar -z)\GG^{(1)}(z,\zbar)={1\over{[(1-z)(1-\zbar)]^{\DL -1}}}{\DL \over 120}\left({{f_{3}(z)-f_{3}(\zbar)}}\right).
}
Expanding \tchanmu\ near $\zbar \to 1$ leads to
\eqn\tchanmuexp{\eqalign{(\zbar -z)\GG^{(1)}(z,\zbar)={{(1-\zbar)}\over{((1-z)(1-\zbar))^{\DL}}}\Bigg(&-\DL\left({3\over 4}(1+z)+{{1+z (z+4)}\over{4 (1-z)}}\log(z)\right)\cr
&-\sum_{p=1}^{\infty}{{\DL (p-2) (p-1) (1-z)}\over{4p (p+1) \left(p+2\right)}}(1-\zbar)^{p}\Bigg).
}}
On the other hand, we expand the integrand of \sChCont\ up to the $\OO(\mu)$, integrate this expansion over $l$, and then expand in the lightcone limit $\zbar \to 1$ to obtain a result of the form
\eqn\schanmu{(\zbar -z)\GG^{(1)}(z,\zbar)={1\over{(1-\zbar)^{\DL-1}}}\sum_{p=0}^\infty\left(\sum_{n=0}^\infty r_{n,p}(z)z^n(1-\zbar)^{p} \right).
}
\noindent The functions $r_{n,p}(z)$ can be  explicitly calculated. Here $r_{n,0}(z)$, $r_{n,1}(z)$ and $r_{n,2}(z)$ are given by
\eqn\rone{\eqalign{r_{n,0}(z)&={{\Gamma(\DL +n-1)}\over {2 \Gamma(\DL) \Gamma(n+1)}}\left(2P^{(1,0)}_{n}+\log(z)\gamma^{(1,0)}_{n}\right),\cr
r_{n,1}(z)&={{\Gamma(\DL +n-1)}\over {2 \Gamma(\DL) \Gamma(n+1)(\DL -2)}}\Big(2(P^{(1,0)}_{n}+P^{(1,1)}_{n})-(\DL -2)\gamma^{(1,0)}_{n}\cr
&\qquad + \log(z)(\gamma^{(1,0)}_{n}+\gamma^{(1,1)}_{n})\Big),\cr
r_{n,2}(z)&={{\Gamma(\DL +n-1)}\over {2(\DL -2)(\DL -3) \Gamma(\DL) \Gamma(n+1)}}\Big(2(\DL +n-1)P^{(1,0)}_{n}+2(\DL +n)P^{(1,1)}_{n}\cr
&+2P^{(1,2)}_{n}-{1\over 2}(\DL -3)(\DL \gamma^{(1,0)}_{n}+2 \gamma^{(1,1)}_{n})+\log(z)((\DL +n-1)\gamma^{(1,0)}_{n} \cr 
&+(\DL +n)\gamma^{(1,1)}_{n}+\gamma^{(1,2)}_{n})\Big).
}}

\noindent Similarly, one can calculate any $r_{n,p}(z)$ for arbitrary $p$. In each $r_{n,p}(z)$ the $z$-dependence enters only through a single logarithmic term as in \rone . In order to extract the OPE data we match \tchanmuexp\ and \schanmu\ and obtain the following relations
\eqn\matching{\eqalign{
\sum_{n=0}^\infty z^n r_{n,0}(z)&=-{\DL \over {(1-z)^{\DL}}}\left({3\over 4}(1+z)+{{1+z (z+4)}\over{4 (1-z)}}\log(z)\right),\cr
\sum_{n=0}^\infty z^n r_{n,p}(z)&=-{\DL \over {(1-z)^{\DL}}}{{(p-2)(p-1)(1-z)}\over{4p(p+1)(p+2)}},
}}
\noindent for $p\geq 1$. To solve these equations, we start from the first line, expand the right-hand side in  $z \to 0$ limit and match term by term on both sides. From terms with $\log(z)$ we extract the $\gamma^{(1,0)}_{n}$ and from terms without $\log(z)$, we extract the $P^{(1,0)}_{n}$. We move on to $p=1$ case, where we again expand the right-hand side of the second line in \matching\ in $z \to 0$ limit. Using $\gamma^{(1,0)}_{n}$ and $P^{(1,0)}_{n}$, we extract $\gamma^{(1,1)}_{n}$ and $P^{(1,1)}_{n}$. Straightforwardly, one can continue this process and extract OPE data for any value of $p$.

By proceeding with this calculation to high enough values and $p$ one can notice that there is a simple expression for $\gamma^{(1,p)}_{n}$ given by
\eqn\gammaonefull{\gamma^{(1,p)}_{n}=(-1)^{p+1}\left({{1}\over{2}} (\DL - 1) \DL + 3 n^2-3(1- \DL) n\right),}
\noindent for all $p\geq 0$ and $n\geq 0$. 
Note that for $p=0$ this expression agrees with the one in \LiZBA.
There is no similar expression for $P^{(1,p)}_{n}$ so we list results for first $p$-s:
\eqn\ponefull{\eqalign{P^{(1,0)}_{n}=&-{{3}\over{4}} (\DL -1) \DL -{{3 \DL n}\over{2}},\cr
					   P^{(1,1)}_{n}=&3 (n-1) n-{{1}\over{4}} \Delta _L \left(\Delta _L \left(\Delta _L+6 n-6\right)+6 (n-4) n+5\right),\cr
					   P^{(1,2)}_{n}=&{{1}\over{8}} (\Delta _L (\Delta _L (\Delta _L^2+8 n \Delta _L+6 n (3 n-1)-13)+2 (n (3 n (2 n-5)-25)+6))\cr 
					   &-12 n (2 n^2+n-3)),\cr
					   P^{(1,3)}_{n}=&{{1}\over{120}} (180 n (n (3-(n-3) n)+5)-234) \Delta _L+3 n (n^3+n^2-2)\cr
					   &+{{1}\over{120}} \Delta _L^2 (-\Delta _L (\Delta _L \left(11 \Delta _L+90 n-20)+90 n (3 n-1)+55\right)\cr
					 &+90 (3-4 n) n^2+280).
}}

After the calculation of the OPE data at $\OO(\mu)$, one can fix the coefficients in the ansatz \subg\ by expanding the integrand of \sChCont\ up to $\OO(\mu^{2})$ and then integrating the obtained expression over $l$. The result of the integration is expanded near $\zbar \to 1$ and we collect the term that behaves as $(1-\zbar)^{-\DL +3}$. It depends on $z$, $n$ and OPE data $P^{(k,p)}_{n}$ and $\gamma^{(k,p)}_{n}$ for $k=1,2$ and $p=0,1$, but we are interested only in the part of this term that contains $\log^{2}(z)$. This part only depends on OPE data at $\OO(\mu)$, so it will be completely determined. We collect terms that behave as $(1-\zbar)^{-\DL +3} \log^{2}(z) z^{m} $. By expanding the ansatz \subg\ near $z \to 0$ we can collect terms that behave as $\log^{2}(z) z^{m}$ and by matching these to the ones calculated through S-channel, we obtain a system of linear equations for the coefficients in the ansatz. This system will be over-determined by taking $m$ to be large enough. Solving it for $m\leq 20$, we obtain

\eqn\sol{\eqalign{a_{33}&={{(\Delta_{L}-4) (\Delta_{L}-3) \Delta_{L}}\over{28800 (\Delta_{L}-2)}},\cr 
                  a_{24}&={{(\Delta_{L}-8) \Delta_{L}}\over{13440 (\Delta_{L}-2)}},\cr
                  a_{15}&={{\Delta_{L} (\Delta_{L}+1)}\over{5040 (\Delta_{L}-2)}},\cr
                  c_{16}&={25\over 396}b_{14}+{{\Delta _L \left(\Delta _L \left(\Delta _L \left(83-7 \Delta _L\right)+158\right)+108\right)}\over{3193344 \left(\Delta _L-3\right) \left(\Delta _L-2\right)}},\cr
                  c_{25}&=-{1\over 12}b_{14}+{{\Delta _L \left(\Delta _L \left(\Delta _L \left(\Delta _L+19\right)-146\right)-108\right)}\over{1451520 \left(\Delta _L-3\right) \left(\Delta _L-2\right)}},\cr
                  c_{34}&={{\left(\Delta _L-4\right) \Delta _L \left(11 \left(\Delta _L-4\right) \Delta _L-27\right)}\over{2419200 \left(\Delta _L-3\right) \left(\Delta _L-2\right)}}.
}
}

As expected, the coefficients $a_{mn}$ are identical to those in \secondCorrIntro . We are left with one undetermined coefficient. This is perhaps not surprising since we know from \FitzpatrickZQZ\ that the OPE coefficients of the subleading twist multi-stress tensor operators are not universal. This non-universality is introduced in our correlator through coefficient $b_{14}$. One can check that after inserting \sol\ to \subg\ the term that multiplies the unknown coefficient $b_{14}$ corresponds to the lightcone limit of the conformal block of the operator with dimension $\Delta=8$ and spin $s=2$. 
We thus conclude that $b_{14}$ is  the OPE coefficient of $:T_{\mu\alpha}{T^{\alpha}}_{\nu}:$,
\eqn\co{b_{14}=P^{(2)}_{8,2}.
}

Now, using \iden\ we can write the T-channel OPE coefficients for the remaining double-stress tensor operators with twist $\tau_{2,1}=6$ and conformal spin $\Delta + s \geq 14$. Explicitly, these are found to be given by

\eqn\opecnew{\eqalign{P_{10+2\ell, 4+2\ell}^{(2)}&={{ \sqrt{\pi} 2^{-4 \ell -17} \Gamma (2 \ell +7)}\over{ (\ell +4) (\ell +5) (\ell +6) (2 \ell +1) (2 \ell +3) (2 \ell +5) \Gamma \left(2 \ell +{{13}\over{2}}\right)}}\cr &\times {{\DL \over{(\Delta_{L}-3) (\Delta_{L}-2)}}}(a_{1,\ell }\Delta_L^3 + b_{1,\ell}\Delta_L^{2} + c_{1,\ell}\DL + d_{1,\ell}),
}}
\noindent where
\eqn\cone{\eqalign{a_{1,\ell}&= (\ell+2) (2 \ell+9) (\ell (2 \ell+13)+9),\cr
b_{1,\ell}&= 144-2 \ell (2 \ell+13) (\ell (2 \ell+13)+12),\cr
c_{1,\ell}&= \ell (2 \ell+13) (\ell (2 \ell+13)+33)+558,\cr
d_{1,\ell}&= 216.
}}
\noindent Here $\ell\geq 0$ and $P_{\Delta, s}^{(2)}$ is the sum of OPE coefficients of all operators with conformal dimension $\Delta$ and spin $s$. There is no way to distinguish operators with the same quantum numbers $\Delta$ and $s$ at this level in the large $C_{T}$ expansion. This type of degeneracy occurs for each conformal spin greater than 10 for twist $\tau_{2,1}=6$.  
Also, perfect agreement between \opecnew\ and all the OPE coefficients of double-stress tensor operators of twist $\tau_{2,1} = 6$ and spin $s>2$ calculated in \FitzpatrickZQZ\ is observed. Note that $P_{8,2}^{(2)}$ can not be found from \opecnew\ by setting $\ell=-1$, this would not agree with the result in \FitzpatrickZQZ. 
In Section 6 we rederive \opecnew\ using the Lorentzian inversion formula.

\subsec{Twist-eight double-stress tensors}

We follow the same logic as in the previous Section in order to write the subsubleading part of the stress tensor sector of the HHLL correlator in the lightcone limit at $\OO(\mu^{2})$. This part scales as $(1-\zbar)^{-\DL +4}$. Here, we include contributions coming from operators with twist $\tau_{2,2} = 8$. These operators can be grouped in three families and they are schematically written as  $:T_{\mu\nu}(\partial^{2})^{2}\partial_{\alpha_{1}}\ldots \partial_{\alpha_{2\ell}}T_{\rho\sigma}:$ with $\Delta=12+2\ell$ and $s=4+2\ell$, $:T_{\mu\beta}\partial^{2}\partial_{\alpha_{1}}\ldots\partial_{\alpha_{2\ell}}{T^{\beta}}_{\nu}:$ with $\Delta=10+2\ell$ and $s=2+2\ell$ and finally $:T_{\beta\gamma}\partial_{\alpha_{1}}\ldots \partial_{\alpha_{2\ell}}{T^{\beta \gamma}}:$ with $\Delta=8+2\ell$ and $s=2\ell$. Subtleties with regard to the contributions of the different families are discussed  in  Section 2.1. 

Once more, we need to include the contributions of lower twist operators, i.e. by expanding their conformal blocks as $\zbar \to 1$ up to order $(1-\zbar)^{4}$ and collect the additional $z$ dependence. Accordingly, we write the following ansatz 
\eqn\gsubsub{\eqalign{\GG^{(2,2)}(z) &= {1\over{\left(1-z\right)^{\DL}}} \Bigg(\left({{z (2 z-7)+11}\over{6 (z-1)^2}}\right)(a_{33}f_{3}^{2}+a_{24}f_{2}f_{4}+a_{15}f_{1}f_{5}   )    \cr &+\left({{2-z}\over{1-z}}\right)(b_{14}f_{1}f_{4}+c_{16}f_{1}f_{6}+c_{25}f_{2}f_{6}+c_{34}f_{3}f_{4})\cr &+(d_{17}f_{1}f_{7}+d_{26}f_{2}f_{6}+d_{35}f_{3}f_{5}+d_{44}f_{4}^{2}+e_{15}f_{1}f_{5}+e_{24}f_{2}f_{4}+e_{33}f_{3}^{2} \cr & + g_{13}f_{1}f_{3} + g_{22}f_{2}^{2})\Bigg),
}}
\noindent where $f_{a}$ means $f_{a}(z)$. Coefficients $a_{mn}$ and $c_{mn}$ are already calculated, while $b_{14}$ is  undetermined from the bootstrap. The linear dependence between certain products of $f_{a}(z)$ functions (for more details see Appendix A, in particular (A.2)) allows us to set three coefficients to zero, e.g., $g_{22}=0$, $e_{33}=0$ and $e_{24}=0$.

To fix the unknown coefficients in \gsubsub\ we match terms that behave as $(1-\zbar)^{-\DL +4}z^{m}\log^{2}z$ from the S-channel calculation of the correlator to terms with the same behavior in \gsubsub\  for small $z$. For the S-channel calculation, we need the OPE data at $\OO(\mu)$ up to $p=2$, given by \gammaonefull\ and \ponefull . We obtain an over-constrained system of linear equations, whose solution is
\eqn\coeftwo{\eqalign{d_{17}&={{9 e_{15}}\over{143}}+{{5 g_{13}}\over{4004}}+{{\Delta _L \left(\Delta _L \left(\Delta _L \left(\Delta _L \left(232-17 \Delta _L\right)+1009\right)+1908\right)+1008\right)}\over{115315200 \left(\Delta _L-4\right) \left(\Delta _L-3\right) \left(\Delta _L-2\right)}},\cr
					  d_{26}&=-{{e_{15}}\over{12}}+{{5 g_{13}}\over{1386}}-{{\Delta _L \left(\Delta _L \left(\left(\Delta _L-7\right) \Delta _L \left(11 \Delta _L-179\right)+3636\right)+2736\right)}\over{119750400 \left(\Delta _L-4\right) \left(\Delta _L-3\right) \left(\Delta _L-2\right)}},\cr
					  d_{35}&=-{{g_{13}}\over{180}}+{{\Delta _L \left(\Delta _L \left(\left(\Delta _L-7\right) \Delta _L \left(37 \Delta _L-13\right)+1332\right)+3312\right)}\over{108864000 \left(\Delta _L-4\right) \left(\Delta _L-3\right) \left(\Delta _L-2\right)}},\cr
					  d_{44}&={{\left(\Delta _L-6\right) \Delta _L \left(\Delta _L+2\right)}\over{9408000 \left(\Delta _L-2\right)}}.
}}

The undetermined coefficients $g_{13}$ and $e_{15}$ are related to the T-channel OPE coefficients $P_{8,0}^{(2)}$ and $P_{10,2}^{(2)}$ by the following relations
\eqn\rell{\eqalign{g_{13}&=P_{8,0}^{(2)},\cr
				   e_{15}&=P_{10,2}^{(2)}-{5\over 252}P_{8,0}^{(2)}.
}}
Here \noindent $P_{8,0}^{(2)}$ is the T-channel OPE coefficient of the operator of the schematic form $:T_{\alpha\beta}T^{\alpha\beta}:$, while $P_{10,2}^{(2)}$ is related to the OPE coefficients of the operators $:T_{\alpha\beta}\pa_{\mu_{1}}\pa_{\mu_{2}}T^{\alpha\beta}:$ and $:T_{\mu\alpha}\pa^{2}{T^{\alpha}}_{\nu}:$ which have the same quantum numbers $\Delta$ and $s$ and are thus indistinguishable at this order in large $C_T$ expansion. After inserting \rell\ and \coeftwo\ into \gsubsub\ one can  check that both $P_{8,0}^{(2)}$ and $P_{10,2}^{(2)}$ will be multiplied by the relevant lightcone conformal blocks.

Exactly as in the previous section, we can now extract the OPE coefficients $P_{\Delta, s}^{(2)}$ for operators with twist $\tau_{2,2}=8$ and $\Delta = 12+2\ell$, $s=4+2\ell$, for $\ell\geq 0$\foot{For each $\Delta=12+2\ell$ and $s=4+2\ell$ with $\ell\geq 0$ there is a triple degeneracy, because all three families of operators with twist $\tau_{2,2}=8$ will be mixed.} 
\eqn\opee{\eqalign{P_{12+2\ell, 4+2\ell}^{(2)}=&{{\sqrt{\pi } \DL 2^{-4 \ell-19}\Gamma (2 \ell+7)}\over{3 (\DL -4) (\DL -3) (\DL -2) (\ell+4) (\ell+5) }}\cr &\times{{a_{2,\ell} \Delta_L^{4} + b_{2,\ell}\Delta_L^{3} +c_{2,\ell}\Delta_L^{2}+d_{2,\ell}\DL + e_{2,\ell}   }\over{(\ell+6) (\ell+7) (2 \ell+1) (2 \ell+3) (2 \ell+5) \Gamma \left(2 \ell+{{15}\over{2}}\right)}},
}}
where
\eqn\coeftwoo{\eqalign{a_{2,\ell}&=\ell (2 \ell+15) (\ell (2 \ell+15) (\ell (2 \ell+15)+59)+1084)+6012,\cr
b_{2,\ell}&=14004-2 \ell (2 \ell+15) (\ell (2 \ell+15) (\ell (2 \ell+15)+32)-131),\cr
c_{2,\ell}&=\ell (2 \ell+15) (\ell(2 \ell+15) (\ell (2 \ell+15)+113)+4594)+60984,\cr
d_{2,\ell}&=216 (11 \ell (2 \ell+15)+302),\cr
e_{2,\ell}&=864 (\ell (2 \ell+15)+34).
}}
\noindent It is quite remarkable that these OPE coefficients are fixed purely by the bootstrap.

\subsec{Twist-ten double-stress tensors}

Now we want to go one step further and analyze the subsubsubleading contribution to the stress tensor sector of the HHLL correlator. This contribution scales as $(1-\zbar)^{-\DL +5}$ in the lightcone limit. We have to take in to account the double-stress tensor operators of twist $\tau_{2,3}=10$ in order to calculate this contribution. These operators can again be grouped in three families of the schematic form $:T_{\mu\nu}(\partial^{2})^{3}\partial_{\alpha_{1}}\ldots \partial_{\alpha_{2\ell}}T_{\rho\sigma}:$ with $\Delta=14+2\ell$ and $s=4+2\ell$, $:T_{\mu\beta}(\partial^{2})^{2}\partial_{\alpha_{1}}\ldots\partial_{\alpha_{2\ell}}{T^{\beta}}_{\nu}:$ with $\Delta=12+2\ell$ and $s=2+2\ell$ and finally $:T_{\beta\gamma}\partial^{2}\partial_{\alpha_{1}}\ldots \partial_{\alpha_{2\ell}}{T^{\beta \gamma}}:$ with $\Delta=10+2\ell$ and $s=2\ell$. 

In order to include contributions from lower twist operators we have to expand their conformal blocks up to $(1-\zbar)^{5}$ for $\zbar \to 1$. The ansatz takes the following form
\eqn\gsubsubsub{\eqalign{\GG^{(2,3)}(z) &= {1\over{\left(1-z\right)^{\DL}}} \Bigg(\left(  {{z ((13-3 z) z-23)+25}\over{12 (1-z)^3}}\right)   (a_{33}f_{3}^{2}+a_{24}f_{2}f_{4}+a_{15}f_{1}f_{5}   )    \cr &+  \left({{1}\over{(1-z)^2}}+{{1}\over{1-z}}+{{9}\over{10}}\right)    (b_{14}f_{1}f_{4}+c_{16}f_{1}f_{6}+c_{25}f_{2}f_{5}+c_{34}f_{3}f_{4})\cr &+ \left(  {{1}\over{1-z}}+ {{3}\over{2}}  \right)     (d_{17}f_{1}f_{7}+d_{26}f_{2}f_{6}+d_{35}f_{3}f_{5}  +d_{44}f_{4}^{2}+e_{15}f_{1}f_{5} + g_{13}f_{1}f_{3})
\cr &-g_{13} f_{3} + (h_{18}f_{1}f_{8}+h_{27}f_{2}f_{7}+h_{36}f_{3}f_{6}+h_{45}f_{4}f_{5}+j_{16}f_{1}f_{6}+j_{25}f_{2}f_{5} \cr & +j_{34}f_{3}f_{4} + i_{14}f_{1}f_{4} +i_{23}f_{2}f_{3})  \Bigg),
}}
\noindent with $h_{mn}$, $j_{mn}$ and $i_{mn}$, coefficients that we need to determine, and with $b_{14}$, $e_{15}$ and $g_{13}$ undetermined from the bootstrap. The term $g_{13}f_{3}(z)$ in the next-to-last line of the previous equation has its origin in the correction to the conformal block of operator $:T_{\alpha\beta}T^{\alpha\beta}:$. This operator has $\beta = \tau_{2,2} = 8$ which implies that both lines in the following expansion of the conformal block 
\eqn\conblotwoo{\eqalign{g_{8, 0}^{(0,0)}(1-z,1-\zbar)= &(1-\zbar)^{4}\left(1+(1-\zbar) \left({{3}\over{2}}+{{1}\over{1-z}}\right) + \OO\left((1-\zbar)^{2}\right)\right)f_{4}(z)\cr
&-(1-\zbar)^{5}\left(1+(1-\zbar)\left(2+{{1}\over{1-z}}\right)+\OO(\left(1-\zbar\right)^{2})\right)f_{3}(z)
}}
\noindent contribute. The contribution from the first line of \conblotwoo\ is included in the third line of \gsubsubsub , while we had to explicitly add the contribution from the second line. Using (A.1) and (A.3) we set $i_{23}=0$, $j_{34}=0$ and $j_{25}=0$.

From the S-channel calculation, we collect the terms in the correlator which behave as $(1-\zbar)^{-\DL +5}\log^{2}(z)z^{m}$ and are fixed in terms of OPE data at $\OO(\mu)$ for $p\leq 3$. By expanding \gsubsubsub\ near $z \to 0$ we obtain terms with the same behavior as linear functions of unknown coefficients and by matching them with the terms from the S-channel, we determine the unknown coefficients. These are

\eqn\solll{\eqalign{h_{18}&={{49 i_{14}}\over{38610}}+{{49 j_{16}}\over{780}}-{{\Delta _L \left(\Delta _L+1\right) \left(\Delta _L \left(\Delta _L \left(\Delta _L \left(47 \Delta _L-721\right)-5182\right)-15204\right)-13680\right)}\over{4942080000 \left(\Delta _L-5\right) \left(\Delta _L-4\right) \left(\Delta _L-3\right) \left(\Delta _L-2\right)}},\cr
				    h_{27}&={{5 i_{14}}\over{1404}}-{{j_{16}}\over{12}}-{{\Delta _L \left(\Delta _L+1\right) \left(\Delta _L \left(\Delta _L \left(\Delta _L \left(8 \Delta _L-229\right)+1097\right)+7224\right)+10080\right)}\over{1383782400 \left(\Delta _L-5\right) \left(\Delta _L-4\right) \left(\Delta _L-3\right) \left(\Delta _L-2\right)}},\cr
				    h_{36}&=-{{i_{14}}\over{180}}+{{\Delta _L \left(\Delta _L+1\right) \left(\Delta _L \left(\Delta _L \left(\Delta _L \left(34 \Delta _L-137\right)-1829\right)+5712\right)+23040\right)}\over{2661120000 \left(\Delta _L-5\right) \left(\Delta _L-4\right) \left(\Delta _L-3\right) \left(\Delta _L-2\right)}},\cr
				    h_{45}&={{\left(\Delta _L-6\right) \Delta _L \left(\Delta _L+1\right) \left(\Delta _L+2\right)}\over{62720000 \left(\Delta _L-3\right) \left(\Delta _L-2\right)}}.
					}}
\noindent Our approach does not allow us to determine the coefficients $j_{16}$ and $i_{14}$. These are related to the T-channel OPE coefficients of operators with twist $\tau_{2,3}=10$ and minimal conformal spin by
\eqn\relll{\eqalign{i_{14}&=P_{10,0}^{(2)},\cr
					j_{16}&=P_{12,2}^{(2)}-{2\over 99}P_{10,0}^{(2)}.
}}
Notice that, despite the fact that the $h_{mn}$ depend on the undetermined OPE data, we are able to extract all the OPE coefficients of double-stress tensors with twist $\tau_{2,3}=10$ and conformal spin $\Delta + s \geq 18$. Explicitly, they are given by:

\eqn\opeee{\eqalign{P_{14+2\ell,4+2\ell}^{(2)}=&{{\sqrt{\pi } 2^{-4 \ell-22}   \Gamma (2 \ell+9)  }\over{5 (2 \ell+1) (2 \ell+3) (2 \ell+5) (2 \ell+7)\Gamma \left(2 \ell+{{17}\over{2}}\right)}} \cr
                              & \times {{\DL (\DL +1)(a_{3,\ell}\Delta_L^4 +b_{3,\ell} \Delta_L^{3} + c_{3,\ell} \Delta_L^{2} +d_{3,\ell} \DL  + e_{3,\ell}  )}\over{(\ell+5) (\ell+6) (\ell+7) (\ell+8)(\DL -5) (\DL -4) (\DL -3) (\DL -2)}},
}}
where
\eqn\coefthr{\eqalign{ a_{3,\ell}= & \ell (2 \ell+17) (\ell (2 \ell+17) (\ell (2 \ell+17)+70)+1513)+9756  ,\cr
                              b_{3,\ell}=&  38232-2 (\ell-1) \ell (2 \ell+17) (2 \ell+19) (\ell (2 \ell+17)+44),\cr
                              c_{3,\ell}=& 196164 + \ell (17 + 2 \ell (11647 + \ell (17 + 2 \ell) (196 + \ell (17 + 2 \ell))) ,\cr
                              d_{3,\ell}=&  504 (647 + 19 \ell (17 + 2 \ell)),\cr
                              e_{3,\ell}=&  4320 (53 + \ell (17 + 2 \ell)).
}}

We expect that a similar picture is true for all subleading twist double-stress tensor operators. At $\OO(\mu^2)$, the ansatz for $\GG^{(2,m)}(z)$ will naturally include products of the type $f_{a}(z)f_{b}(z)$, such that $a+b=6+m$, together with $f_{1}(z)f_{3+m}(z)$ and $f_{1}(z)f_{1+m}(z)$. The coefficients of the latter two will be left undetermined from the lightcone bootstrap at every order in the lightcone expansion. Such coefficients will be related to the non-universal OPE coefficients of double-stress tensors with spin $s=0,2$ for a given twist. On the other hand, the coefficients of the products $f_{a}(z)f_{b}(z)$, with $a+b=6+m$, once determined, will allow us to extract the OPE coefficients of all double-stress tensors with conformal spin $\beta\geq 12+2m$. We expect them to be universal, despite the fact that the coefficients of the products $f_{a}(z)f_{b}(z)$, with $a+b=6+m$, will be plagued by the ambiguities present in the determination of the OPE coefficients of operators spin $s=0,2$ -- just as herein.

\newsec{Triple-stress tensors in four dimensions}

In this Section, we consider the stress tensor sector of the HHLL correlator at $\OO(\mu^{3})$ in $d=4$. The operators which contribute in the T-channel are triple-stress tensors. Since we are interested in the lightcone limit $1-\zbar \ll 1$, we consider contributions of operators with low twist. 
Triple-stress tensors with minimal twist can be written in the schematic form $:T_{\mu\nu}T_{\rho\sigma}\partial_{\alpha_{1}}\ldots\partial_{\alpha_{2\ell}}T_{\eta\xi}:$. These operators have twist $\tau_{3,{\rm min}}=6$ and their contribution to the HHLL correlator in the lightcone limit was found in \KarlssonDBD:
\eqn\gt{\eqalign{
  &\GG^{(3,0)}(z) = {1\over{(1-z)^{\Delta_{L}}}} \Big(a_{117} f_{1}(z)^2 f_7(z) +a_{126}f_1(z)f_2(z)f_6(z) \cr 
  &  +a_{135}f_1(z)f_3(z)f_5(z) +a_{225}f_2(z)^2f_5(z)+a_{234}f_2(z)f_3(z)f_4(z)+a_{333}f_3(z)^3\Big),
}}
where the coefficients $a_{ikl}$ are 
\eqn\ResultTripleFourD{\eqalign{
    a_{117} &= {5\DL(\DL+1)(\DL+2)\over 768768(\DL-2)(\DL-3)},\cr
    a_{126} &= {5\DL( 5\Delta_L^2-57\DL-50)\over 6386688(\DL-2)(\DL-3)},\cr
    a_{135} &= {\DL (2\Delta_L^2-11\DL-9)\over 1209600 (\DL-3)},\cr 
    a_{225} &= -{\DL(7\Delta_L^2-51\DL-70)\over 2903040(\DL-2)(\DL-3)},\cr 
    a_{234} &= {\DL(\DL-4)(3\Delta_L^2-17\DL+4)\over 4838400(\DL-2)(\DL-3)},\cr 
    a_{333} &= {\DL(\DL-4)(\Delta_L^3-16\Delta_L^2+51\DL+24)\over 10368000(\DL-2)(\DL-3)}.
}}

\subsec{Twist-eight triple-stress tensors}

We now consider the subleading contributions at $\OO(\mu^3)$ coming from triple-stress tensor operators with twist $\tau_{3,1}=8$.
There are two families of such operators, these can be schematically written as $:T_{\mu\nu}T_{\rho\alpha}\partial_{\alpha_{1}}\ldots\partial_{\alpha_{2\ell}} {T^{\alpha}}_{\xi}:$ with $\Delta = 12+2\ell$ and spin $s=4+2\ell$ and $:T_{\mu\nu}T_{\rho\sigma}\partial^{2}\partial_{\alpha_{1}}\ldots\partial_{\alpha_{2\ell}}T_{\eta\xi}:$ with $\Delta=14+2\ell$ and spin $s=6+2\ell$. 
The conformal spins of these  families are $\beta=16+4\ell$ and $\beta = 20+4\ell$, respectively, so we expect products of three $f_{a}(z)$ functions such that their indices add up to 8 and 10. The contribution to the correlator of these operators scales as $(1-\zbar)^{-\DL +4}$ for $\zbar \to 1$. This implies that one needs to include the contribution from the minimal twist triple-stress tensor operators (due to corrections to their conformal blocks).

Our ansatz takes the form
\eqn\gtsub{\eqalign{
  \GG^{(3,1)}(z) &= {1\over{(1-z)^{\Delta_{L}}}} \Bigg( \left({{2-z}\over{1-z}}\right)  (a_{117} f_{1}^2 f_7 +a_{126}f_1 f_2 f_6 +a_{135}f_1 f_3 f_5 +a_{225}f_2 ^2 f_5  \cr 
  &  +a_{234}f_2 f_3 f_4 +a_{333}f_3 ^3) +(b_{116} f_6 f_1^2+b_{134} f_3 f_4 f_1+b_{125} f_2 f_5 f_1 +b_{233} f_2 f_3^2\cr 
  &+b_{224} f_2^2 f_4+c_{118} f_8 f_1^2  +c_{145} f_4 f_5 f_1+c_{136} f_3 f_6 f_1+c_{127} f_2 f_7 f_1+c_{244} f_2 f_4^2\cr 
  & +c_{334} f_3^2 f_4 +c_{235} f_2 f_3 f_5+c_{226} f_2^2 f_6)     \Bigg),
}}
where $a_{jkl}$ are given in \ResultTripleFourD . The linear dependence between products of three $f_{a}$ functions, with explicit relations given in Appendix A, allows us to set the following coefficients to zero
\eqn\zero{b_{125}=b_{134}=b_{224}=b_{233}=c_{136}=0.
}

To fix the coefficients $b_{116}$ and $c_{jkl}$ we perform an S-channel calculation up to $\OO(\mu^3)$. The relevant terms now scale as $(1-\zbar)^{-\DL +4}\log^{3}(z) z^{m}$ and
$(1-\zbar)^{-\DL + 4}\log^{2}(z)z^{m}$ when $\zbar \to 1$ and $z\to 0$. 

We fix the S-channel OPE data at $\OO(\mu^{2})$ using the results of the previous Section, specifically eqs. \subg , \gsubsub\ and \gsubsubsub . 
Since the OPE coefficients of double-stress operators of spin $0$ and $2$ are left undetermined, the S-channel OPE data is fixed in terms of these. 
Concretely, $\gamma^{(2,0)}_{n}$ and $P^{(2,0)}_{n}$ are completely determined since the leading-twist OPE coefficients are known and universal, while $\gamma^{(2,1)}_{n}$ and $P^{(2,1)}_{n}$ depend on $b_{14}$, $\gamma^{(2,2)}_{n}$ and $P^{(2,2)}_{n}$ depend on $b_{14}$, $g_{13}$ and $e_{15}$ and so on.\foot{Explicit expressions for the S-channel OPE data are too cumbersome to quote here.}

We were able to fix all the unknown coefficients in the ansatz \gtsub\ using bootstrap. Crucially, there are no spin $s=0,2$ operators that contribute at this level. Here, we list two of the coefficients while all others can be found in Appendix B. 

\eqn\solts{\eqalign{b_{116}=&-{{\Delta _L \left(\Delta _L+3\right) \left(\Delta _L \left(\Delta _L \left(\Delta _L \left(1001 \Delta _L+387\right)-4326\right)+13828\right)+5040\right)}\over{10378368000 \left(\Delta _L-4\right) \left(\Delta _L-3\right) \left(\Delta _L-2\right)}}\cr
							&+{{b_{14} \left(\Delta _L \left(143 \Delta _L+427\right)+540\right)}\over{17160 \left(\Delta _L-4\right)}},\cr
					c_{118}=&{{7 \left(\Delta _L+3\right) \left(604800 b_{14} \left(\Delta _L^2-5 \Delta _L+6\right)+\Delta _L \left(-21 \Delta _L^3+229 \Delta _L^2+414 \Delta _L+284\right)\right)}\over{856627200 \left(\Delta _L^3-9 \Delta _L^2+26 \Delta _L-24\right)}}.
}}

\noindent Notice that they depend on $b_{14}$. This is because the anomalous dimensions at $\OO(\mu^{2})$, $\gamma^{(2,2)}_{n}$ depend on it. Moreover, no OPE coefficient of triple-stress tensors with twist $\tau_{3,1}=10$ is universal since all of them depend on $b_{14}$. These OPE coefficients can be written in the form of a finite sum, similarly to what happens for the OPE coefficients of leading twist triple-stress tensor, given in \KarlssonDBD . We define $i_{1}(r,q)$ and $i_{2}(r,q)$ as
\eqn\io{\eqalign{i_{1}(r ,q)&= b_{116} p(1,1,r) p(2 r+2,6,q),
}}
\noindent and
\eqn\itt{\eqalign{i_{2}(r,q)&= c_{118} p(1,1,r) p(2 r+2,8,q) + c_{127} p(1,2,r) p(2 r+3,7,q) \cr
& + c_{145} p(1,4,r) p(2 r+5,5,q) + c_{226} p(2,2,r) p(2 r+4,6,q)\cr
& + c_{235} p(2,3,r) p(2 r+5,5,q) + c_{244} p(2,4,r) p(2 r+6,4,q)\cr
& + c_{334} p(3,3,r) p(2 r+6,4,q),
}}
\noindent where $p(a,b,\ell)$ are given by \deffp . The OPE coefficients can be written as
\eqn\koeficijenti{P^{(3)}_{14+2\ell,6+2\ell}=\sum_{r=0}^{\ell+1}i_1(r,\ell+1-r)+\sum_{r=0}^{\ell}i_2(r,\ell-r),
}
\noindent for $k\geq 0$, while $P^{(3)}_{12,4}=i_{1}(0,0)=b_{116}$. We give the explicit expressions for some OPE coefficients in Appendix D. 

\subsec{Twist-ten triple-stress tensors}

Here, we consider the contribution of triple-stress tensor operators of twist $\tau_{3,2}=10$. These operators can be divided in three families of the schematic form $:T_{\mu\nu}T_{\alpha\beta}\pa_{\mu_{1}}\ldots\pa_{\mu_{2\ell}}(\pa^{2})^{2}T_{\rho\sigma}:$ with conformal dimension $\Delta = 16+2\ell$ and spin $s=6+2\ell$, $:T_{\mu\nu}T_{\alpha\beta}\pa_{\mu_{1}}\ldots\pa_{\mu_{2\ell}}\pa^{2} {T^{\beta}}_{\rho}:$ with $\Delta=14+2\ell$ and $s=4+2\ell$ and finally $:T_{\mu\alpha}T_{\nu\beta}\pa_{\mu_{1}}\ldots\pa_{\mu_{2\ell}}T^{\alpha\beta}:$ with $\Delta=12+2\ell$ and $s=2+2\ell$. One can see that in the last family an operator of spin $s=2$ is included.

An appropriate ansatz in this case is
\eqn\gtsubsub{\eqalign{
  &\GG^{(3,2)}(z,\zbar) = {{1}\over{(1-z)^{\Delta_{L}}}} \Bigg( \left({{144 z^2-448 z+464}\over{160 (z-1)^2}}\right)  (a_{117} f_{1}^2 f_7 +a_{126}f_1 f_2 f_6  +a_{135}f_1 f_3 f_5 \cr 
  & +a_{225}f_2 ^2 f_5  +a_{234}f_2 f_3 f_4 +a_{333}f_3 ^3)+\left({{1}\over{1-z}}+{{3}\over{2}}\right)(b_{116} f_6 f_1^2+c_{118} f_8 f_1^2  +c_{145} f_4 f_5 f_1\cr 
  &+c_{127} f_2 f_7 f_1+c_{244} f_2 f_4^2+c_{334} f_3^2 f_4+c_{235} f_2 f_3 f_5+c_{226} f_2^2 f_6)   +(d_{117}f_{1}^{2}f_{7}+e_{115}f_{1}^{2}f_{5}\cr
  &+g_{119}f_{1}^{2}f_{9}+g_{128}f_{1}f_{2}f_{8}+g_{155}f_{1}f_{5}^{2}+g_{227}f_{2}^{2}f_{7}+g_{236}f_{2}f_{3}f_{6}+g_{245}f_{2}f_{4}f_{5}+g_{335}f_{3}^{2}f_{5}\cr
  &+g_{344}f_{3}f_{4}^{2})  \Bigg),\cr 
}}
\noindent where $f_{a}=f_{a}(z)$ and we have included only the linearly independent products of these functions.

The lightcone bootstrap fixes all coefficients except $e_{115}$. One can check that this is exactly the OPE coefficient $P_{12,2}^{(3)}$ of the spin-2 operator $:T_{\mu\alpha}T_{\nu\beta}T^{\alpha\beta}:$ with $\Delta=12$ and spin $s=2$ 
\eqn\coefff{
e_{115}=P_{12,2}^{(3)}.
}
All other coefficients can be found in Appendix B. Notice that all coefficients depend on $b_{14}$, $g_{13}$ and $e_{15}$ because the S-channel OPE data at $\OO(\mu^{2})$ depend on them. 

Again, we write the OPE coefficients for all triple-stress tensor operators with twist $\tau_{3,2}=10$ and $\beta\geq 18$ in the form of a finite sum. We define $j_{1}(r,q)$, $j_{2}(r,q)$ and $j_{3}(r,q)$ as
\eqn\jo{j_{1}(r,q)=e_{115} p(1,1,r) p(2 r+2,5,q),
}
\eqn\jtw{j_{2}(r,q)=d_{117} p(1,1,r) p(2 r+2,7,q)
}
\noindent and
\eqn\jth{\eqalign{j_{3}(r,q)&= g_{119} p(1,1,r) p(2 r+2,9,q) + g_{128} p(1,2,r) p(2 r+3,8,q)\cr
& +g_{155} p(1,5,r) p(2 r+6,5,q)+g_{227} p(2,2,r) p(2 r+4,7,q)\cr
& +g_{236} p(2,3,r) p(2 r+5,6,q)+g_{245} p(2,4,r) p(2 r+6,5,q)\cr
& +g_{335} p(3,3,r) p(2 r+6,5,q)+g_{344} p(3,4,r) p(2 r+7,4,q),
}}
\noindent where $p(a,b,\ell)$ is given by \deffp . The OPE coefficients can now be written as
\eqn\koeficijentii{P^{(3)}_{16+2\ell,6+2\ell}=\sum_{r=0}^{\ell+2}j_1(r,\ell+2-r)+\sum_{r=0}^{\ell+1}j_2(r,\ell+1-r)+\sum_{r=0}^{\ell}j(r,\ell-r),
}
\noindent for $\ell\geq 0$, while 
\eqn\kk{
P^{(3)}_{14,4}=j_1(0,1) +j_1(1,0)+j_2(0,0).
}

Finally, we conclude that the stress tensor sector of the HHLL correlator to all orders in $\mu$ and in the lightcone expansion will take a similar form in terms of products of $f_{a}$ functions. 
One should be able to completely fix the coefficients, except for terms that correspond to the OPE coefficients of multi-stress tensor operators with spin $s=0,2$, using the lightcone bootstrap. 

\newsec{Holographic phase shift and multi-stress tensors}

In this Section, we demonstrate how to calculate the T-channel OPE coefficients of spin-2 operators (up to undetermined spin-0 data) which are left undetermined after the lightcone bootstrap, using a gravitational calculation of the scattering phase shift. We are interested in the scattering phase shift -- or eikonal phase -- resulting from the eikonal resummation of graviton exchanges when a fast particle is scattered by a black hole\foot{For CFT approach to the Regge scattering of scalar particles in pure AdS see \refs{\CornalbaXK\CornalbaXM\CornalbaZB\CornalbaFS\CostaCB\FitzpatrickEFK-\MeltzerPYL}.}. 
Seeking to explore the universality properties of the undetermined OPE coefficients of the previous section, we perform the calculation in Gauss-Bonnet gravity extending the results of \KulaxiziDXO\ to this case. We argue that the phase shift in the large impact parameter limit is independent of higher-derivative corrections to the dual gravitational lagrangian. This is consistent with the universality of the minimal-twist multi-stress tensor sector in the dual CFT. On the other hand, we observe that the subleading OPE data of spin-2 multi-stress tensors depend explicitly on the Gauss-Bonnet coupling $\lambda_{\rm GB}$.

The computation involves performing an inverse Fourier transform of the exponential of the phase shift in the large impact parameter expansion, to obtain the HHLL correlator in position space\foot{Recall that the exponential of the phase shift corresponds to the Regge limit of HHLL four-point function in momentum space \KulaxiziDXO.}. This is done  following the approach of \KarlssonTXU. Comparison with the expressions for the HHLL correlator in the lightcone limit requires analytically continuing the results of Sections 3 and 4 and taking the limit $z \to 1$. Identifying terms in the HHLL four-point function with the same large impact parameter and 
$z \to 1$ behavior allows us to extract the spin-2 OPE coefficients of the double- and triple-stress tensor operators
(up to undetermined spin zero data). 

\subsec{Universality of the phase shift in the large impact parameter limit}

In this subsection, we consider Gauss-Bonnet gravity in $(d+1)$--dimensions and argue that the phase shift obtained by a highly energetic particle traveling in a spherical AdS-Schwarzschild background is independent of the Gauss-Bonnet coupling $\lambda_{\rm GB}$ in the large impact parameter limit. 

The action of Gauss-Bonnet gravity in $(d+1)$-dimensional spacetime is  
\eqn\GBAction{
  S={1\over 16\pi G}\int d^{d+1}\sqrt{-g}\left(R+{d(d-1)\over \ell^2}+{{\tilde{\lambda}_{\rm GB}}\over {(d-2)(d-3)}}(R_{\mu\nu\gamma\delta}R^{\mu\nu\gamma\delta}-4R_{\mu\nu}R^{\mu\nu}+R^2)\right),
}
where the coupling parameter $\tilde{\lambda}_{\rm GB}$ is measured in units of the cosmological constant $\ell$: $\tilde{\lambda}_{\rm GB}=\lambda_{\rm GB} \ell^2$, with $\lambda_{\rm GB}$ being a dimensionless coefficient.
The AdS-Schwarzschild black hole metric which is a solution of the Gauss-Bonnet theory is given by \refs{\BoulwareWK-\CaiDZ}:
\eqn\Metric{
  ds^2=-r_{AdS}^{2} f(r)dt^2+{dr^2\over f(r)}+r^2d\Omega_{d-1}^2,
}
where 
\eqn\f{
  f(r)= 1+{r^2\over 2\lambda_{\rm GB}}\left(1-\sqrt{1-4\lambda_{\rm GB}(1 -{\tmu\over r^{d}}})\right),
}
with
\eqn\tmudef{
    \tmu={16\pi G M\over (d-1)\Omega_{d-1}\ell^{d-2}}, \qquad {\mu}={\tmu \over {r_{AdS}^{d-2}\sqrt{1-4\lambda_{\rm GB}}}},
}
and
\eqn\radsdef{r_{AdS}=\Big({1\over 2}(1+\sqrt{1-4\lambda_{\rm GB}})\Big)^{1/2}}
where $\Omega_{d-1}$ is the surface area of a $(d-1)$-dimensional unit sphere embedded in $d$-dimensional Euclidean space. The metric is normalized such that the speed of light is equal to $1$ at the boundary (i.e. $g_{tt}/g_{\phi \phi}\rightarrow 1$ as $r\rightarrow \infty$) and all dimensionful parameters are measured in units of $\ell$. The product $(\ell r_{AdS})$ is the radius of the asymptotic Anti-de Sitter space. 

The two conserved charges along the geodesics, $p^t$ and $p^{\phi}$, are
\eqn\Charges{\eqalign{
    p^t &= r_{AdS}^{2} f(r){dt\over d\lambda},\cr
    p^{\phi} &=r^2{d\phi\over d\lambda}.
}}
where $\lambda$ denotes an affine parameter. Null geodesics are described by the following equation,
\eqn\NullGeodesic{
  {1\over2}\left({dr\over d\lambda}\right)^2+{(p^\phi)^2\over 2r^2}f(r)={1\over 2}{(p^t)^2\over r_{AdS}^{2}}\,.
}
similarly to Einstein gravity. 

A light particle, starting from the boundary, traversing the bulk and reemerging on the boundary experiences a time delay and a path deflection given by :
\eqn\TimeDelayAndDeflectionAngle{\eqalign{
    \Delta t = 2\int_{r_0}^\infty {dr \over r_{AdS} f(r) \sqrt{1-{\alpha^2 {r_{AdS}^{2} \over r^2}}f(r)}},\cr
    \Delta \phi = 2 \alpha \, r_{AdS}\int_{r_0}^\infty {dr\over r^2\sqrt{1-{\alpha^2 {r_{AdS}^{2} \over r^2}}f(r)}},\cr
}}
where $\alpha=p^\phi/p^t$ and $r_0$ the impact parameter determined by ${dr\over d\lambda}|_{r(\lambda)=r_0}=0$, {\it i.e.}, 
\eqn\ImpactParameter{
  1-\alpha^{2}{r_{AdS}^{2}\over r_0^2}f(r_0)=0.
}
Defining the phase shift as $\delta=-p\cdot \Delta x=p^t\Delta t-p^\phi \Delta\phi$, we find that
\eqn\PhaseShift{
  \delta = 2{p^t\over r_{AdS}}\int_{r_0}^\infty {dr\over f(r)}\sqrt{1-{\alpha^2 {r_{AdS}^{2} \over r^2}}f(r)}.
}

Just as in \KulaxiziDXO, we are interested in expanding the phase shift order by order in $\mu$. It is easy to see that in terms of CFT data $\mu$ can be expressed as
\eqn\aP{
    \mu ={4\over (d-1)^2}{\Gamma(d+2)\over \Gamma(d/2)^{2} }{\Delta_H\over C_T},
}
which is consistent with \mudef. Here $C_T$ is the central charge of the dual conformal theory \BuchelSK :
\eqn\CentralCharge{
  C_T = {\pi^{{d\over 2}-1}\over 2(d-1)}{\Gamma(d+2)\over \Gamma(d/2)^{3} G} ({r_{AdS}\ell})^{d-1} \sqrt{1-4\lambda_{\rm GB}},\,
}
and $\Delta_H=M\ell  r_{AdS}$.

In order to calculate the phase shift, we introduce a new variable $y$, given by $y={r_{0}\over r}$. Using this variable \PhaseShift\ can be written as:
\eqn\phaseshiftone{
  \delta = 2{p^t r_{0} \over r_{AdS}}\int_{0}^{1} {dy\over y^{2}f({r_{0}\over y})}\left(1-{\alpha^2 {r_{AdS}^{2} y^2 \over {r_{0}}^2}}f({r_{0}\over y})\right)^{1/2}.
} 
Expanding the phase shift  
\eqn\phs{\delta = \sum_{k=0}^{\infty}\mu^{k}\delta^{(k)},
}
and solving \ImpactParameter\ perturbatively in ${\mu}$ reads
\eqn\rzero{r_{0}=b-{b^{3-d}\over {2r_{AdS}^{2-d}}}\mu + {b^{3-2d}\over {8r_{AdS}^{4-2d}}} \left(b^{2}(3-2d)+{4\lambda_{\rm GB} \over {\sqrt{1-4\lambda_{\rm GB}}}}\right)\mu^{2}+\OO(\mu^{3}).
}

\noindent Generically, we get an expansion of the form
\eqn\turningp{r_{0}=b+\sum_{k=1}^{\infty} a_{k}{\mu}^{k},}
where the $a_{k}$, which depend on $b$, in the large impact parameter limit ($b\to \infty$) behave as 
\eqn\aklargeb{a_{k}\propto b \left({r_{AdS}\over b}\right)^{k(d-2)}\,.} 
Notice that there is no explicit $\lambda_{\rm GB}$ dependence in the leading term\foot{Except the overall dependence on $r_{AdS}$.}, since the  metric \Metric\ approaches the one in pure GR. 

To study the leading behavior of the phase shift for large impact parameters it is convenient to define a function $g(x)$ as
\eqn\deftwo{
g(x)=r_{AdS}^{2} {f(x)\over x^{2}},
} 
\noindent with $f$ given by \f , and denote the integrand of \phaseshiftone\ by $h\left(g\left({r_{0}\over y}\right)\right)$, with
\eqn\newh{h(x)={1 \over x}\sqrt{1-\alpha^{2} x},
}
to express \phaseshiftone\ as
\eqn\phsh{\delta=2p^{t}\left({r_{AdS}\over r_{0}}\right)\int_{0}^{1} h\left(g\left({r_{0}\over y}\right)\right) dy.
}
In practice, to calculate the phase shift in the large impact parameter limit, we first expand the integrand of \phsh\ in powers of $\mu$, perform the integration with respect to $y$, and then expand the result in powers of $b$. The $b$-dependence of $\delta^{(k)}$ is therefore fixed before the integration and the integral just determines the overall numerical factor (assuming that it is convergent). 


We can immediately see that $g\left({r_{0}\over y}\right)$ depends on $\mu$ explicitly and implicitly through $r_{0}(\mu)$ in \rzero . In order to make this clear we write $g\left({r_{0}\over y},\mu\right)$ instead of just $g\left({r_{0}\over y}\right)$. Defining $g^{(n,m)}\left({b\over y},0\right)$ as 
\eqn\derdef{
g^{(n,m)}\left({b\over y},0\right)={{\partial^{n}\partial^{m}}\over{\partial r_{0}^{n}\partial \mu^{m}}}g\left({r_{0}\over y},\mu\right)\Big|_{r_{0}=b, \mu=0} .
}
allows us to write the following expansion for $h\left(g\left({{r_{0}\over y},\mu}\right)\right)$:
\eqn\expantwo{\eqalign{ h\left(g\left({r_{0}/y,\mu}\right)\right) =& h(g(b/y,0)) + \mu h'(g(b/y,0))\left(g^{(0,1)}(b/y,0) + a_1 g^{(1,0)}(b/y,0)\right) \cr & + {\mu^{2}\over 2}h''(g(b/y,0))\left(g^{(0,1)}(b/y,0) + a_1 g^{(1,0)}(b/y,0)\right)^{2}  \cr & + {\mu^{2}\over 2}h'(g(b/y,0))\Big(g^{(0,2)}(b/y,0)+2 a_2 g^{(1,0)}(b/y,0) \cr
&+ 2 a_1 g^{(1,1)}(b/y,0)+a_1^{2} g^{(2,0)}(b/y,0)    \Big) + \OO(\mu^{3}),
}}
\noindent where $a_{k}$ are the coefficients appearing in \turningp . It is clear that at each order in the $\mu$-expansion we will have a sum of products composed from derivatives of $h(x)$ and sums of the form 
\eqn\sumbigps{\sum\limits_{  \{ \scriptscriptstyle k_{i}:  \sum\limits_{i=1}^{p}k_{i}\leqslant n\}}^{\quad} a_{k_{1}}a_{k_{2}}\ldots a_{k_{p}} g^{(p,n-\sum_{i=1}^{p}k_{i})}(b/y,0)\,.}

Notice first that $g(b/y,0)$, $g^{(m,0)}(b/y,0)$ and $g^{(m,1)}(b/y,0)$ do not depend on $\lambda_{\rm GB}$ as can be  seen from \deftwo. The same is true for $h^{(n)}(g(b/y,0))$ for any $n$ as follows from \newh . On the contrary, $g^{(m,n)}(b/y,0)$ with $n\geq 2$ depend explicitly on $\lgb$. It is then evident that any dependence on $\lgb$ will come from terms like the ones in parenthesis in \expantwo\ which are of the type \sumbigps. We will now show that all the terms in such sums which contain $\lgb$, are subleading in the large impact parameter limit.

Recall that $a_{k}\propto b^{1-k(d-2)}$ for $k\geq 1$. Using \deftwo\ one can  check that $g^{(m,n)}(b/y,0)\propto b^{-m-nd}$ for $n>0$ and $g^{(m,0)}(b/y,0)\propto b^{-m-2}$. 
We thus need to spearately consider two cases: products of the form $a_{k_{1}}a_{k_{2}}\ldots a_{k_{p}} g^{(p,n-q)}(b/y,0)$, with $q=\sum_{i=1}^{p}k_{i}$ and $q<n$ and products of the form $a_{k_{1}}a_{k_{2}}\ldots a_{k_{p}}g^{(p,0)}(b/y,0)$ for which $q=n$. 

The former behave as
\eqn\countt{a_{k_{1}}a_{k_{2}}\ldots a_{k_{p}} g^{(p,n-q)}(b/y,0)\propto {1\over b^{nd-2q}}.
}
\noindent Clearly, the leading behavior in the large impact parameter regime corresponds in this case to $q=n-1$, recall, however, that $g^{(p,1)}$ does not depend on $\lgb$. The behavior of the latter terms is 
\eqn\counttt{a_{k_{1}}a_{k_{2}}\ldots a_{k_{p}}g^{(p,0)}(b/y,0)\propto {1\over b^{nd-2(n-1)}}.
}
\noindent which is again independent of $\lgb$. The conclusion is that the leading behavior in the large impact parameter regime comes from terms containing $g^{(p,0)}(b/y,0)$ and $g^{(p,1)}(b/y,0)$ that do not contain $\lgb$. 

One can extend these considerations straightforwardly to any gravitational theory that contains a spherical black hole with a metric given by
\eqn\generalMetric{
  ds^2 = -(1+r^2\tilde{f}(r))dt^2+{dr^2\over 1+r^2\tilde{h}(r)}+r^2d\Omega_{d-1}^2
}
where the functions $\tilde{f}(r)$ and $\tilde{h}(r)$ admit an expansion of the following form in the large $r$ limit:
\eqn\fandh{\eqalign{
  \tilde{f}(r) &= 1-\sum_{n=0}^\infty {\tilde{f}_{nd}\over r^{nd}}=1-{\tilde{f}_0\over r^d}-{\tilde{f}_{d}\over r^{2d}}-\ldots\cr
  \tilde{h}(r) &= 1-\sum_{n=0}^\infty {\tilde{h}_{nd}\over r^{nd}}=1-{\tilde{h}_0\over r^d}-{\tilde{h}_{d}\over r^{2d}}-\ldots,
}}
for some constants $\tilde{f}_{nd}$ and $\tilde{h}_{nd}$ (these are the spherical black hole metrics considered in eqs.\ (5.1) and (5.10) in \FitzpatrickZQZ).

\subsec{Spin-2 multi-stress tensor OPE data from the gravitational phase shift}
\noindent The gravitational phase shift in a black hole background is related to the lightcone HHLL four-point function discussed extensively in this article. In the following, we will exploit the precise relationship between the two to extract the OPE data of multi-stress tensor operators of spin-2 in the dual conformal field theory (modulo spin zero data). While the explicit procedure can be worked out for arbitrary multi-stress tensors, we will herein focus on double and triple-stress tensor operators, which control the $\OO(\mu^2)$ and $\OO(\mu^3)$ lightcone behavior of the HHLL correlation function.

{\it\subsubsec{{ The phase shift in Gauss-Bonnet gravity to $\OO(\mu^3)$.}}}
\noindent In this section, we focus on the gravity side and determine the phase shift order by order in $\mu$ up to $\OO(\mu^3)$ relevant for this article. Starting from $\OO(\mu^0)$ we consider the following expression
\eqn\deltazero{\delta^{(0)}=2 b \,p^{t}\, r_{AdS}\sqrt{1-\alpha^{2}}\int_{0}^{1}{\sqrt{1-y^{2}}\over{b^{2}+r_{AdS}^{2}y^{2}}}dy.
}
Evaluating this integral and using the following notation $p^{\pm}=p^{t}\pm p^{\phi}$, $-p^{2}=p^{+}p^{-}$, leads to
\eqn\deltazerof{\delta^{(0)}=\pi p^{-}.
} 
This is of course none other but the ``phase shift'' in pure AdS space.

At $\OO(\mu)$ the result is the same as in \KulaxiziDXO, where Einstein gravity was considered, 
\eqn\deltaone{\delta^{(1)}=\sqrt{-p^{2}}\left({b\over r_{AdS}}\right)^{1-d}\left({{d-1}\over 2}\right) B\Big[{{d-1}\over 2},{3\over 2}\Big]{}_2F_{1}\Big(1,{{d-1}\over 2},{d\over 2}+1,-{{r_{AdS}^{2}}\over b^{2}}\Big).
}
At this order, the phase shift depends only on the single graviton exchange, which is unaffected by the higher derivative terms in the gravitational action. 
According to the holographic dictionary, the exchange of a single graviton is related to the exchange of a single stress tensor in the T-channel.
The corresponding OPE coefficient is fixed by the Ward identity, so it does not depend on the details of the theory.

We now consider the phase shift at higher orders in $\mu$. For convenience herein all results are presented in $d=4$. At $\OO(\mu^{2})$, using the technique presented in the previous subsection, we find that:
\eqn\secondOrderdFourD{\eqalign{
    \delta^{(2)}=&{7\pi \over 8}\sqrt{-p^2}\Bigg[5{b\over r_{AdS}}(\sqrt{1+{r_{AdS}^2\over b^2}}-1)-{5\over 2}{r_{AdS}\over b}+{5\over 4}{r_{AdS}^3\over b^{3}}\cr 
    &+{\lgb\over{r_{AdS}^{2} \sqrt{1-4\lgb}}}\Big(4{b\over r_{AdS}}(\sqrt{1+{r_{AdS}^2\over b^2}}-1)-2{r_{AdS}\over b}+{1\over 2}{r_{AdS}^3\over b^{3}}-{1\over 4}{r_{AdS}^5\over b^{5}}\Big)\Bigg].
}}
\noindent In the lightcone limit ($b\to \infty$) this reduces to 
\eqn\DeltaLC{
  \delta^{(2)}\btoinf {35\pi \sqrt{-p^2}r_{AdS}^{5}\over 128b^5}-{35\pi \sqrt{-p^2}r_{AdS}^{7}\over 1024 b^7}\left(5+{4\lgb\over{r_{AdS}^2\sqrt{1-4\lgb}}}\right)+\ldots .}
\noindent We explicitly see that the leading contribution does not depend on $\lgb$, while the subleading does. 

Let us denote $\delta^{(2)}_{\rm GR}$ to be equal to \secondOrderdFourD\ when $\lgb=0$,
\eqn\GrResult{
  \delta^{(2)}_{\rm GR} ={35\pi r_{AdS}^{5}\sqrt{-p^2}\over 128b^5}{}_2F_1(1,{5\over 2},4,-{r_{AdS}^2\over b^2}),
}
which is the pure Einstein gravity result for the phase shift at $\OO(\mu^2)$.
Then $\delta^{(2)}$ can be written as 
\eqn\SecondRewr{
  \delta^{(2)} = \delta^{(2)}_{\rm GR}\left(1+{4\lgb\over 5r_{AdS}^2\sqrt{1-4\lgb}}\right)-{7\pi \sqrt{-p^2}\lgb\over 32r_{AdS}^{2}\sqrt{1-4\lgb}}\left({r_{AdS}\over b}\right)^{5}.
}

The phase shift at $\OO(\mu^3)$ is given by
\eqn\deltath{\eqalign{\delta^{(3)}=&\delta^{(3)}_{GR}\left(1+{{12\lgb}\over{7 r_{AdS}^{2}\sqrt{1-4\lgb}}}+{{16\lgb^{2}}\over{21 r_{AdS}^{4}(1-4\lgb)}}\right)\cr
&-\sqrt{-p^{2}}\left({r_{AdS}\over b}\right)^{7}\left({{495\pi \lgb}\over{512 r_{AdS}^{2}\sqrt{1-4\lgb}}} + {{55\pi \lgb^{2}}\over{128 r_{AdS}^{4}(1-4\lgb)}} \right)\cr
&+\sqrt{-p^{2}}\left({r_{AdS}\over b}\right)^{9}{{77\pi \lgb^2}\over{256 r_{AdS}^{4}(1-4\lgb)}},
}}
\noindent where 
\eqn\dtgr{\delta^{(3)}_{GR}={{231 r_{AdS}^{7}}\over{16 b^{7}}}\sqrt{-p^{2}}B\left({7\over 2},{3\over 2}\right){}_{2}F_{1}(1,{7\over 2},5,-{r_{AdS}^{2}\over b^{2}}).
}

\noindent By expanding \deltath\ in the large impact parameter limit, one again explicitly sees that the leading term does not depend on $\lgb$.

{\it\subsubsec{{ Inverse Fourier transform of the phase shift at $\OO(\mu^2)$.}}}
\noindent To make contact with the position space HHLL correlation function, one needs to perform a Fourier transform of the phase shift. 
According to \KulaxiziDXO , the HHLL four-point function in the Regge limit $\sqrt{-p^2}\gg 1$ is given by
\eqn\ft{\tilde{\GG}(x) = \int {d^{d}p\over(2\pi)^{d}}e^{ipx}\BB (p),
} 
\noindent where $\tilde{\GG}(x)=\langle\OO_{H}(x_{1})\OO_{L}(x_{2})\OO_{L}(x_{3})\OO_{H}(x_{4}) \rangle_{\rm Regge \ limit}$ and $\BB(p)=\BB_{0}(p)e^{i\delta}$. The factor $\BB_{0}(p)$ reproduces the disconnected correlator and it is given by 
\eqn\bzero{\BB_{0}(p)=C(\DL)\theta(p^{0})\theta(-p^{2})e^{i\pi \DL}(-p^{2})^{\DL-{d\over 2}},
}
\noindent with normalization
\eqn\norm{C(\DL)={{2^{d+1-2\DL}\pi^{1+{d\over 2}}}\over{\Gamma(\DL)\Gamma(\DL -{d\over 2}+1)}}.
}

\noindent We expand the integrand of \ft\ in powers of $\mu$ using \phs , explicitly
\eqn\expan{\eqalign{\BB(p)=\BB_{0}(p)\Bigg(& 1+\mu i \delta^{(1)} +\mu^{2} \left(i\delta^{(2)}-{1\over 2}{\delta^{(1)}}^{2} \right)\cr
  										   &+ \mu^{3}\left(i\delta^{(3)}-\delta^{(1)}\delta^{(2)}-{{i}\over{6}}{\delta^{(1)}}^{3}  \right) + \OO(\mu^{4})\Bigg).
}}
\noindent This generates an expansion for $\tilde{\GG}(x)$ from \ft\ as
\eqn\expg{\tilde{\GG}(x)=\sum_{k=0}^{\infty}\mu^{k}\tilde{\GG}^{(k)}(x).}

\noindent Let us start by studying the correlator at $\OO(\mu^{2})$. The imaginary part of the correlator in the Regge limit at this order comes from $i\delta^{(2)}$ in \expan\ while the real part comes from $-{1\over 2}{\delta^{(1)}}^{2}$.

Consider first the imaginary part. To perform the inverse Fourier transform it is convenient to first expand $\delta^{(2)}$  as follows:
\eqn\expandt{\eqalign{\delta^{(2)}=7\pi^{2}\sqrt{-p^2}\Bigg(&{5\over 2}\Pi_{5,3}(L)+\left({15\over 4}-{5\lgb \over{r_{AdS}^{2}\sqrt{1-4\lgb}}}\right)\Pi_{7,3}(L)\cr
											&+\left(5-{16\lgb \over{r_{AdS}^{2}\sqrt{1-4\lgb}}}\right)\Pi_{9,3}(L)+  \ldots \Bigg).
}}
In  \expandt{\ $b/r_{AdS} = \sinh(L)$ and 
\eqn\propagator{
\Pi_{\Delta-1;d-1}(x) = { \pi^{1 - {d \over 2}} \Gamma(\Delta-1) \over 2 \Gamma(\Delta - {d-2\over 2})} \ e^{-(\Delta-1)x} \ _{2} F_1({d \over 2} - 1, \Delta-1, \Delta- {d-2\over 2} , e^{- 2 x})  \,, }
the three-dimensional hyperbolic space propagator of a massive particle with mass square equal to $(\Delta-1)^2$. 
The dots in \expandt\ stand for  terms with hyperbolic space propagators  with $\Delta>10$.
We can now perform the inverse Fourier transform of \expandt\ with the help of eqs. (3.23) in \KulaxiziDXO\ and (3.4) in \KarlssonTXU .

The term which contains $\Pi_{5,3}(L)$ includes (after the inverse Fourier transform) the contribution of double-stress tensors with minimal twist $\tau=4$. As we have already shown it does not depend on $\lgb$, which we can also explicitly see in \expandt . The next term, that contains $\Pi_{7,3}(L)$, includes the contribution from the double-stress tensor operators of twist $\tau_{2,1}=6$. We can use this term to fix the coefficient $b_{14}$ which was left undetermined in \subg . Similar reasoning applies to all the higher-order terms in the large impact parameter expansion of \expandt. Namely, the term proportional to $\Pi_{2m+1,3}(L)$ is related to double-stress tensor operators of twist $\tau=2m$. 

Performing the inverse Fourier transform following \KarlssonTXU\ leads to
\eqn\ftone{\eqalign{&i{\rm Im}\left(\tilde{\GG}^{(2)}(\sigma,\rho)\right)=\int{d^{4}p\over{(2\pi)^{4}}}e^{ipx}\BB_{0}(p)i\delta^{(2)}={2i\over{\Gamma(\DL)\Gamma(\DL -1) \sigma^{2\DL +1}}}\cr 
     &\times\Big(a_{1} \Pi_{5,3}(\rho)\Gamma(\DL-2)\Gamma(\DL +2)+b_{1} \Pi_{7,3}(\rho)\Gamma(\DL-3)\Gamma(\DL +3) \cr
     &+c_{1} \Pi_{9,3}(\rho)\Gamma(\DL-4)\Gamma(\DL +4)+  \ldots \Big)+\ldots,
}}
\noindent where $a_{1}={35\over 2}\pi^{2}$, $b_{1}=7\pi^{2}\left({15\over 4}-{5\lgb \over{r_{AdS}^{2}\sqrt{1-4\lgb}}}\right)$ and $c_{1}=7\pi^{2}\left(5-{16\lgb \over{r_{AdS}^{2}\sqrt{1-4\lgb}}}\right)$. 
The ellipses outside the parenthesis in \ftone\ denote contributions due to double-trace operators in the T-channel that are not important for studying the stress tensor sector. The position space coordinates $\sigma$ and $\rho$ are defined as 
\eqn\coord{z=1-\sigma e^{\rho}, \qquad \zbar = 1-\sigma e^{-\rho}.} 
\noindent after the analytic continuation $z \to z e^{-2i\pi }$. Once more, notice that the dominant contribution in the large impact parameter regime, $\rho \to \infty$, comes from the factor $\Pi_{5,3}(\rho)$ in \ftone\ which exactly matches the imaginary part of the correlator \secondCorrIntro\ in \KulaxiziDXO .

\bigskip

{\it\subsubsec{{ Comparison with the HHLL correlation function in the lightcone limit at $\OO(\mu^2)$.}}}
\noindent A few simple steps are required before we can finally relate \ftone\ with the results of Section 3 and determine the OPE coefficients of the spin-2 double-stress tensor operators. As explained in \KulaxiziDXO, one has to analytically continue $\GG^{(2,1)}$, $\GG^{(2,2)}$ and $\GG^{(2,3)}$ (defined in Section 2) around the origin by taking $z \to z e^{-2i\pi }$ and expand the result in the vicinity of $\sigma\rightarrow 0$. The relevant term, which corresponds to the imaginary part of the correlator \subg\ as $\sigma \to 0$, reads:
\eqn\isubgr{\eqalign{i{\rm Im}\left((\sigma e^{-\rho})^{3-\DL}\GG^{(2,1)}(1-\sigma e^{\rho})\right)=& 7 i \pi  {e^{-7 \rho }\over \sigma ^{2 \Delta _L+1}} \Big(12600 b_{14}\cr 
&+{{\Delta _L \left(\Delta _L \left(\Delta _L \left(123-7 \Delta _L\right)+78\right)-12\right)}\over{16 \left(\Delta _L-3\right) \left(\Delta _L-2\right)}}\Big). }}
Comparing this with the subleading term of \ftone\ as $\rho \to \infty$, {\it i.e.},
\eqn\subi{\eqalign{i{\rm Im}\left(\tilde{\GG}^{(2)}(\sigma,\rho)\right)|_{e^{-7\rho}}&=-{{35 i \pi  e^{-7 \rho } \Delta _L \left(\Delta _L+1\right)  \left(8 \lgb +\Delta _L \left(4 \lgb -5 \sqrt{1-4 \lgb } r_{AdS}^2\right)\right)}\over{4 \sigma ^{2 \Delta _L+1} \sqrt{1-4 \lgb } r_{AdS}^2 \left(\Delta _L^2-5 \Delta _L+6\right)}}\cr
&+\ldots,}}
with the ellipses again denoting double-trace operators, allows one to obtain the following expression for the unknown parameter $b_{14}$:
\eqn\bof{\eqalign{b_{14}=P_{8,2}^{(2)}=&{{\Delta _L \left(\Delta _L \left(\Delta _L \left(7 \Delta _L-23\right)+22\right)+12\right)}\over{201600 \left(\Delta _L-3\right) \left(\Delta _L-2\right)}}\cr
&-{{\lgb  \Delta _L \left(\Delta _L+1\right) \left(\Delta _L+2\right)}\over{2520 \sqrt{1-4 \lgb } r_{AdS}^2 \left(\Delta _L-3\right) \left(\Delta _L-2\right)}}.
}}
Note that this precisely matches the OPE coefficient of the double trace operator of conformal dimension $\Delta=8$ and $s=2$ calculated in \FitzpatrickZQZ\ from gravity by other means. As expected, the OPE coefficient in \bof\ explicitly depends on $\lgb$.

Let us now go one step further and fix $P_{10,2}^{(2)}$ contributing to $\GG^{(2,2)}(z)$ through \rell. Analytically continuing \gsubsub\ and taking the limit $\sigma \to 0$, yields 
\eqn\issg{\eqalign{&i{\rm Im}\left((\sigma e^{-\rho})^{4-\DL}\GG^{(2,2)}(1-\sigma e^{\rho})\right)=i{{49}\over{400}}{{\pi  e^{-9 \rho }}\over{\sigma ^{2 \Delta _L+1}}}\Bigg(720000 b_{14}+11404800 {P_{10,2}^{(2)}\over \mu^{2}}\cr
&+{{\Delta _L \left(\Delta _L \left(\Delta _L \left(\Delta _L \left(6327-362 \Delta _L\right)+749\right)+12888\right)+12288\right)}\over{7 \left(\Delta _L-4\right) \left(\Delta _L-3\right) \left(\Delta _L-2\right)}}\Bigg).
}}
For reasons that will be explained later, we only consider here the imaginary part of the subsubleading term in the correlator. To extract the OPE data we need to compare \issg\ with the subsubleading contribution in the large impact parameter limit of \ftone, which is
\eqn\subsubi{\eqalign{i{\rm Im}\left(\tilde{\GG}^{(2)}(\sigma,\rho)\right)&|_{e^{-9\rho}}=i{7\over 4}{{\pi  e^{-9 \rho }}\over{\sigma ^{2 \Delta _L+1}}}\Bigg({{10 \Delta _L \left(\Delta _L+1\right)}\over{\Delta _L-2}}\cr
&-{{7 \Delta _L \left(\Delta _L+1\right) \left(\Delta _L+2\right) \left(16 \lgb +\Delta _L \left(12 \lgb -5 \sqrt{1-4 \lgb } r_{AdS}^2\right)\right)}\over{\sqrt{1-4 \lgb } r_{AdS}^2 \left(\Delta _L-4\right) \left(\Delta _L-3\right) \left(\Delta _L-2\right)}}\Bigg).
}}
Substituting \bof\ in \issg\ and matching to \subsubi\  enables us to determine the OPE coefficient $P_{10,2}^{(2)}$, 
\eqn\opetwo{\eqalign{P_{10,2}^{(2)}=&{{\Delta _L \left(\Delta _L \left(\Delta _L \left(\Delta _L \left(187 \Delta _L-552\right)+901\right)+1012\right)+912\right)}\over{79833600 \left(\Delta _L-4\right) \left(\Delta _L-3\right) \left(\Delta _L-2\right)}}\cr
&-{{\lgb  \Delta _L \left(\Delta _L+1\right) \left(\Delta _L+2\right) \left(\Delta _L+3\right)}\over{12474 \sqrt{1-4 \lgb } r_{AdS}^2 \left(\Delta _L-4\right) \left(\Delta _L-3\right) \left(\Delta _L-2\right)}}.
}}
This precisely matches the one calculated in \FitzpatrickZQZ . 

Similarly, one can match the CFT expression for ${\rm Im}\left((\sigma e^{-\rho})^{5-\DL}\GG^{(2,3)}(1-\sigma e^{\rho})\right)$ in \gsubsubsub, to its gravitational  counterpart ${\rm Im}\left(\GG^{(2)}(x)\right)|_{e^{-11\rho}}$, by expanding \expandt\ and \ftone\ up to $\OO(e^{-11\rho})$. This allows one to additionally determine $P_{12,2}^{(2)}$ in \relll\
\eqn\opet{\eqalign{P_{12,2}^{(2)}=&{{\Delta _L \left(\Delta _L+1\right) \left(\Delta _L \left(\Delta _L \left(\Delta _L \left(6721 \Delta _L-15603\right)+46474\right)+100828\right)+143760\right)}\over{44396352000 \left(\Delta _L-5\right) \left(\Delta _L-4\right) \left(\Delta _L-3\right) \left(\Delta _L-2\right)}}\cr
&-{{5 \lgb  \Delta _L \left(\Delta _L+1\right) \left(\Delta _L+2\right) \left(\Delta _L+3\right) \left(\Delta _L+4\right)}\over{453024 \sqrt{1-4 \lgb } r_{AdS}^2 \left(\Delta _L-5\right) \left(\Delta _L-4\right) \left(\Delta _L-3\right) \left(\Delta _L-2\right)}}.
}}

Notice that we did not use the real part of $\tilde{\GG}^{(2)}(\sigma,\rho)$, which comes from the term $-{1\over 2}{\delta^{(1)}}^{2}$ in \expan\ and behaves as $\sigma^{-2\DL-2}$ for $\sigma \to 0$. This term matches the corresponding term with the same $\sigma$ behavior in the correlator. It does not give us any new information, because it is independent of the OPE coefficients of operators with spin $s=0,2$.

{\it\subsubsec{{ Extracting OPE data from the gravitational phase shift at $\OO(\mu^3)$.}}}
\noindent Let us now consider the $\OO(\mu^{3})$ terms in the correlator. Focusing on the gravity side, we start by performing an inverse Fourier transform. \expan\ instructs us to consider three terms $i\delta^{(3)}$, $\delta^{(1)}\delta^{(2)}$ and $i(\delta^{(1)})^{3}$, which give rise to terms that behave as $\sigma^{-2\DL-1}$, $\sigma^{-2\DL-2}$ and $\sigma^{-2\DL-3}$, respectively. Performing the relevant computations, we observe that $\delta^{(1)}\delta^{(2)}$ and $i(\delta^{(1)})^{3}$ do not provide additional information because the corresponding terms in the correlators are already fixed by bootstrap (these terms simply give us an extra consistency check). Focusing on the inverse Fourier transform of $i\delta^{(3)}$, we expand \deltath\  in terms of the hyperbolic space propagators, $\Pi_{m,3}(L)$,
\eqn\exptd{\delta^{(3)}=\sqrt{-p^2}\Big(a_{2} \Pi_{7,3}(L) + b_{2} \Pi_{9,3}(L) + c_{2} \Pi_{11,3}(L) + \ldots \Big), 
}
where
\eqn\coefftwo{\eqalign{a_{2}&={1155\over 8}\pi^{2},\cr
b_{2}&=231 \pi ^2 \left(-{{3 \lgb }\over{r_{AdS}^2\sqrt{1-4 \lgb }}}+2\right),\cr
c_{2}&={{231 \pi ^2}\over{8}} \left({{32 \lgb ^2}\over{r_{AdS}^4(1-4 \lgb)}}-{{120 \lgb }\over{r_{AdS}^2\sqrt{1-4 \lgb }}}+35\right),
}}
which leads to
\eqn\fttwo{\eqalign{&i{\rm Im}\left(\tilde{\GG}^{(3)}(\sigma,\rho)\right)\Big|_{{1\over{\sigma^{2\DL +1}}}}=\int{d^{4}p\over{(2\pi)^{4}}}e^{ipx}\BB_{0}(p)i\delta^{(3)}={2i\over{\Gamma(\DL)\Gamma(\DL -1) \sigma^{2\DL +1}}}\cr 
     &\times\Big(a_{2} \Pi_{7,3}(\rho)\Gamma(\DL-3)\Gamma(\DL +3)+b_{2} \Pi_{9,3}(\rho)\Gamma(\DL-4)\Gamma(\DL +4) \cr
     &+c_{2} \Pi_{11,3}(\rho)\Gamma(\DL-5)\Gamma(\DL +5)+  \ldots \Big) +{\rm double \ traces},
}}
The leading and subleading contributions in the large impact parameter limit $\rho \to \infty$ come from $\Pi_{7,3}(\rho)$ and $\Pi_{9,3}(\rho)$ and behave as ${{i\pi e^{-7\rho}}\over{\sigma^{2\DL +1}}}$ and ${{i\pi e^{-9\rho}}\over{\sigma^{2\DL +1}}}$, respectively. They are precisely matched by the relevant terms in \gt\ in the vicinity of $\sigma\to 0$ after analytic continuation \KarlssonTXU . This is another sanity check of the procedure described herein, since these terms do not incorporate contributions from spin-2 operators.

To extract further OPE data, we proceed to match the subsubleading correction of \fttwo\ in the large impact parameter limit to the term in \gtsubsub\ which behaves as $\sim {{i\pi e^{-11\rho}}\over{\sigma^{2\DL +1}}}$. This allows us to determine the coefficient $e_{115}=P_{12,2}^{(3)}$  in \gtsubsub\ which corresponds to the OPE coefficient of the triple-stress tensors of spin $s=2$ with conformal dimension $\Delta=12$: 
\eqn\eoof{\eqalign{e_{115}=&-{117\Delta_L^6-439\Delta_L^5+407\Delta_L^4+859\Delta_L^3+202\Delta_L^2+696\Delta_L\over 172972800(\DL-2)(\DL-3)(\DL-4)(\DL-5)}\cr
&-{\lgb(143\Delta_L^6-231\Delta_L^5-3597\Delta_L^4-9489\Delta_L^3-11186\Delta_L^2-4920\Delta_L)\over 43243200r_{AdS}^2\sqrt{1-4\lgb}(\DL-2)(\DL-3)(\DL-4)(\DL-5)}\cr
&+{\lgb^2\DL(\DL+1)(\DL+2)(\DL+3)(\DL+4)\over 24024r_{AdS}^4(1-4\lgb)(\DL-2)(\DL-3)(\DL-4)(\DL-5)}\cr
&+P_{8,0}^{(2)}{76+{400\over \DL-5}+11\DL\over 1320}.
}}
Notice that $e_{115}$ is not completely determined by the above procedure since the spin-0 OPE data, $P_{8,0}^{(2)}$, is not fixed.
Summarising, we conclude that we are able to fix all coefficients in the ansatz except those that correspond to the OPE coefficients of operators of spin-0. However, using the expression for $P_{8,0}^{(2)}$ found in \FitzpatrickZQZ\ one finds
\eqn\ptwelvetwo{\eqalign{
  P_{12,2}^{(3)} &=\cr
  &{1001\Delta_L^7-6864\Delta_L^6+12615\Delta_L^5-3980\Delta_L^4-6156\Delta_L^3-11736\Delta_L^2-1440\Delta_L\over 3459456000 (\DL-2)(\DL-3)(\DL-4)(\DL-5)}\cr
  &-{\lgb(143\Delta_L^6-206\Delta_L^5-1631\Delta_L^4-3622\Delta_L^3-3540\Delta_L^2-1200\Delta_L)\over 28828800r_{AdS}^2\sqrt{1-4\lgb}(\DL-2)(\DL-3)(\DL-4)(\DL-5)}\cr
  &+{\lgb^2\DL(\DL+1)(\DL+2)(\DL+3)(\DL+4)\over 24024r_{AdS}^4(1-4\lgb)(\DL-2)(\DL-3)(\DL-4)(\DL-5)}.
}}

\newsec{Lorentzian inversion formula}
\noindent It was recently shown in \LiZBA\ that one can obtain the OPE coefficients of minimal twist double and triple-stress tensors using the Lorentzian inversion formula. Here, we review this method and show how it can be generalized to extract the OPE coefficients of twist-six double-stress tensors. In principle, it can also be generalized to multi-stress tensors of arbitrarily high twist.

\subsec{Twist-four double-stress tensors}
\noindent Consider the correlation function
\eqn\corrr{(w\wbar)^{-\DL}G(w,\wbar)=\langle \OH(\infty)\OH(1)\OL(w,\wbar)\OL(0)\rangle .
}
The Lorentzian inversion formula is given by \refs{\HuotLIF ,\WittenLIF}
\eqn\opei{\eqalign{c(\tau , \beta)={{1+(-1)^{{{\beta - \tau}\over 2}}}\over 2}&\kappa_{\beta}\int_{0}^{1}dwd\wbar \mu^{(0,0)}(w,\wbar)\cr
&\times g_{-\tau + 2(d-1),{{\beta + \tau}\over 2} -d+1}^{(0,0)}(w,\wbar){\rm dDisc}[G(w,\wbar)],
}}
where
\eqn\muu{\mu^{(0,0)}(w,\wbar)={{{|w-\wbar |}^{d-2}}\over{(w\wbar)^{d}}},
}
\eqn\kapa{\kappa_{\beta}={{\Gamma({\beta\over 2})^4}\over{2\pi^{2}\Gamma(\beta)\Gamma(\beta -1)}},
}
\noindent where $\tau = \Delta -s$ and $\beta = \Delta +s$. Here $g_{\tau, s}^{(0,0)}$ is a conformal block given with $\Delta\rightarrow s+d-1$ and $s\rightarrow\Delta-d+1$ and  in $d=4$ is given  by \blocksFourD . Moreover, dDisc denotes the double-discontinuity of
$G(w, \bar w)$  in \corrr, which is equal to the correlator of a double commutator, and it is given by
\eqn\dD{{\rm dDisc}[G(w,\wbar)]=G(w,\wbar)-{1\over 2}G^{\circlearrowleft}({w,\wbar})-{1\over 2}G^{\circlearrowright}({w,\wbar})\,.
}
Here $G^{\circlearrowleft}$ and $G^{\circlearrowright}$ correspond to the same correlator analytically continued in two different ways around $w=1$, namely $(1-w)\to (1-w)e^{\pm 2\pi i}$. The OPE data, $P_{ {\tau'+\beta\over 2},{ \beta-\tau'\over 2}}$, can be extracted from $c(\tau, \beta)$ via\foot{In principle there is an extra term in this relation when $\tau-d=0,1,2,\ldots$ \CaronHuotVEP, however, it vanishes in the cases considered.}
\eqn\ree{P_{  {\tau'+\beta\over 2},{\beta-\tau'\over 2} } =-{\rm Res}_{\tau = \tau '}c(\tau , \beta),
}
\noindent where $\tau'$ and $\beta$ denote the twist and conformal spin of operators in the physical spectrum of the theory exchanged in the channel $\OL\times\OL\to\OO_{\tau',J'}\to\OH\times\OH$.

We would like to apply the Lorentzian inversion formula to the HHLL correlator to extract the OPE data of the double-stress tensors. To this end, we will use information of the correlator from the channel where $\OO_{H}\OO_{L}$ merge. The function $G(z,\zbar)$ can be obtained from $\GG(z,\zbar)$ via
\eqn\lifdd{G(w,\wbar)=(w\wbar)^{\DL}\GG(1-w,1-\wbar).
}

To apply the Lorentzian inversion formula we first need to calculate $\GG(z,\zbar)$ using the S-channel operator product expansion \sChannel . First, let us start with the leading contribution of $\GG(z,\zbar)$ in the lightcone limit $\zbar \to 1$ at $\OO(\mu^{2})$. These give the leading contributions when $\wbar \to 0$ in $G(w,\wbar)$. After the integration with respect to $\wbar$ in \opei , these contributions fix the position of the pole and residue of $c(\tau, \beta)$ that corresponds to lowest-twist double-stress tensors. Subleading contributions in $\zbar \to 1$ (or $\wbar \to 0$) only create new poles, without changing the residue of existing ones, therefore, they do not affect the OPE coefficients of lowest-twist operators. The leading contribution in the $(1-\zbar)$-expansion comes from the leading contribution of the $1/l$-expansion of the S-channel OPE data. Only the term proportional to $\log^{2}(z)$ contributes to the double-discontinuity and we denote it by $\GG^{(2)}(z,\zbar)\big|_{\log^{2}(z)}$. The number in the superscript denotes the power of $\mu$ in which we are working.  Substituting in to \sChCont\ equations \expansion , \mfto , \pfunc\ and \SpinBehavior, we find that
\eqn\gtz{\eqalign{\GG^{(2)}(z,\zbar)\big|_{\log^{2}(z)}=\log ^2(z \zbar)\int_{0}^{\infty}dl\sum_{n=0}^{\infty}&{{(z \zbar)^n l^{\Delta _L-3} \left(z^{l+1}-\zbar^{l+1}\right)  \Gamma \left(n+\Delta _L-1\right)}\over{8(z-\zbar) \Gamma (n+1)  \Gamma \left(\Delta _L-1\right) \Gamma \left(\Delta _L\right)}}\times\cr
&\left(\left(\gamma^{(1,0)}_{n}\right)^{2}+\OO\left({1\over l}\right)\right).
}}
\noindent In the lightcone limit, the dominant contribution to this expression comes from operators with large spin $l\gg 1$, we can, therefore, approximate the sum over $l$ by an integral. Note that only $\OO(\mu)$ OPE data, {\it i.e.}, $\gamma_{n}^{(1,0)}$, appears in \gtz. Using \gammaonefull\ we evaluate \gtz\ and collect the leading term as $\zbar \to 1$,
\eqn\gtzz{\eqalign{&\GG^{(2)}(z,\zbar)\big|_{\log^{2}(z)}=\log^{2}(z){{(1-\zbar)^{2-\Delta _L}(1-z)^{-\Delta _L-4}}\over{32 \left(\Delta _L-2\right)}}\times \cr
&\Delta _L \left(\Delta _L \left((z (z+4)+1)^2 \Delta _L+z (z (54-(z-28) z)+28)-1\right)+72 z^2\right)+\OO\left((1-\zbar)^{3-\DL}\right).
}}
With the help of \lifdd\ one obtains 
\eqn\gg{\eqalign{&G^{(2)}(w,\wbar)\big|_{\log^{2}(1-w)}={{\DL \wbar^{2} \log^{2}(1-w)}\over{32w^{4}(\DL -2)}}\times\cr
&\left(\Delta _L \left(((w -6) w +6)^2 \Delta _L-w (w (w (w +24)-132)+216)+108\right)+72 (w -1)^2\right) + \OO(\wbar^{3}),
}}
\noindent which agrees with (4.12) in \LiZBA . Now, it is easy to see that 
\eqn\ddo{\eqalign{&{\rm dDisc}[G^{(2)}(w,\wbar)]={{\pi \wbar^{2} \DL}\over{8 w^{4}(\DL -2)}}\times \cr
& \left(\Delta _L \left(((w -6) w +6)^2 \Delta _L-w (w (w (w +24)-132)+216)+108\right)+72 (w -1)^2 \right)+ \OO(\wbar^{3}).
}}

\noindent To compute the integral \opei\ we substitute
\eqn\im{\mu^{(0,0)}(w,\wbar)={1\over{w^{2}\wbar^{4}}} + \OO \left({1\over\wbar^{3}}\right),
}
\eqn\fcb{g_{-\tau + 2(d-1),{{\tau + \beta}\over 2} -d+1}^{(0,0)}(w,\wbar)=\wbar^{3-{{\tau}\over 2}}\left(f_{{\beta}\over2}(1-w)+\OO(\wbar)\right),
}
\noindent valid in the lightcone limit $\wbar \to 0$ (or $\zbar \to 1$), and set $(-1)^{{{\beta - \tau}\over 2}}=1$ since only even-spin operators contribute. Combining the above we arrive at the following expression for $c(\tau, \beta)$
\eqn\cone{\eqalign{&c_{0}(\tau, \beta)=-{{\sqrt{\pi } 2^{-\beta +1} \Delta _L \Gamma \left({{\beta}\over{2}}\right)}\over{(\tau -4) (\beta -10) (\beta -6) (\beta -2) \beta (\beta +4)}}\times\cr
&\Bigg({{384 \left(\Delta _L-7\right) \Delta _L+4608}\over{(\beta +8) \left(\Delta _L-2\right) \Gamma \left({{1}\over{2}} (\beta -1)\right)}}+{{(\beta -2) \beta \Delta _L \left((\beta -2) \beta \left(\Delta _L-1\right)-56 \Delta _L+200\right)}\over{(\beta +8) \left(\Delta _L-2\right) \Gamma \left({{1}\over{2}} (\beta -1)\right)}}\Bigg),
}}
where the subscript denotes that this result is obtained in the leading order of the lightcone expansion. The OPE coefficients of the minimal-twist double-stress tensors are given by
\eqn\re{P^{(2)}_{ {\beta\over 2}+2,{\beta\over 2}-2}=-{\rm Res}_{\tau = 4}c_{0}(\tau , \beta),
}
\noindent where $\beta=12+4 \ell$, $\ell\geq 0$, and are in precise agreement with (1.6) in \KulaxiziTKD\ and (4.15) in \LiZBA.

\subsec{Twist-six double-stress tensors}

Here we use the same method to obtain the OPE coefficients of double-stress tensors with twist $\tau_{2,1}=6$. We first need to compute the subleading contribution in the lightcone limit to eqs. \ddo , \im\ and \fcb . Specifically, the integration measure 
\eqn\imm{\mu^{(0,0)}(w,\wbar)={1\over{w^{2}\wbar^{4}}}-{2\over{w^{3}\wbar ^3}}+\OO\left(\wbar^{-2}\right),
}
\noindent and the conformal block,
\eqn\cbww{\eqalign{g_{-\tau + 2(d-1),{{\tau + \beta}\over 2} -d+1}^{(0,0)}(w,\wbar)=&\wbar^{3-{{\tau}\over 2}}f_{{\beta}\over 2}(1-w)\left(1+\wbar\left(1-{{\tau}\over{4}}+{1\over w}\right)+\OO(\wbar^{2})\right),
}}
were obtained from the explicit expressions given in \muu\ and \blocksFourD .

To evaluate the subleading term in ${\rm dDisc}[G^{(2)}(w,\wbar)]$ we reconsider the S-channel computation. Similarly to the case of leading twist, only the part of the correlator with $\log^{2}(z)$ contributes to the discontinuity. However, we now have to include the subleading corrections in the $1/l$-expansion of the S-channel OPE data. With the help of \sChCont , \expansion , \pfunc , \SpinBehavior\ and \mfto\ one finds that 
\eqn\ggg{\eqalign{&\GG^{(2)}(z,\zbar)\big|_{\log^{2}(z)}={{\log^{2}(z \zbar)}\over{16(z-\zbar)\Gamma(\DL)\Gamma(\DL -1)}}\sum_{n=0}^{\infty}(z\zbar)^{n}{{\Gamma(\DL -1 +n)}\over{\Gamma(n+1)}}\int_{0}^{\infty}dl\cr
&l^{\Delta _L-6} \left(z^{l+1}-\zbar^{l+1}\right) \left(2 (l-2 n)+\Delta _L \left(\Delta _L+2 n-1\right)\right)\left(l\gamma^{(1,0)}_{n}+\gamma^{(1,1)}_{n} \right)^{2}+\OO\left(l^{\Delta_L-7}\right) \,.
}}
To proceed, one evaluates \ggg\ using \gammaonefull\ and collects the leading and subleading contributions as $\zbar \to 1$, which behave as $(1-\zbar)^{2-\DL}$ and $(1-\zbar)^{3-\DL}$ respectively. Using  \lifdd\  it is then simple to obtain $G^{(2)}(w,\wbar)\big|_{\log^{2}(1-w)}$ up to $\OO(\wbar ^4)$ and evaluate its double-discontinuity:
\eqn\ddd{\eqalign{&{\rm dDisc}[G^{(2)}(w,\wbar)]=-{{\pi ^2 \wbar^2 \Delta _L}\over{8 w ^5 \left(\Delta _L-3\right) \left(\Delta _L-2\right)}}\Big(-3 w^5 \Delta _L-72 w^4 \Delta _L+324 w^3 \Delta _L\cr
&-504 w^2 \Delta _L+252 w \Delta _L+216 w^3-432 w^2+216 w + 4 w^5 \Delta _L^2-12 w^4 \Delta _L^2+12 w^3 \Delta _L^2\cr
&-36 w \Delta _L^3-w^5 \Delta _L^3+12 w^4 \Delta _L^3-48 w^3 \Delta _L^3+72 w^2 \Delta _L^3 + \wbar(-144 \Delta _L+612  w \Delta _L+216  w^3\cr
& -432  w^2+216  w -w^5 \Delta _L-52 w^4 \Delta _L+324 w^3 \Delta _L-744 w^2 \Delta _L   +  540 w \Delta _L^2-216 \Delta _L^2 \cr
&-72 \Delta _L^3+w^5 \Delta _L^2-18 w^4 \Delta _L^2+156 w^3 \Delta _L^2-456 w^2 \Delta _L^2+144 w \Delta _L^3-2 w^4 \Delta _L^3+24 w^3 \Delta _L^3\cr
&-96 w^2 \Delta _L^3)\Big)+\OO(\wbar^{4})\,.
}}
Substituting \imm, \cbww\ and \ddd\ in \opei\ and integrating leads to an analytic expression for $c(\tau, \beta)$. The relevant part of this expression -- the one with non-zero residue at $\tau=6$ -- turns out to be:
\eqn\cccc{\eqalign{c_{1}(\tau,\beta)=&-{{ 2^{4-\beta }\sqrt{\pi } \Gamma \left({{\beta }\over{2}}\right) \Delta _L}\over{(\beta -12) (\beta -8) (\beta -4)  (\tau -10) (\tau -8) (\tau -6) (\tau -4)}}\cr
&\times\Bigg({{\beta ^4 \Delta _L-4 \beta ^3 \Delta _L-68 \beta ^2 \Delta _L-960 \beta  \Delta _L^2+144 \beta  \Delta _L-14976 \Delta _L^2}\over{(\beta +2) (\beta +6) (\beta +10)\Gamma \left({{\beta -1}\over{2}}\right) \left(\Delta _L-3\right) \left(\Delta _L-2\right)}}\cr
&+{{\beta ^4 \Delta _L^3-2 \beta ^4 \Delta _L^2-4 \beta ^3 \Delta _L^3+8 \beta ^3 \Delta _L^2-116 \beta ^2 \Delta _L^3+472 \beta ^2 \Delta _L^2}\over{(\beta +2) (\beta +6) (\beta +10)\Gamma \left({{\beta -1}\over{2}}\right) \left(\Delta _L-3\right) \left(\Delta _L-2\right)}}\cr
&+{{240 \beta  \Delta _L^3+2304 \Delta _L^3+19584 \Delta _L+13824  }\over{(\beta +2) (\beta +6) (\beta +10)\Gamma \left({{\beta -1}\over{2}}\right) \left(\Delta _L-3\right) \left(\Delta _L-2\right)}}\Bigg)+\ldots,
}}
\noindent where the ellipsis stands for the terms with zero residue at $\tau=6$ and $1$ in the subscript denotes that this expression is obtained in the subleading order of the lightcone expansion.

It is now straightforward to read off the OPE coefficients of double-stress tensors with twist $\tau_{2,1}=6$ from
\eqn\opeeee{P_{{\beta\over 2}+3,{\beta\over 2}-3}^{(2)}=-{\rm Res}_{\tau = 6}c_1(\tau, \beta).
} 

\noindent For $\beta=14+4\ell$ \opecnew\ is reproduced. It is already stated in Section 3 that this formula does not reproduce the right OPE coefficient $P_{8,2}^{(2)}$ for $\ell=-1$. Thus, we explicitly see that the Lorentzian inversion formula does not allow us to obtain the OPE data of spin-2 double-stress tensors with twist $\tau=6$. 

In general, to determine for which operators at $\OO(\mu^{k})$ the Lorentzian inversion formula can be applied, one has to consider the behavior of the correlator in the Regge limit. At $\OO(\mu^{k})$ the correlator in the Regge limit behaves like $1/\sigma^{2 \Delta_L+k}$. Therefore, the Lorentzian inversion formula correctly produces the OPE coefficients of multi-stress tensor operators with spin $s>k+1$. Accordingly, already at order $\OO(\mu^{3})$, fixing the OPE coefficients by combining an ansatz for the correlator with the crossing symmetry (or Lorentzian inversion formula) appears more powerful than the Lorentzian inversion formula alone. Namely, we were able to fix the OPE coefficients of spin-4 operators and the one with twist $\tau=8$ is given by (D.1), while using the Lorentzian inversion formula one can only fix the OPE coefficients of operators with spin $s>4$.


\newsec{Discussion}
In this paper, we consider the stress tensor sector of a four-point function of pairwise identical scalars in a class of CFTs with a large central charge.
It is completely determined by the OPE coefficients of multi-stress tensor operators, which can 
be read off the result for a heavy-heavy-light-light correlator.
The stress tensor sector of the HHLL correlator is naturally expanded perturbatively in $\mu\sim{\DH\over C_T}$, where $\DH$ is the scaling dimension of the heavy operator. The power of $\mu$ counts the number of stress tensors within the exchanged multi-stress tensor operators. 
By further expanding the HHLL stress tensor sector in the lightcone limit, the multi-stress tensor operators can be organized into sectors of different twists. 
Similarly to the minimal-twist sector, combining an appropriate ansatz with the lightcone bootstrap, we show that the contribution from the non-minimal twist multi-stress tensors is almost completely determined.
Unlike the minimal twist case, a few coefficients are not fixed by the bootstrap -- these correspond to the OPE coefficients of multi-stress tensors with spin $s=0,2$. 

An extra check is provided by applying the Lorentzian OPE inversion formula (see \LiZBA\ for an earlier application of the inversion formula in this context).
 It gives the same results but has less predictive power than the ansatz.

The OPE coefficients for double-stress tensors are particularly simple and we provide closed-form expressions for those with twist $\tau=4,6,8,10$ and any spin greater than $2$. All of these OPE coefficients are 
completely fixed by the bootstrap.
This is related to their independence of the  higher-derivative terms in the dual bulk gravitational Lagrangian.
The OPE coefficients for double-stress tensors with spin $s=0,2$ are not fixed by the bootstrap and do depend on such higher derivative terms. 
It is interesting that at the level of double-stress tensors, only the OPE coefficients with spin $s=0,2$ are not fixed by the bootstrap (non-universal). 
On the other hand,  all non-minimal twist triple-stress tensor OPE coefficients are non-universal\foot{Here we use universality and ``fixed by the bootstrap" terms interchangeably. However, it remains to be determined what is the universality class and whether it  the same as the set of 
unitary holographic theories.} .

Assuming a holographic dual, we show that the OPE coefficients for spin-2 multi-stress tensors can be determined by studying the large impact parameter regime of the Regge limit,
following \refs{\KulaxiziDXO,\KarlssonQFI,\KarlssonTXU}
 (modulo the spin zero OPE data). This is done explicitly in Einstein Hilbert+Gauss-Bonnet gravity. 
 Some of these OPE coefficients are known \FitzpatrickZQZ\ and agree with our results. 

It would be interesting if one could compute the spin zero and spin two multi stress tensor  OPE coefficients with  CFT techniques.
Perhaps the conglomeration approach first discussed in \FitzpatrickDM\ or the more recent work \refs{\CarmiCUB-\BissiKKX} will be useful in this direction. 

The regime of applicability of the ansatz (and the exact meaning of universality) used in this paper remains unsettled
(the ansatz seems to work in holographic CFTs, but does it also apply for other CFTs with a large central charge?).
This question appears already in the leading twist case studied in \KarlssonDBD.
To address this issue, it would be interesting to investigate the OPE coefficients of multi-stress tensors in CFTs with a large central charge, but
not necessarily holographic. 
A related question is  the existence of an infinite-dimensional algebra responsible for the form of the near-lightcone correlator.
In two dimensions the relevant algebra is simply the Virasoro algebra.
The Virasoro vacuum block has been computed in several ways \refs{\FitzpatrickVUA\HijanoRLA\FitzpatrickZHA\HijanoQJA\FitzpatrickFOA\CollierEXN-\BeskenJYW}.
Recently an algebraic way of  reproducing the near lightcone contribution of the stress tensor was discussed in
\HuangYCS\ -- it would be interesting to investigate this further.

Returning to holographic theories, one interesting question would be to understand the critical behavior of geodesics
in the vicinity of the circular light orbit, recently studied in \BianchiDES, from the CFT point of view.
This corresponds to the situation where the deflection angle is very large.
The deflection angle $\varphi$ in asymptotically flat Schwarzschild geometries is supposed to be related to the eikonal
phase $\delta$ via
\eqn\angleeik{   2 \sin {\varphi \over 2} = -{1\over E}  {\p \delta \over \p b}    }
where $E$ is the incoming particle energy and $b$ is the impact parameter (see e.g. \BernGJJ\ for a recent discussion).
This agrees with eq. (E.1) for small deflection angles, but deviations might occur for  large deflection angles.
It would be interesting to investigate this further.

\bigskip
\bigskip

\noindent {\bf Acknowledgments}: 
We thank G.S. Ng, K. Sen  and A. Zhiboedov for useful discussions.
The work of R.K. and A.P. is supported in part by an Irish Research Council Laureate Award. The work of P.T. is supported in part by an Ussher Fellowship Award. 

\appendix{A}{Linear relations between products of $f_{a}(z)$ functions}

Here we list some linear relations between products of the $f_{a}(z)$ functions used in the main text.
\eqn\relap{f_{1}(z)f_{4}(z)+{1\over 15}f_{3}(z)f_{4}(z)-{4\over 63}f_{2}(z)f_{5}(z)-f_{2}(z)f_{3}(z)=0,
}
\eqn\lindeptwoap{\eqalign{ {308\over 25}f_{2}^{2}(z)-{308\over 25}f_{1}(z)f_{3}(z)+{5929\over 375}f_{3}^{2}(z)-{2673\over 2500}f_{4}^{2}(z)-{396\over 25}f_{1}(z)f_{5}(z)+f_{2}(z)f_{6}(z)&=0,\cr 
245f_{2}^{2}(z)-245f_{1}(z)f_{3}(z)-{7\over 12}f_{3}^{2}(z)-{81\over 80}f_{4}^{2}(z)+f_{3}(z)f_{5}(z)&=0,\cr
{140\over 9}f_{2}^{2}(z)-{140\over 9}f_{1}(z)f_{3}(z)-{28\over 27}f_{3}^{2}(z)+f_{2}(z)f_{4}(z)&=0,
}}
\eqn\reltap{\eqalign{  {{3991680}\over{16000}}f_2(z) f_3(z)-{{99}\over{125}} f_4(z) f_3(z)+f_6(z) f_3(z)-{{6237}\over{25}} f_1(z) f_4(z)-{{891}\over{875}} f_4(z) f_5(z) &=  0, \cr f_2(z) f_7(z)+ {{7007}\over{500}} f_2(z) f_3(z)+{{39611}\over{2500}} f_4(z) f_3(z) - {{7007}\over{500}} f_1(z) f_4(z)-{{4719}\over{4375}}f_4(z) f_5(z)& \cr  -{{143}\over{9}} f_1(z) f_6(z) &= 0,
}}
\eqn\relo{\eqalign{&-{{1}\over{15}} f_6(z) f_2(z){}^2+{{297}\over{4375}}f_4(z){}^2 f_2(z)+f_1(z) f_5(z) f_2(z)+{{44}\over{625}} f_3(z) f_5(z) f_2(z)\cr
&+{{9}\over{143}} f_1(z) f_7(z) f_2(z)-{{44}\over{625}} f_3(z){}^2 f_4(z)-{{297}\over{4375}}f_1(z) f_4(z) f_5(z)-f_1(z) f_1(z) f_6(z)=0,
}}
\eqn\reltw{\eqalign{&-f_6(z) f_1(z){}^2+f_3(z) f_4(z) f_1(z)-{{297}\over{4375}}f_4(z) f_5(z) f_1(z)+{{9}\over{143}} f_2(z) f_7(z) f_1(z)\cr
&+{{9}\over{2500}}f_2(z) f_4(z){}^2-{{7}\over{1875}}f_3(z){}^2 f_4(z)+{{7}\over{1875}}f_2(z) f_3(z) f_5(z)-{{7 }\over{1980}}f_2(z){}^2 f_6(z)=0,
}}
\eqn\relth{\eqalign{&-f_6(z) f_1(z){}^2+{{9}\over{143}} f_2(z) f_7(z) f_1(z)-{{297}\over{4375}}f_4(z) f_5(z) f_1(z)+{{297}\over{4375}}f_2(z) f_4(z){}^2\cr
&+f_2(z){}^2 f_4(z)-{{44}\over{625}} f_3(z){}^2 f_4(z)+{{7}\over{1875}}f_2(z) f_3(z) f_5(z)-{{7 }\over{1980}}f_2(z){}^2 f_6(z)=0,
}}
\eqn\relfo{\eqalign{&-f_6(z) f_1(z){}^2+{{9}\over{143}} f_2(z) f_7(z) f_1(z)-{{297}\over{4375}} f_4(z) f_5(z) f_1(z)+f_2(z) f_3(z){}^2 \cr
&+{{9}\over{2500}}f_2(z) f_4(z){}^2-{{44}\over{625}} f_3(z){}^2 f_4(z)+{{2647}\over{39375}} f_2(z) f_3(z) f_5(z)-{{7}\over{1980}} f_2(z){}^2 f_6(z)=0,
}}
\eqn\relfi{\eqalign{-f_6(z) f_2(z){}^2+{{891}\over{875}} f_4(z){}^2 f_2(z)&+{{132}\over{125}} f_3(z) f_5(z) f_2(z)-{{132}\over{125}} f_3(z){}^2 f_4(z)\cr
&-{{891}\over{875}} f_1(z) f_4(z) f_5(z)+f_1(z) f_3(z) f_6(z)=0,
}}

\appendix{B}{Coefficients in $\GG^{(3,1)}(z)$}
Here we list the coefficients in $\GG^{(3,1)}(z)$:
\eqn\solts{\eqalign{b_{116}=&-{{\Delta _L \left(\Delta _L+3\right) \left(\Delta _L \left(\Delta _L \left(\Delta _L \left(1001 \Delta _L+387\right)-4326\right)+13828\right)+5040\right)}\over{10378368000 \left(\Delta _L-4\right) \left(\Delta _L-3\right) \left(\Delta _L-2\right)}}\cr
							&+{{b_{14} \left(\Delta _L \left(143 \Delta _L+427\right)+540\right)}\over{17160 \left(\Delta _L-4\right)}},\cr
					c_{118}=&{{7 \left(\Delta _L+3\right) \left(604800 b_{14} \left(\Delta _L^2-5 \Delta _L+6\right)+\Delta _L \left(-21 \Delta _L^3+229 \Delta _L^2+414 \Delta _L+284\right)\right)}\over{856627200 \left(\Delta _L^3-9 \Delta _L^2+26 \Delta _L-24\right)}},\cr
					c_{127}=&{{\Delta _L \left(\Delta _L \left(\Delta _L \left(\Delta _L \left(\Delta _L \left(14 \Delta _L-15\right)+6040\right)-36125\right)-75814\right)-49620\right)}\over{2306304000 \left(\Delta _L-4\right) \left(\Delta _L-3\right) \left(\Delta _L-2\right)}}\cr
							&-{{3 b_{14} \left(\Delta _L \left(2 \Delta _L+3\right)+135\right)}\over{11440 \left(\Delta _L-4\right)}},\cr
					c_{145}=&{{\Delta _L \left(\Delta _L \left(\Delta _L \left(\Delta _L \left(\left(32680-1183 \Delta _L\right) \Delta _L-183605\right)+34900\right)+570808\right)+436440\right)}\over{47040000000 \left(\Delta _L-4\right) \left(\Delta _L-3\right) \left(\Delta _L-2\right)}}\cr
							&+{{3 b_{14} \left(\Delta _L \left(257 \Delta _L-2227\right)+510\right)}\over{700000 \left(\Delta _L-4\right)}},\cr
					c_{226}=&{{\Delta _L \left(\Delta _L \left(\Delta _L \left(\Delta _L \left(\left(40020-1337 \Delta _L\right) \Delta _L-274845\right)+96350\right)+2323212\right)+1910160\right)}\over{71850240000 \left(\Delta _L-4\right) \left(\Delta _L-3\right) \left(\Delta _L-2\right)}}\cr
							&+{{b_{14} \left(\Delta _L \left(22 \Delta _L-267\right)+960\right)}\over{39600 \left(\Delta _L-4\right)}},\cr
					c_{235}=&{{b_{14} \left(\left(10283-1153 \Delta _L\right) \Delta _L-5790\right)}\over{900000 \left(\Delta _L-4\right)}}+{{\Delta _L \left(51463 \Delta _L^5-846480 \Delta _L^4+1320405 \Delta _L^3\right)}\over{1632960000000 \left(\Delta _L^3-9 \Delta _L^2+26 \Delta _L-24\right)}}\cr
							&+{{\Delta _L \left(22381100 \Delta _L^2-46886088 \Delta _L-46446840\right)}\over{1632960000000 \left(\Delta _L^3-9 \Delta _L^2+26 \Delta _L-24\right)}},\cr
					c_{244}=&{{\Delta _L \left(\Delta _L \left(\Delta _L \left(\Delta _L \left(\Delta _L \left(1337 \Delta _L-32145\right)+160095\right)+19525\right)-266712\right)-182160\right)}\over{70560000000 \left(\Delta _L-4\right) \left(\Delta _L-3\right) \left(\Delta _L-2\right)}}\cr
							&+{{9 b_{14} \left(\Delta _L \left(71-11 \Delta _L\right)+270\right)}\over{175000 \left(\Delta _L-4\right)}},\cr
					c_{334}=&{{\Delta _L \left(\Delta _L \left(\Delta _L \left(\Delta _L \left(\Delta _L \left(509 \Delta _L-1515\right)+83415\right)-808325\right)+823116\right)+902880\right)}\over{90720000000 \left(\Delta _L-4\right) \left(\Delta _L-3\right) \left(\Delta _L-2\right)}}\cr
							&+{{b_{14} \left(\Delta _L \left(11 \Delta _L-71\right)-270\right)}\over{18750 \left(\Delta _L-4\right)}}.
}}

\appendix{C}{Coefficients in $\GG^{(3,2)}(z)$}
Here we list the coefficients in $\GG^{(3,2)}(z)$:
\eqn\coefthree{\eqalign{g_{119}&={{g_{13} \left(7 \Delta _L \left(128-77 \Delta _L\right)+6720\right)}\over{16409250 \left(\Delta _L-5\right)}}+{{49 b_{14} \left(\Delta _L \left(\Delta _L \left(170-11 \Delta _L\right)+981\right)+1620\right)}\over{16409250 \left(\Delta _L-5\right) \left(\Delta _L-4\right)}}\cr
&+{{196 e_{115}}\over{49725}}+{{539 \Delta _L^7-15386 \Delta _L^6+54215 \Delta _L^5+951510 \Delta _L^4+2911426 \Delta _L^3}\over{472586400000 \left(\Delta _L-5\right) \left(\Delta _L-4\right) \left(\Delta _L-3\right) \left(\Delta _L-2\right)}}\cr
&+{{98 e_{15} \left(\Delta _L+4\right)}\over{16575 \left(\Delta _L-5\right)}}+{{3737076 \Delta _L^2+1779120 \Delta _L}\over{472586400000 \left(\Delta _L-5\right) \left(\Delta _L-4\right) \left(\Delta _L-3\right) \left(\Delta _L-2\right)}},\cr
g_{128}&=-{{7 g_{13} \left(\Delta _L \left(4 \Delta _L-469\right)+930\right)}\over{12355200 \left(\Delta _L-5\right)}}-{{7 b_{14} \left(\Delta _L \left(22 \Delta _L^2-64 \Delta _L+4197\right)+11745\right)}\over{6177600 \left(\Delta _L-5\right) \left(\Delta _L-4\right)}}\cr
&+{{462 \Delta _L^7-24203 \Delta _L^6+1044630 \Delta _L^5-3466005 \Delta _L^4-24181012 \Delta _L^3-39855972 \Delta _L^2}\over{1779148800000 \left(\Delta _L-5\right) \left(\Delta _L-4\right) \left(\Delta _L-3\right) \left(\Delta _L-2\right)}}\cr
&-{{49 e_{15} \left(\Delta _L \left(\Delta _L+2\right)+102\right)}\over{93600 \left(\Delta _L-5\right)}}-{{61201 \Delta _L}\over{4942080000 \left(\Delta _L-5\right) \left(\Delta _L-4\right) \left(\Delta _L-3\right) \left(\Delta _L-2\right)}},\cr
g_{155}&={{11 e_{15} \left(\Delta _L \left(278 \Delta _L-2789\right)+126\right)}\over{2756250 \left(\Delta _L-5\right)}}+{{11 g_{13} \left(\Delta _L \left(2279 \Delta _L-7400\right)-8370\right)}\over{231525000 \left(\Delta _L-5\right)}}\cr
&-{{3146 e_{115}}\over{275625}}+{{b_{14} \left(12063 \Delta _L^3-88048 \Delta _L^2-131165 \Delta _L+196110\right)}\over{77175000 \left(\Delta _L-5\right) \left(\Delta _L-4\right)}}\cr
&+{{-244401285 \Delta _L^4+853023786 \Delta _L^3+2178372216 \Delta _L^2+1399907880 \Delta _L}\over{233377200000000 \left(\Delta _L-5\right) \left(\Delta _L-4\right) \left(\Delta _L-3\right) \left(\Delta _L-2\right)}}\cr
&+{{-1406986 \Delta _L^7+28367309 \Delta _L^6-123035140 \Delta _L^5}\over{233377200000000 \left(\Delta _L-5\right) \left(\Delta _L-4\right) \left(\Delta _L-3\right) \left(\Delta _L-2\right)}},\cr
g_{227}&={{e_{15} \left(\Delta _L \left(52 \Delta _L-751\right)+3234\right)}\over{93600 \left(\Delta _L-5\right)}}-{{e_{115}}\over{240}}+{{g_{13} \left(\Delta _L \left(1051 \Delta _L-12370\right)-52530\right)}\over{86486400 \left(\Delta _L-5\right)}}\cr
&+{{b_{14} \left(\Delta _L \left(\Delta _L \left(3131 \Delta _L-33896\right)-62985\right)+1236870\right)}\over{86486400 \left(\Delta _L-5\right) \left(\Delta _L-4\right)}}\cr
&+{{-213549 \Delta _L^7+6031106 \Delta _L^6-23990385 \Delta _L^5-205647690 \Delta _L^4}\over{87178291200000 \left(\Delta _L-5\right) \left(\Delta _L-4\right) \left(\Delta _L-3\right) \left(\Delta _L-2\right)}}\cr
&+{{853227874 \Delta _L^3+2135805744 \Delta _L^2+1445776920 \Delta _L}\over{87178291200000 \left(\Delta _L-5\right) \left(\Delta _L-4\right) \left(\Delta _L-3\right) \left(\Delta _L-2\right)}},\cr
g_{236}&={{e_{15} \left(\left(15074-1223 \Delta _L\right) \Delta _L-39816\right)}\over{6804000 \left(\Delta _L-5\right)}}+{{g_{13} \left(\Delta _L \left(186926 \Delta _L-1951295\right)+5891220\right)}\over{6286896000 \left(\Delta _L-5\right)}}\cr
&+{{143 e_{115}}\over{340200}}+{{b_{14} \left(\Delta _L \left(\Delta _L \left(23001 \Delta _L-469741\right)+3383740\right)-7782480\right)}\over{1047816000 \left(\Delta _L-5\right) \left(\Delta _L-4\right)}}\cr
&-{{9324749 \Delta _L^7-433851406 \Delta _L^6+5233472135 \Delta _L^5-21967190310 \Delta _L^4}\over{6337191168000000 \left(\Delta _L-5\right) \left(\Delta _L-4\right) \left(\Delta _L-3\right) \left(\Delta _L-2\right)}}\cr
&-{{10644674676 \Delta _L^3+72859312056 \Delta _L^2+65903302080 \Delta _L}\over{6337191168000000 \left(\Delta _L-5\right) \left(\Delta _L-4\right) \left(\Delta _L-3\right) \left(\Delta _L-2\right)}},
}}

\eqn\coeftt{\eqalign{g_{245}&=-{{99 e_{15} \left(\Delta _L \left(83 \Delta _L-754\right)-1064\right)}\over{4900000 \left(\Delta _L-5\right)}}+{{g_{13} \left(73 \Delta _L \left(275-274 \Delta _L\right)+170060\right)}\over{137200000 \left(\Delta _L-5\right)}}\cr
&+{{5577 e_{115}}\over{245000}}+{{b_{14} \left(\Delta _L \left(\Delta _L \left(79801-14981 \Delta _L\right)+410980\right)-55320\right)}\over{68600000 \left(\Delta _L-5\right) \left(\Delta _L-4\right)}}\cr
&+{{1300313 \Delta _L^7-22489422 \Delta _L^6+63989995 \Delta _L^5+399569530 \Delta _L^4}\over{138297600000000 \left(\Delta _L-5\right) \left(\Delta _L-4\right) \left(\Delta _L-3\right) \left(\Delta _L-2\right)}}\cr
&+{{-690996588 \Delta _L^3-2276065528 \Delta _L^2-1491467040 \Delta _L}\over{138297600000000 \left(\Delta _L-5\right) \left(\Delta _L-4\right) \left(\Delta _L-3\right) \left(\Delta _L-2\right)}},\cr
g_{335}&={{1144 e_{115}}\over{5315625}}+{{g_{13} \left(\Delta _L \left(6426275-894839 \Delta _L\right)+685170\right)}\over{17860500000 \left(\Delta _L-5\right)}}\cr
&-{{11 e_{15} \left(\Delta _L \left(11143 \Delta _L-143659\right)+451206\right)}\over{212625000 \left(\Delta _L-5\right)}}\cr
&-{{b_{14} \left(\Delta _L \left(\Delta _L \left(446853 \Delta _L-4788638\right)+4992635\right)+44234910\right)}\over{5953500000 \left(\Delta _L-5\right) \left(\Delta _L-4\right)}}\cr
&+{{43544683 \Delta _L^7-877022702 \Delta _L^6+4877336920 \Delta _L^5-1356232020 \Delta _L^4}\over{9001692000000000 \left(\Delta _L-5\right) \left(\Delta _L-4\right) \left(\Delta _L-3\right) \left(\Delta _L-2\right)}}\cr
&+{{-28767381333 \Delta _L^3-34411007748 \Delta _L^2-12217009140 \Delta _L}\over{9001692000000000 \left(\Delta _L-5\right) \left(\Delta _L-4\right) \left(\Delta _L-3\right) \left(\Delta _L-2\right)}},\cr
g_{344}&={{11 e_{15} \left(\Delta _L \left(278 \Delta _L-2789\right)+126\right)}\over{2625000 \left(\Delta _L-5\right)}}+{{g_{13} \left(\Delta _L \left(17194 \Delta _L-10525\right)-249570\right)}\over{220500000 \left(\Delta _L-5\right)}}\cr
&-{{1573 e_{115}}\over{131250}}+{{b_{14} \left(\Delta _L \left(\Delta _L \left(9438 \Delta _L-48673\right)-325415\right)+511110\right)}\over{73500000 \left(\Delta _L-5\right) \left(\Delta _L-4\right)}}\cr
&+{{-1593347 \Delta _L^7+27045868 \Delta _L^6-6670280 \Delta _L^5-1193221320 \Delta _L^4}\over{444528000000000 \left(\Delta _L-5\right) \left(\Delta _L-4\right) \left(\Delta _L-3\right) \left(\Delta _L-2\right)}}\cr
&+{{1878076947 \Delta _L^3+5698801932 \Delta _L^2+3877115760 \Delta _L}\over{444528000000000 \left(\Delta _L-5\right) \left(\Delta _L-4\right) \left(\Delta _L-3\right) \left(\Delta _L-2\right)}}.
}}

\eqn\doos{\eqalign{d_{117}&=-{9\over 220}e_{115}+{{84+\DL (53+13\DL)}\over{1560(\DL -5)}}e_{15}+{{13 \Delta _L \left(209 \Delta _L+409\right)+8340}\over{7207200 \left(\Delta _L-5\right)}}g_{13}\cr
&-{{4641 \Delta _L^7+22727 \Delta _L^6+44901 \Delta _L^5+67569 \Delta _L^4+519742 \Delta _L^3}\over{290594304000 \left(\Delta _L-5\right) \left(\Delta _L-4\right) \left(\Delta _L-3\right) \left(\Delta _L-2\right)}}\cr
&-{{828876 \Delta _L^2+333648 \Delta _L}\over{290594304000 \left(\Delta _L-5\right) \left(\Delta _L-4\right) \left(\Delta _L-3\right) \left(\Delta _L-2\right)}}\cr
&+{{\Delta _L \left(\Delta _L \left(5317 \Delta _L+18140\right)+68763\right)+69660}\over{7207200 \left(\Delta _L-5\right) \left(\Delta _L-4\right)}}b_{14}.
}}

\appendix{D}{OPE coefficients of twist-eight triple-stress tensors}
Here we list a few OPE coefficients of twist-eight triple-stress tensors which are found using \koeficijenti:
\eqn\opt{\eqalign{P^{(3)}_{12,4}&={{P_{8,2}^{(2)} \left(\Delta _L \left(143 \Delta _L+427\right)+540\right)}\over{17160 \left(\Delta _L-4\right)}}\cr
&-{{1001 \Delta _L^6+3390 \Delta _L^5-3165 \Delta _L^4+850 \Delta _L^3+46524 \Delta _L^2+15120 \Delta _L}\over{10378368000 \left(\Delta _L-4\right) \left(\Delta _L-3\right) \left(\Delta _L-2\right)}},
}}
\eqn\optt{\eqalign{&P^{(3)}_{14,6}={{9 P_{8,2}^{(2)} \left(\Delta _L \left(13 \Delta _L+11\right)+12\right)}\over{544544 \left(\Delta _L-4\right)}}\cr
&+{{7917 \Delta _L^6+38174 \Delta _L^5+140795 \Delta _L^4+266390 \Delta _L^3+253908 \Delta _L^2+97776 \Delta _L}\over{548900352000 \left(\Delta _L-4\right) \left(\Delta _L-3\right) \left(\Delta _L-2\right)}},
}}
\eqn\opttt{\eqalign{&P^{(3)}_{16,8}={{5 P_{8,2}^{(2)} \left(\Delta _L \left(17 \Delta _L+2\right)+6\right)}\over{9876048 \left(\Delta _L-4\right)}}\cr
&+{{362593 \Delta _L^6+881129 \Delta _L^5+2782307 \Delta _L^4+4155839 \Delta _L^3+3518084 \Delta _L^2+1198176 \Delta _L}\over{438022480896000 \left(\Delta _L-4\right) \left(\Delta _L-3\right) \left(\Delta _L-2\right)}},
}}
\eqn\optttt{\eqalign{P^{(3)}_{18,10}=&{{P_{8,2}^{(2)} \left(\Delta _L \left(323 \Delta _L-77\right)+54\right)}\over{823727520 \left(\Delta _L-4\right)}}+{{17413253 \Delta _L^6+23717684 \Delta _L^5+79039447 \Delta _L^4}\over{377794389772800000 \left(\Delta _L-4\right) \left(\Delta _L-3\right) \left(\Delta _L-2\right)}}\cr
&+{{92754344 \Delta _L^3+73231064 \Delta _L^2+22535496 \Delta _L}\over{377794389772800000 \left(\Delta _L-4\right) \left(\Delta _L-3\right) \left(\Delta _L-2\right)}}.
}}

Assuming Einstein-Hilbert + Gauss-Bonnet gravity in the bulk, the OPE coefficient $P_{8,2}^{(2)}$ was derived in \bof\ and can be inserted in \opt-\optttt.  

\appendix{E}{Derivation of the deflection angle from the phase shift.}

\noindent Here we simply show that the bulk phase shift, defined as $\delta=p^t (\Delta t)-p^\phi (\Delta \phi)$ in \KulaxiziDXO\  is consistent with the standard equation relating the eikonal phase and the scattering angle 
\eqn\defanglea{{\p\delta\over\p b}= - p^t\,\Delta\phi}
obtained with the use of the stationary phase approximation for small scattering angles. Our discussion is focused on asymptotically flat space. In this case, the formulas in classical gravity which provide the deflection angle and the time delay are:
\eqn\deltatphidef{\eqalign{\Delta t=2\int_{r_0}^\infty {dr\over f\sqrt{1-{b^2 f\over r^2}}}\cr
\Delta \phi=2 b \int_{r_0}^\infty {dr\over r^2\sqrt{1-{b^2 f\over r^2}}}\,.
}}
They can be obtained from eq.(2.9) in \KulaxiziDXO\  with the substitution ${p^\phi\over p^t}=b$ (and the appropriate definition of the blackening factor $f(r)$).
Note that the equation for the turning point of the geodesic, $r_0$, reduces in Schwarzchild geometry to:
\eqn\rnotdef{1-{b^2\over r^2 f(r_0)}=0}
Defining the bulk phase shift via $\delta=p^t (\Delta t)-p^\phi (\Delta \phi)$, leads to
\eqn\derphaseshift{\delta=p^t (\Delta t)-p^\phi (\Delta \phi)=p^t \left(\Delta t-b \Delta \phi\right)=2 p^t  \int_{r_0}^\infty {dr\over f} \sqrt{1-{b^2 f\over r^2}} 
}
Differentiating the bulk phase shift with respect to the impact parameter yields:
\eqn\derivativeb{{\p \delta\over\p b}=-2 p^t\, b \int_{r_0}^\infty {dr\over r^2\sqrt{1-{b^2 f\over r^2}}}-2 p^t {1\over f(r_0)} \sqrt{1-{b^2 f(r_0)\over r_0^2}} = - p^t (\Delta \phi)\,,
}
where to arrive at the last equality we used the equation satisfied by the turning point $r_0$. Hence,
\eqn\deltadifb{\Delta\phi=-{1\over p^t}{\p \delta\over \p b}\,.}
Finally note that assuming the classical relation $J\equiv p_\phi=b \, p^t$, the deflection angle can also be computed through
\eqn\deltaphiang{\Delta\phi=-{\p\delta\over\p J}\,.}

\appendix{F}{Anomalous dimensions and phase shift at $\OO(\mu^2)$}

We give explicit expressions for $\gamma^{(2,0)}_{n}$, $\gamma^{(2,1)}_{n}$ and $\gamma^{(2,2)}_{n}$ from \SpinBehavior\
\eqn\gammatz{\eqalign{\gamma^{(2,0)}_{n}=&-{{1}\over{8}} \left(\Delta _L-1\right) \Delta _L \left(4 \Delta _L+1\right)-{{51}\over{4}} n^2 \left(\Delta _L-1\right)\cr
&+{{1}\over{4}} n \left(3 \left(11-7 \Delta _L\right) \Delta _L-17\right)-{{17}\over{2}}n^3,
}}

\eqn\gammaaa{\eqalign{\gamma^{(2,1)}_{n}&={{1}\over{8 \sqrt{1-4 \lgb } r_{AdS}^2}}\Big(\lgb(4 \Delta _L^4+8 \Delta _L^3-4 \Delta _L^2-8 \Delta _L+560 n^3 \Delta _L+360 n^2 \Delta _L^2\cr
&-600 n^2 \Delta _L+80 n \Delta _L^3-120 n \Delta _L^2+200 n \Delta _L+280 n^4-560 n^3+440 n^2-160 n)\cr
&r_{AdS}^2\sqrt{1-4\lgb}(-\Delta _L^4+6 \Delta _L^3-5 \Delta _L^2-140 n^3 \Delta _L-90 n^2 \Delta _L^2+354 n^2 \Delta _L\cr
&-20 n \Delta _L^3+114 n \Delta _L^2-182 n \Delta _L-70 n^4+276 n^3-314 n^2+108 n)\Big),
}}

\eqn\gammaaaa{\eqalign{\gamma^{(2,2)}_{n}&={{1}\over{8 \sqrt{1-4 \lgb } r_{AdS}^2}}\Big(\lgb(16 \Delta _L^3-16 \Delta _L+840 n^4 \Delta _L+720 n^3 \Delta _L^2-2880 n^3 \Delta _L\cr
&+240 n^2 \Delta _L^3-1440 n^2 \Delta _L^2+3720 n^2 \Delta _L+24 n \Delta _L^4-192 n \Delta _L^3+888 n \Delta _L^2-1536 n \Delta _L\cr
&+336 n^5-1680 n^4+3440 n^3-3120 n^2+1024 n)+r_{AdS}^2\sqrt{1-4\lgb}(3 \Delta _L^4-10 \Delta _L^3\cr
&+6 \Delta _L^2+\Delta _L+420 n^3 \Delta _L+270 n^2 \Delta _L^2-876 n^2 \Delta _L+60 n \Delta _L^3-264 n \Delta _L^2+420 n \Delta _L\cr
&+210 n^4-704 n^3+756 n^2-262 n)\Big),
}}
\noindent where we use the expression for $P_{8,0}^{(2)}$, found in \FitzpatrickZQZ , to fix $\gamma^{(2,2)}_n$. If one considers limit $1\ll l,n\ll\DH$ one gets
\eqn\gammamm{\gamma^{(2)}_{n,l}\lntoinf -{{17 n^3}\over{2 l^2}}-{{35 n^4}\over{4 l^3}}\left(1-{{4\lgb}\over{r_{AdS}^2\sqrt{1-4\lgb}}}\right)+{{42 \lgb n^5}\over{\sqrt{1-4 \lgb } l^4 r_{AdS}^2}}+\ldots ,
}
\noindent where $\ldots$ denote terms that come from $\gamma^{(2,m)}_{n}$ for $m>2$ and they have higher powers of $l$ (and $n$) as well as terms that are subleading in the given limit and behave as $\OO(1)$.

By using the following relations from \refs{\KulaxiziDXO,\KarlssonQFI}
\eqn\relations{\sinh(L)={b\over r_{AdS}},\quad \cosh(L)={{p^{+}+p^{-}}\over{2\sqrt{-p^2}}},
}
\noindent with
\eqn\relone{-p^{2}=p^{+}p^{-}, \quad p^{+}=2h, \quad p^{-}=2\bar{h},
}
\noindent where
\eqn\reltwo{h=n+l,\quad \bar{h}=n,
}
\noindent one obtains $\delta^{(2)}$ from \SecondRewr\ in terms of the S-channel variables $n$ and $l$
\eqn\dtll{\delta^{(2)} ={{7 \pi  n^3}\over{4 l^5}}\left(10 l^3+5 l^2 n-{{4 n (5 l^2 + 6 l n + 2 n^2)\lgb}\over{r_{AdS}^2\sqrt{1-4\lgb}}}\right).
}

From \gammaonefull\ and \SpinBehavior\ one concludes that the leading behavior in the large-$l$ and large-$n$ limit ($1\ll n,l\ll \Delta_{H}$) of $\gamma^{(1)}_{n,l}$ is
\eqn\gammaonlll{\gamma^{(1)}_{n,l} \lntoinf -{{3  n^2}\over{l}}+\OO\left(1\right).
}
Now, one can evaluate (1.5) from \KarlssonQFI\ using \dtll\ and \gammaonlll\
\eqn\eee{\eqalign{\gamma^{(2)}_{n,l}\lntoinf &-{\delta^{(2)}\over \pi}+{1\over 2}\gamma^{(1)}_{n,l}\partial_{n}\gamma^{(1)}_{n,l}\cr
\lntoinf &-{{17 n^3}\over{2 l^2}}-{{35 n^4}\over{4 l^3}}\left(1-{{4\lgb}\over{r_{AdS}^2\sqrt{1-4\lgb}}}\right)+{{42 \lgb n^5}\over{\sqrt{1-4 \lgb } l^4 r_{AdS}^2}}\cr
&+{{14 \lgb  n^6}\over{\sqrt{1-4 \lgb } l^5 r_{AdS}^2}}+\OO(1).
}}
\noindent We see that first three terms in \eee\ precisely matches with terms in \gammamm , which explicitly confirms the validity of relation (1.5) in \KarlssonQFI . One would expect that term ${{14 \lgb  n^6}\over{\sqrt{1-4 \lgb } l^5 r_{AdS}^2}}$ is due to ${{\gamma^{(2,3)}_{n}}\over {l^{5}}}$ in \SpinBehavior , while all other ${{\gamma^{(2,k)}_{n}}\over{l^{2+k}}}$, for $k>3$, should behave as $\OO(1)$ in $1\ll n,l\ll \Delta_{H}$ limit for (1.5) from \KarlssonQFI\ to be true.

\bigskip
\bigskip

\listrefs

\bye